\documentclass[journal]{IEEEtran}
\usepackage{amssymb}
\usepackage{booktabs}

\usepackage{cite}
\ifCLASSINFOpdf
  \usepackage[pdftex]{graphicx}
\else
  \usepackage[dvips]{graphicx}
\fi
\usepackage{amsmath}
\usepackage{rotating}
\usepackage{multirow}
\usepackage{url}
\usepackage{color}
\usepackage{diagbox}

\usepackage{algorithm, algpseudocode}

\newcommand{\argmin}{\mathop{\rm argmin}\limits}

\begin{document}
\title{A General Destriping Framework for Remote Sensing Images Using Flatness Constraint}

\author{Kazuki~Naganuma,~\IEEEmembership{Student~Member,~IEEE,}
        Shunsuke~Ono,~\IEEEmembership{Member,~IEEE,}
\thanks{K. Naganuma is with the Department of Computer Science, Tokyo Institute of Technology, Yokohama, 226-8503, Japan (e-mail: naganuma.k.aa@m.titech.ac.jp).}% <-this % stops a space
\thanks{S. Ono is with the Department of Computer Science, Tokyo Institute of Technology, Yokohama, 226-8503, Japan (e-mail: ono@c.titech.ac.jp).}% <-this % stops a space
\thanks{This work was supported in part by JST CREST under Grant JPMJCR1662 and JPMJCR1666, in part by JST PRESTO under Grant JPMJPR21C4, and in part by JSPS KAKENHI under Grant 20H02145, 19H04135, and 18H05413.}}


\markboth{IEEE TRANSACTIONS ON GEOSCIENCE AND REMOTE SENSING}%
{Shell \MakeLowercase{\textit{et al.}}: Bare Demo of IEEEtran.cls for IEEE Journals}



\maketitle

\begin{abstract}
Removing stripe noise, i.e., destriping, from remote sensing images is an essential task in terms of visual quality and subsequent processing. Most existing destriping methods are designed by combining a particular image regularization with a stripe noise characterization that cooperates with the regularization, which precludes us to examine and activate different regularizations to adapt to various target images. To resolve this, two requirements need to be considered: a general framework that can handle a variety of image regularizations in destriping, and a strong stripe noise characterization that can consistently capture the nature of stripe noise, regardless of the choice of image regularization. To this end, this paper proposes a general destriping framework using a newly-introduced stripe noise characterization, named \textit{flatness constraint}, where we can handle various regularization functions in a unified manner. Specifically, we formulate the destriping problem as a nonsmooth convex optimization problem involving a general form of image regularization and the flatness constraint. The constraint mathematically models that the intensity of each stripe is constant along one direction, resulting in a strong characterization of stripe noise. For solving the optimization problem, we also develop an efficient algorithm based on a diagonally preconditioned primal-dual splitting algorithm (DP-PDS), which can automatically adjust the stepsizes. The effectiveness of our framework is demonstrated through destriping experiments, where we comprehensively compare combinations of a variety of image regularizations and stripe noise characterizations using hyperspectral images (HSI) and infrared (IR) videos.
\end{abstract}


% \fi
\IEEEpeerreviewmaketitle

\section{Introduction}
	\IEEEPARstart{R}{emote Sensing Images} such as hyperspectral images (HSIs) and infrared (IR) videos offer various applications, including mineral detection, earth observation, agriculture, astronomical imaging, automatic target recognition, and video surveillance~\cite{HSI_unmixing_review,IR_astronomy,TIV_visual_analysis}. Such data, however, are often contaminated by \textit{stripe noise}, which is mainly due to differences in the nonuniform response of individual detectors, calibration error, and dark currents~\cite{LRMRSSTV,cause_stripe_2,IR_destriping_2020}. Stripe noise not only degrades visual quality but also seriously affects subsequent processing, such as hyperspectral unmixing~\cite{HSI_unmixing_review,unmixing2014}, HSI classification~\cite{classification2016,classification2017,classification_Gao,classification_Hong}, and IR video target recognition~\cite{target_recognition_Li2020}. Therefore, stripe noise removal, i.e., destriping, has been an important research topic in remote sensing and related fields.

	In the past decades, a large number of destriping methods have been proposed. Filtering-based approaches are widely used due to their simplicity~\cite{Filtering_based_destriping_2009,Filtering_based_destriping_2016,Filtering_based_destriping_2020}. They effectively remove periodic stripe noise by truncating the specific stripe components in a Fourier or wavelet data domain. However, these approaches are limited in use since they assume that stripe noise is periodic and can be identified from the power spectrum. Deep learning-based approaches have also been studied~\cite{Learning_based_Destriping_of_IR_Image_2017,Learning_based_Destriping_of_IR_Image_2019_1,Learning_based_Destriping_of_IR_Image_2019_2,unsupervised_based,Sidorov_2019_ICCV}. They can automatically extract the nature of desirable data to remove stripe noise by learned neural networks, but have difficulties, such as domain dependence, a lack of a learning dataset, and excessive removal of image structures (e.g., textures and singular features)~\cite{P_Liu_2019,T_X_Jiang_2022}.

Among many destriping techniques, optimization-based approaches have received much attention. In these approaches, desirable data and stripe noise are modeled by functions that capture their nature, and then both are simultaneously estimated by solving an optimization problem involving the functions. These approaches adopt some form of regularization to characterize desirable data, including piecewise smoothness~\cite{HTV,SSTV,ASSTV,TVGSC,l0l1HTV,video_TV}, low-rankness~\cite{LRMR,LRTR,NonLRMA,S_Ye_2019,3DTNN}, self-similarity~\cite{NLR_CPTD_2019}, sparse representation~\cite{Shen_2014,Wang_2017}, and combinations of these regularizations~\cite{GLSSTV,SSTV_LRTD}. 

The characterization of stripe noise is as essential as image regularization in destriping. Existing stripe noise characterizations can be roughly classified into a sparsity-based model~\cite{LRMR,LRTR,S_Ye_2019,GS_CHEN_2017,3DTNN}, a low-rank-based model~\cite{NN_char,LR_Hu_2021}, and a total variation (TV) model~\cite{TV_Liu_2016,Kind_of_Stripes,gradient_constraint}. The first model relies on the fact that stripe noise in observed data is (group) sparsely distributed. The second model characterizes stripe noise as low rankness since stripe noise has a strong low-rank structure~\cite{NN_char}. The third model captures the vertical (or horizontal) smoothness of stripe noise using TV regularization.

Many of the existing destriping methods are designed by combining a particular image regularization with a stripe noise characterization that cooperates with the regularization. Since the function used for image regularization is often also used for stripe noise characterization, these methods carefully select the function used for stripe noise characterization so that it does not conflict with the adopted image regularization. For example, destriping methods using the low-rank based model employ TV as the image regularization~\cite{NN_char}, but in the case of destriping methods with the TV model, only the horizontal TV is used to regularize the image~\cite{gradient_constraint,Kind_of_Stripes} because the vertical TV is used to characterize the stripe noise.

On the other hand, it would be very beneficial to establish a destriping framework that can handle various image regularizations in a unified manner, so that we can select a regularization that matches each target image of different nature. In fact, a number of image regularization techniques have been proposed for remote sensing images. Typical examples are hyperspectral image regularization techniques based on spatio-spectral smoothness and correlation~\cite{HTV,SSTV,LRMR,LRTR,3DTNN}. In the case of video data, there are also many regularization techniques that consider moving objects~\cite{moving_object1,moving_object2,moving_object3}. Combining multiple regularizations is also a promising strategy~\cite{SSTV_LRTD,video_TV_NN}.

\begin{figure}[!t]
	\begin{minipage}{0.3\hsize}
		\centerline{\includegraphics[width=\hsize]{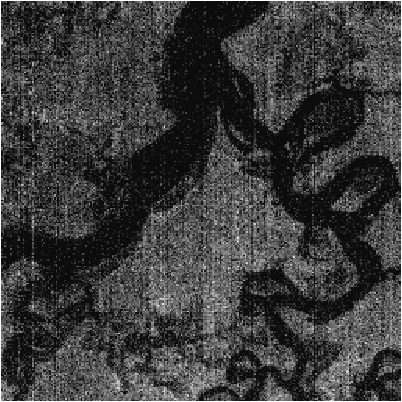}} %画像
	\end{minipage}
	\begin{minipage}{0.3\hsize}
		\centerline{\includegraphics[width=\hsize]{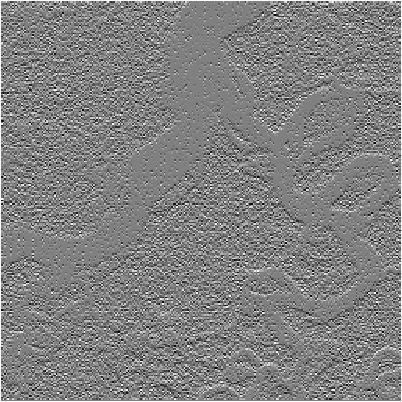}} %画像
	\end{minipage}
	\begin{minipage}{0.3\hsize}
		\centerline{\includegraphics[width=\hsize]{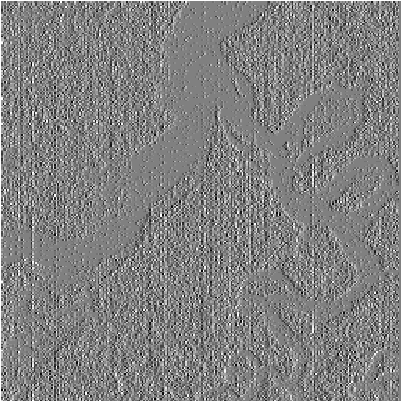}} %画像
	\end{minipage}
	
	\vspace{1mm}
	
	\begin{minipage}{0.3\hsize}
		\centerline{\small{(a1)}}
	\end{minipage}
	\begin{minipage}{0.3\hsize}
		\centerline{\small{(a2)}}
	\end{minipage}
	\begin{minipage}{0.3\hsize}
		\centerline{\small{(a3)}}
	\end{minipage}
	
	\vspace{1mm}
	
	\begin{minipage}{0.3\hsize}
		\centerline{\includegraphics[width=\hsize]{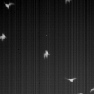}} %画像
	\end{minipage}
	\begin{minipage}{0.3\hsize}
		\centerline{\includegraphics[width=\hsize]{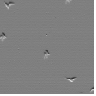}} %画像
	\end{minipage}
	\begin{minipage}{0.3\hsize}
		\centerline{\includegraphics[width=\hsize]{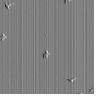}} %画像
	\end{minipage}
	
	\vspace{1mm}
	
	\begin{minipage}{0.3\hsize}
		\centerline{\small{(b1)}}
	\end{minipage}
	\begin{minipage}{0.3\hsize}
		\centerline{\small{(b2)}}
	\end{minipage}
	\begin{minipage}{0.3\hsize}
		\centerline{\small{(b3)}}
	\end{minipage}
	
	\caption{Spatial flatness of stripe noise on HSI and IR video data. (a1) Striped HSI data. (a2) Vertical gradient. (a3) Horizontal gradient. (b1) Striped IR video data. (b2) Vertical gradient. (b3) Horizontal gradient.}
	\label{fig:stripe_noise_analysis}
\end{figure}

In order to achieve the aforementioned unified framework, two requirements need to be considered: 1) a general formulation and algorithm that can handle a variety of image regularizations, and 2) a strong stripe noise characterization that can consistently capture the nature of stripe noise, regardless of the choice of image regularization.

Based on the above discussion, this paper proposes a general destriping framework for remote sensing images. First, we formulate destriping as a constrained convex optimization problem involving a general form of image regularization and a newly introduced strong stripe noise characterization. Second, we develop an efficient algorithm based on the diagonally-preconditioned primal-dual splitting algorithm (DP-PDS) [37]-[39], which can automatically determine the appropriate stepsizes for solving this problem.

\begin{figure}[!t]
	\begin{minipage}{0.3\hsize}
		\centerline{\includegraphics[width=\hsize]{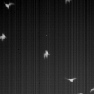}} %画像
	\end{minipage}
	\begin{minipage}{0.3\hsize}
		\centerline{\includegraphics[width=\hsize]{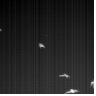}} %画像
	\end{minipage}
	\begin{minipage}{0.3\hsize}
		\centerline{\includegraphics[width=\hsize]{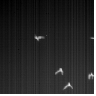}} %画像
	\end{minipage}
	
	\vspace{1mm}
	
	\begin{minipage}{0.3\hsize}
		\centerline{\small{(a1)}}
	\end{minipage}
	\begin{minipage}{0.3\hsize}
		\centerline{\small{(a2)}}
	\end{minipage}
	\begin{minipage}{0.3\hsize}
		\centerline{\small{(a3)}}
	\end{minipage}
	
	\vspace{1mm}
	
	\begin{minipage}{0.3\hsize}
		\centerline{\includegraphics[width=\hsize]{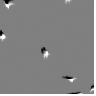}} %画像
	\end{minipage}
	\begin{minipage}{0.3\hsize}
		\centerline{\includegraphics[width=\hsize]{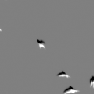}} %画像
	\end{minipage}
	\begin{minipage}{0.3\hsize}
		\centerline{\includegraphics[width=\hsize]{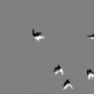}} %画像
	\end{minipage}
	
	\vspace{1mm}
	
	\begin{minipage}{0.3\hsize}
		\centerline{\small{(b1)}}
	\end{minipage}
	\begin{minipage}{0.3\hsize}
		\centerline{\small{(b2)}}
	\end{minipage}
	\begin{minipage}{0.3\hsize}
		\centerline{\small{(b3)}}
	\end{minipage}
	
	\caption{Temporal invariance of stripe noise on IR video data. (a1), (a2), and (a3) Frames of a raw data. (b1), (b2), and (b3) Differences in the frames.}
	\label{fig:TDFPN_of_IRvideo}
\end{figure}

The main contributions of the paper are as follows:
\begin{itemize}
	\item (General framework) Our framework incorporates image regularization as a general form represented by a sum of (possibly) nonsmooth convex functions involving linear operators. This enables us to leverage various image regularizations according to target images.
	\item (Effective characterization of stripe noise) The most common type of stripe noise has a strong flat structure in the vertical or horizontal direction. As a typical example, a band of a raw HSI, a frame of a raw IR video, and their vertical and horizontal gradients are shown in Fig.~\ref{fig:stripe_noise_analysis}, where we can see that the stripe component only exists in the horizontal differences. This implies that stripe noise is flat in the vertical direction. Therefore, we can capture the flatness by constraining its vertical gradient to zero, named the flatness constraint. Moreover, stripe noise in videos is often time-invariant. For example, IR videos are corrupted with time-invariant stripe noise due to focal plane arrays~\cite{IRI_TDFPN1,IRI_TDFPN2}. Some frames of a raw IR video and their differences are shown in Fig.~\ref{fig:TDFPN_of_IRvideo}, where we can see that the stripe noise is time-invariant because it does not appear in the differences. For such data, we impose the flatness constraint along the temporal direction in addition to the spatial constraint. Thanks to such a strong characterization, our framework has a marked ability of stripe noise removal that does not so much depend on what image regularization is adopted.
	\item (Automatic stepsize adjustment) Our algorithm can automatically adjust the stepsizes based on the structure of the optimization problem to be solved. In general, the appropriate stepsizes of PDS would be different depending on image regularizations, meaning that we have to manually adjust them many times. Our algorithm is free from such a troublesome task.
\end{itemize}

\begin{table*}[t]
	\begin{center}
		\caption{Notations and Definitions}
		\label{tab:notations}
		\begin{tabular}{ccc}
			\toprule
			Line number  & Notation & Terminology \\
			\cmidrule(lr){2-3} 
			
			1 & $\mathbb{R}$ and $\mathbb{R}_{++}$ & Real and positive real numbers \\ \cmidrule(lr){2-2} \cmidrule(lr){3-3} 
			
			2 & $\Pi_{i=1}^{M}\mathbb{R}^{n_{i,1}\times \cdots\times n_{i,N_{i}}}$, $\Pi_{i=1}^{M}\mathbb{R}_{++}^{n_{i,1}\times \cdots\times n_{i,N_{i}}}$ & $M$ $N_{i}$th-order tensor/positive-element-tensor product space\footnotemark \\  \cmidrule(lr){2-2} \cmidrule(lr){3-3}
			
			3 & $\mathcal{X}$, $(\mathcal{X}_{1},\cdots,\mathcal{X}_{M})$ & Elements of tensor product space \\  \cmidrule(lr){2-2} \cmidrule(lr){3-3}
			
			4 & $\mathcal{X}_{i}(i_{1},\cdots,i_{N_{j}})$ or $[\mathcal{X}_{i}]_{i_{1},\cdots,i_{N_{j}}}$ & $(i_{1},\cdots,i_{N_{j}})$th element of an $i$th tensor of $\mathcal{X}$ \\  \cmidrule(lr){2-2} \cmidrule(lr){3-3}
			
			5 & \begin{tabular}{c} $\|\mathcal{X}\|_{1}$, \\ $\mathcal{X}=(\mathcal{X}_{1},\cdots,\mathcal{X}_{M})\in\Pi_{i=1}^{M}\mathbb{R}^{n_{i,1}\times \cdots\times n_{i,N_{i}}}$ \end{tabular} & \begin{tabular}{c} $\ell_{1}$-norm, \\ $\|\mathcal{X}\|_{1}=\sum_{i}\sum_{i_{1},\cdots,i_{N_{i}}}|\mathcal{X}_{i}(i_{1},\cdots ,i_{M})|$ \end{tabular} \\  \cmidrule(lr){2-2} \cmidrule(lr){3-3}
			
			6 & \begin{tabular}{c} $\langle \mathcal{X},\mathcal{Y}\rangle$, \\ $\mathcal{X}=(\mathcal{X}_{1},\cdots,\mathcal{X}_{M})\in\Pi_{i=1}^{M}\mathbb{R}^{n_{i,1}\times \cdots\times n_{i,N_{i}}}$, \\ $\mathcal{Y}=(\mathcal{Y}_{1},\cdots,\mathcal{Y}_{M})\in\Pi_{i=1}^{M}\mathbb{R}^{n_{i,1}\times \cdots\times n_{i,N_{i}}}$ \end{tabular} & \begin{tabular}{c} Inner product, \\ $\langle \mathcal{X},\mathcal{Y}\rangle = \sum_{i}\sum_{i_{1},\cdots,i_{N_{i}}}\mathcal{X}_{i}(i_{1},\cdots ,i_{M}) \mathcal{Y}_{i}(i_{1},\cdots,i_{N_{i}})$ \end{tabular} \\  \cmidrule(lr){2-2} \cmidrule(lr){3-3}
			
			7 & \begin{tabular}{c} $\|\mathcal{X}\|_{F}$, \\ $\mathcal{X}\in\Pi_{i=1}^{M}\mathbb{R}^{n_{i,1}\times \cdots\times n_{i,N_{i}}}$ \end{tabular} & \begin{tabular}{c} Frobenius norm, \\ $\|\mathcal{X}\|_{F}=\sqrt{\langle \mathcal{X}, \mathcal{X} \rangle}$ \end{tabular} \\  \cmidrule(lr){2-2} \cmidrule(lr){3-3}
			
			8 & \begin{tabular}{c} $\mathcal{X}\odot\mathcal{Y}\in\Pi_{i=1}^{M}\mathbb{R}^{n_{i,1}\times \cdots\times n_{i,N_{i}}}$, \\ $\mathcal{X}=(\mathcal{X}_{1},\cdots,\mathcal{X}_{M})\in\Pi_{i=1}^{M}\mathbb{R}^{n_{i,1}\times \cdots\times n_{i,N_{i}}}$, \\ $\mathcal{Y}=(\mathcal{Y}_{1},\cdots,\mathcal{Y}_{M})\in\Pi_{i=1}^{M}\mathbb{R}^{n_{i,1}\times \cdots\times n_{i,N_{i}}}$ \end{tabular} & \begin{tabular}{c} Hadamard product, \\ 
				$\mathcal{Z}_{i}(i_{1},\cdots,i_{N_{i}})=\mathcal{X}_{i}(i_{1},\cdots,i_{N_{i}})\mathcal{Y}_{i}(i_{1},\cdots,i_{N_{i}})$, \\
				$\begin{cases} \mathcal{Z}=\mathcal{X}\odot\mathcal{Y}, \\ \forall i_{k}\in\{1,\cdots,n_{i,N_{k}}\}, \\ \forall i\in\{1,\cdots,M\}, \\ \forall k\in\{1,\cdots,M\}\end{cases}$ \end{tabular}\\  \cmidrule(lr){2-2} \cmidrule(lr){3-3}
			
			9 & $\mathcal{I}=(\mathcal{I}_{1},\cdots,\mathcal{I}_{M})$ & \begin{tabular}{c} Identity tensor product element with the Hadamard product, \\ $\mathcal{I}_{i}(i_{1},\cdots,i_{N_{i}})=1$, \\ $\begin{cases} \forall i_{k}\in\{1,\cdots,n_{i,N_{k}}\}, \\ \forall i\in\{1,\cdots,M\}, \\ \forall k\in\{1,\cdots,M\}\end{cases}$ \end{tabular} \\  \cmidrule(lr){2-2} \cmidrule(lr){3-3}
			
			10 & $\mathcal{G}^{-1}=(\mathcal{G}^{-1}_{1},\cdots,\mathcal{G}^{-1}_{M})$ & \begin{tabular}{c} Inverse tensor product element of $\mathcal{G}$ with the Hadamard product, \\
				$\mathcal{G}\odot\mathcal{G}^{-1}=\mathcal{I}$ \end{tabular} \\  \cmidrule(lr){2-2} \cmidrule(lr){3-3}
			
			11 & $\begin{matrix}\|\mathcal{X}\|_{F,\mathcal{G}}, \\ \mathcal{X}\in\Pi_{i=1}^{M}\mathbb{R}^{n_{i,1}\times \cdots\times n_{i,N_{i}}}, \\ \mathcal{G}\in\Pi_{i=1}^{M}\mathbb{R}_{++}^{n_{i,1}\times \cdots \times n_{i,N_{i}}} \end{matrix}$ & \begin{tabular}{c}Frobenius norm skewed by the metric induced by $\mathcal{G}$ \\
				$\|\mathcal{X}\|_{F,\mathcal{G}}=\sqrt{\langle \mathcal{G}\odot\mathcal{X}, \mathcal{X}\rangle}$ \end{tabular} \\  \cmidrule(lr){2-2} \cmidrule(lr){3-3}

			12 & \begin{tabular}{c} $\mathrm{prox}_{\mathcal{G},\gamma f}(\mathcal{X})$, \\ $\mathcal{X}\in\Pi_{i=1}^{M}\mathbb{R}^{n_{i,1}\times\cdots\times n_{i,N_{i}}}$,\\ $\mathcal{G}\in\Pi_{i=1}^{M}\mathbb{R}_{++}^{n_{i,1}\times\cdots\times n_{i,N_{i}}}$, \\ 
				$f$ is a proper lower semi-continuous convex function \end{tabular} & \begin{tabular}{c} The proximity operator of $f$ with index $\gamma>0$ \\ within the metric induced by $\mathcal{G}$, \\ $\mathrm{prox}_{\mathcal{G},\gamma f}(\mathcal{X}):=\argmin_{\mathcal{Y}}\frac{1}{2}\|\mathcal{Y}-\mathcal{X}\|_{F,\mathcal{G}}^{2}+\gamma f(\mathcal{Y})$. \end{tabular}\\ \bottomrule
		\end{tabular}
	\end{center}
\end{table*}

We demonstrate the effectiveness of our framework through destriping experiments, where we comprehensively compare combinations of image regularizations and stripe noise characterizations using hyperspectral images (HSI) and infrared (IR) videos.

The remainder of this paper is organized as follows.
The mathematical notations are summarized in Tab.~\ref{tab:notations}. For more detailed and visual understandings of tensor operators,~\cite{tensor_reviewer,tensor_reviewer_Ji_2019} are helpful.
Section~\ref{sec:Destriping_Methods} gives reviews the existing sparsity-based, low-rank-based, and TV-based destriping models. 
Section~\ref{sec:Flatness_Constraint_Destriping_and_Solving_Algorithm} presents the details of the proposed formulation and the solver. Experimental results and discussion are given in Section~\ref{sec:experiments}. Finally, we summarize the paper in Section~\ref{sec:Conclusion}.

The preliminary version of this work, without mathematical details, comprehensive experimental comparison, deeper discussion, or implementation using DP-PDS, has appeared in conference proceedings~\cite{Naganuma_ICASSP2021}.

\footnotetext[1]{If $M=1$, a tensor product space is equivalent to a tensor space.}


\section{Review of Existing Approaches}
\label{sec:Destriping_Methods}

HSI and IR video data can be represented as third-order tensors, where the spatial information lies in the first two dimensions, and the spectral or frame information lies in the third dimension.
To estimate desirable data from the observed data contaminated by stripe noise and random noise, we model the observation data as follows:
\begin{equation}
	\label{eq:observation_model}
	\mathcal{V}=\bar{\mathcal{U}}+\mathcal{S+N},
\end{equation}
where $\bar{\mathcal{U}}\in\mathbb{R}^{n_{1}\times n_{2}\times n_{3}}$ is a desirable data of interest, $\mathcal{S}\in\mathbb{R}^{n_{1}\times n_{2}\times n_{3}}$ is stripe noise, $\mathcal{N}\in\mathbb{R}^{n_{1}\times n_{2}\times n_{3}}$ is random noise, and $\mathcal{V}\in\mathbb{R}^{n_{1}\times n_{2}\times n_{3}}$ is the observed data.

\begin{figure*}
	\centering
	\begin{minipage}{0.80\hsize}
		\centerline{\includegraphics[width=\hsize]{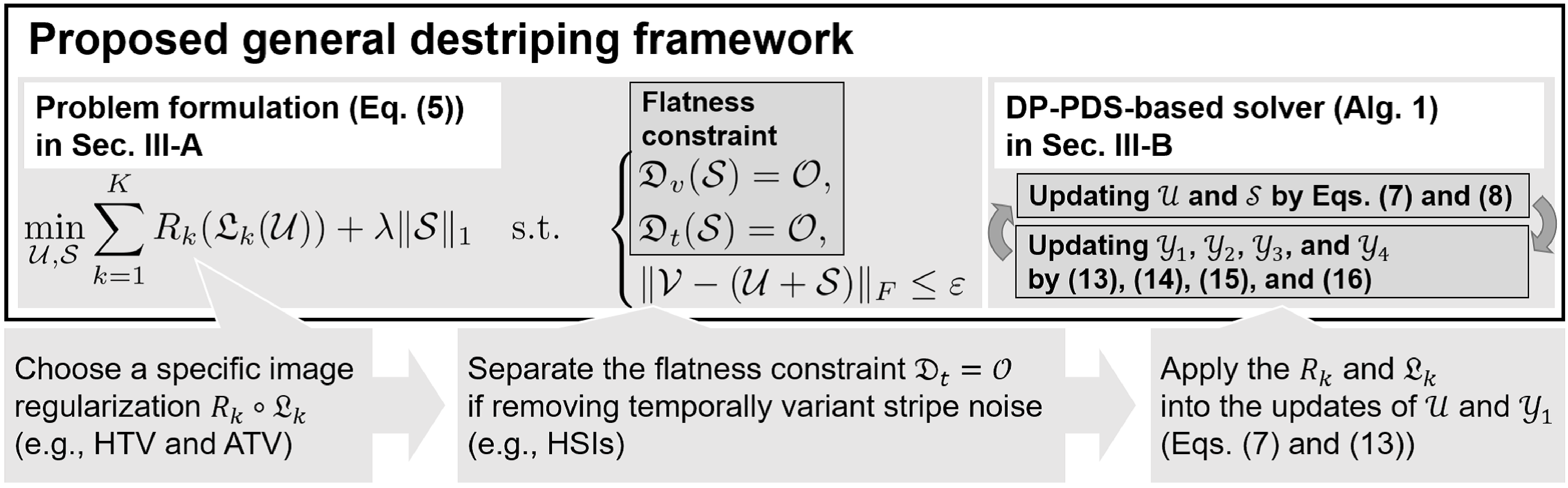}}
	\end{minipage}
	
	\caption{A whole workflow of the proposed general destriping framework.}
	\label{fig:overall_workflow}
\end{figure*}

Under the model in~\eqref{eq:observation_model}, the destriping problem is often formulated as convex optimization problems with the following form: 
\begin{equation*}
	\min_{\mathcal{U,S}}\sum_{k=1}^{K}R_{k}(\mathfrak{L}_{k}(\mathcal{U})) + \lambda_{\mathcal{S}}J(\mathcal{S}) + \frac{\lambda_{\mathcal{N}}}{2}\|\mathcal{V-(U+S)}\|_{F}^{2},
\end{equation*}
where $R_{k}(\mathfrak{L}_{k}(\cdot)):\mathbb{R}^{n_{1}\times n_{2} \times n_{3}}\rightarrow (-\infty, \infty]$ is regularization functions for imaging data with a linear operation $\mathfrak{L}_{k}$ and a function $R_{k}$ $(\forall k = 1,\cdots,K)$, and $J:\mathbb{R}^{n_{1}\times n_{2} \times n_{3}}\rightarrow (-\infty, \infty]$ is a function characterizing stripe noise, respectively.
The positive scalars $\lambda_{\mathcal{S}}$ and $\lambda_{\mathcal{N}}$ are the hyperparameters.
Depending on how $J$ is chosen, destriping models can be classified into the following three categories: the (group-)sparsity-based model, the low-rank-based model, and the TV-based model.

The sparsity-based model has been used in a lot of methods. Among them, the method proposed in~\cite{LRMR} is known as a representative work. This method uses the $\ell_{1}$-norm as $J$, which is a well-known sparsity measure. As mentioned, this model relies on the fact that stripe noise is sparsely distributed in observed data. The method proposed in~\cite{GLSSTV} sets $J$ to the mixed $\ell_{2,1}$-norm since each column of stripe noise is viewed as a group. The mixed $\ell_{2,1}$-norm is the sum of the $\ell_{2}$-norm of each column vector, which groups stripe noise by each column, and thus it is used for the characterization of stripe noise based on group sparsity. The sparsity-based model results in efficient optimization due to its simple modeling, but cannot fully capture the nature of stripe noise. Specifically, its destriping performance strongly depends on image regularization, as will be shown in Section~\ref{ssec:Simulated_experiments}.

\begin{table}[t]
	\begin{center}
		\caption{Stripe Noise Characterizations}
		\label{tab:various_stripe_noise_characterization}
		\begin{tabular}{c|c}
			\hline
			Model & $J(\mathcal{S})$ \\
			\hline \hline
			Sparsity-based model & $\lambda \|\mathcal{S}\|_{1}$ \\
			Group sparsity-based model & $\lambda \sum_{j}^{n_{2}}\sum_{k}^{n_{3}}\|\mathcal{S}(:,j,k)\|_{2}$\\
			Low-rank-based model & $\lambda \sum_{i=1}^{n_{3}}\|\mathcal{S}(:,:,i)\|_{*}$ \\
			TV-based model & $\mu \|\mathfrak{D}_{v}(\mathcal{S})\|_{0 or 1} + \lambda \|\mathcal{S}\|_{1}$ \\
			\hline
		\end{tabular}
	\end{center}
\end{table}

The low-rank-based model has been proposed in~\cite{NN_char}. In~\cite{NN_char}, the authors revealed that stripe noise only exists in the horizontal gradient component and that the rank of stripe noise is one. Based on this observation, they adopted the nuclear norm for $J$, which is a reasonable convex function that can evaluate the low-rankness of a matrix. In general, this model outperforms the sparsity-based model. However, it conflicts with low-rank image regularizations where the nuclear norm is employed~\cite{LRMR,LRTR,NonLRMA,3DTNN}.

The TV-based model~\cite{Kind_of_Stripes,gradient_constraint} adopted a TV term and a sparse term to capture the one-directional smoothness of stripe noise. This model is also superior to the sparsity-based model. However, the TV-based model weakens the TV regularization ability to capture the vertical smoothness, as will be shown in Section~\ref{ssec:Simulated_experiments}.

We summarize the stripe noise characterizations in Tab.~\ref{tab:various_stripe_noise_characterization}.

\section{Proposed Framework}
\label{sec:Flatness_Constraint_Destriping_and_Solving_Algorithm}
The proposed framework involves a general form of regularization term and two types of the flatness constraint. The choice of the specific image regularization and the removal of the temporal flatness constraint are required to fit the nature of an observed image. Depending on image regularization and the temporal flatness constraint, the DP-PDS-based solver needs to be implemented. We illustrate a whole workflow for the proposed framework in Fig.~\ref{fig:overall_workflow}.

\subsection{General Destriping Model with Flatness Constraint}
\label{ssec:Flatness_Constraint_Destriping}

In this section, we propose a general destriping model using the flatness constraint. As mentioned, stripe noise $\mathcal{S}$ has the characteristic that the vertical/temporal gradient is zero, i.e.,
\begin{equation}
	\label{eq:flatness_constraint}
	\begin{cases}\mathfrak{D}_{v}(\mathcal{S})=\mathcal{O}, \\ \mathfrak{D}_{t}(\mathcal{S})=\mathcal{O},\end{cases}
\end{equation}
where $\mathcal{O}\in\mathbb{R}^{n_{1}\times n_{2} \times n_{3}}$ is a zero tensor, i.e., $\mathcal{O}(i,j,k)=0$, $\forall i\in\{1,\cdots,n_{1}\}$, $\forall j\in\{1,\cdots,n_{2}\}$, and $\forall k\in\{1,\cdots,n_{3}\}$. Moreover, $\mathfrak{D}_{v}:\mathbb{R}^{n_{1}\times n_{2}\times n_{3}}\rightarrow \mathbb{R}^{(n_{1}-1)\times n_{2}\times n_{3}}$ and $\mathfrak{D}_{t}:\mathbb{R}^{n_{1}\times n_{2}\times n_{3}}\rightarrow \mathbb{R}^{n_{1}\times n_{2}\times (n_{3}-1)}$ are the vertical/temporal difference operators with the Neumann boundary, which are defined as 
\begin{align}
	& [\mathfrak{D}_{v}(\mathcal{X})]_{i,j,k}:=\mathcal{X}(i,j,k) - \mathcal{X}(i+1,j,k), \nonumber \\
	& \begin{cases}\forall i\in\{1,\cdots,n_{1}-1\},\\ \forall j\in\{1,\cdots,n_{2}\},\\ \forall k\in\{1,\cdots,n_{3}\},\end{cases}
\end{align}
\begin{align}
	& [\mathfrak{D}_{t}(\mathcal{X})]_{i,j,k}:=\mathcal{X}(i,j,k) - \mathcal{X}(i,j,k+1), \nonumber \\
	& \begin{cases}\forall i\in\{1,\cdots,n_{1}\},\\ \forall j\in\{1,\cdots,n_{2}\},\\ \forall k\in\{1,\cdots,n_{3}-1\}.\end{cases}
\end{align}
Using the flatness constraints in Eq.~\eqref{eq:flatness_constraint}, we newly formulate destriping as the following convex optimization problem:
\begin{equation}
	\label{eq:flatness_constraint_vt}
	\min_{\mathcal{U,S}}\sum_{k=1}^{K}R_{k}(\mathfrak{L}_{k}(\mathcal{U})) + \lambda\|\mathcal{S}\|_{1} \quad \mathrm{s.t.} \quad \begin{cases}
		\mathfrak{D}_{v}(\mathcal{S})=\mathcal{O}, \\
		\mathfrak{D}_{t}(\mathcal{S})=\mathcal{O}, \\
		\|\mathcal{V-(U+S)}\|_{F}\leq\varepsilon,
	\end{cases}
\end{equation}
where $\lambda>0$ is a hyperparameter, and $R_{k}(\mathfrak{L}_{k}(\cdot))$ $(k=1,\cdots,K)$ is a regularization term with a proper semi-continuous convex proximable\footnote{
	If an efficient computation of the skewed proximity operator of $f$ is available, we call $f$ skew proximable.
} function $R_{k}$ and a linear operator $\mathfrak{L}_{k}$. 
The vertical and temporal gradients of stripe noise are constrained to zero by the first and second constraint, which captures the vertical/temporal flatness of stripe noise.
Additionally, we impose the $\ell_{1}$-norm on $\mathcal{S}$ to exploit the sparsity of stripe noise.
The third constraint is a Frobenius norm constraint with the radius $\varepsilon$ for data fidelity to $\mathcal{V}$ given in~\eqref{eq:observation_model}.
The data-fidelity constraint has an important advantage over the standard additive data fidelity in terms of facilitating hyperparameter settings, as addressed in~\cite{CSALSA,ConstPoisson2012,EPIpre,ono_2015,ono_2019}. If stripe noise is variant in the third direction such as HSIs, we remove the second constraint.

For data with horizontally featured stripe noise, as in images acquired by whiskbroom scanning~\cite{Kind_of_Stripes}, we rotate the data 90 degrees in the spatial direction before optimization.

\subsection{Diagonally Preconditioned Primal-Dual Splitting Algorithm for Solving the General Destriping model }
In this part, we introduce DP-PDS~\cite{DP-PDS} to solve Prob.~\eqref{eq:flatness_constraint_vt}. DP-PDS (see Appendix), which is a diagonally preconditioned version of the primal-dual splitting algorithm~\cite{PDS_1,PDS_2}, frees us from tedious stepsize settings. Moreover, the convergence speed of DP-PDS is much faster in general than that of the original PDS algorithm.

To solve Prob.~\eqref{eq:flatness_constraint_vt} with DP-PDS, we rewrite it into the following equivalent problem:
\begin{align}
	& \min_{\substack{\mathcal{U,S},\mathcal{Y}_{1,1},\ldots,\mathcal{Y}_{1,K}, \\ \mathcal{Y}_{2},\mathcal{Y}_{3},\mathcal{Y}_{4}}}\lambda\|\mathcal{S}\|_{1}+\sum_{k=1}^{K}R_{k}(\mathcal{Y}_{1,k})+\iota_{\{\mathcal{O}\}}(\mathcal{Y}_{2}) \nonumber \\ 
	& +\iota_{\{\mathcal{O}\}}(\mathcal{Y}_{3})+\iota_{B_{(\mathcal{V},\varepsilon)}}(\mathcal{Y}_{4})\quad\mathrm{s.t.}\quad \begin{cases} \mathcal{Y}_{1,1} = \mathfrak{L}_{1} ( \mathcal{U} ) , \\ \vdots \\ \mathcal{Y}_{1,K} = \mathfrak{L}_{K} ( \mathcal{U} ) , \\ \mathcal{Y}_{2}=\mathfrak{D}_{v}(\mathcal{S}), \\ \mathcal{Y}_{3}=\mathfrak{D}_{t}(\mathcal{S}), \\ \mathcal{Y}_{4}=\mathcal{U+S}, \end{cases}
	\label{eq:DP_PDS_form_of_zero_gradient_optimization_vt}
\end{align}
where $\iota_{\{\mathcal{O}\}}$ and $\iota_{B_{(\mathcal{V},\varepsilon)}}$ are the indicator functions\footnote{For a given nonempty closed convex set $C$, the indicator function of $C$ is defined by $\iota_{C}(\mathcal{X}):=0$, if $\mathcal{X}\in C$; $\infty$, otherwise.} of $\{\mathcal{O}\}$ and $B_{(\mathcal{V},\varepsilon)}:=\{\mathcal{X}\in\mathbb{R}^{n_{1}\times n_{2}\times n_{3}}|\|\mathcal{V-X}\|_{F}\leq\varepsilon\}$, respectively.
DP-PDS computes the solution of Eq.~\eqref{eq:DP_PDS_form_of_zero_gradient_optimization_vt} by updating primal variables ($\mathcal{U}$ and $\mathcal{S}$) and dual variables ($\mathcal{Y}_{1,1},\ldots,\mathcal{Y}_{1,K},\mathcal{Y}_{2},\mathcal{Y}_{3}$, and $\mathcal{Y}_{4}$) alternately.

%%%%%%%%%%%%%%%%%%%%%%%%%%%%%%%%%%%%%%%%%%%%%%%%%%%%%%%%%%%%%%%%%%%%
%% The algorithm solving time invariant case
%%%%%%%%%%%%%%%%%%%%%%%%%%%%%%%%%%%%%%%%%%%%%%%%%%%%%%%%%%%%%%%%%%%%
\begin{algorithm}[t]
	\caption{The DP-PDS algorithm for solving Prob.~\eqref{eq:flatness_constraint_vt}}
	\label{algo:DP_PDS_for_zero_gradient_constraint_vt}
	\begin{algorithmic}[1]
		\Require{An observed image $\mathcal{V}$, a balancing parameter $\lambda$, and a data fidelity parameter $\varepsilon$}
		\Ensure{$\mathcal{U}^{(n)},\mathcal{S}^{(n)}$}
		\State Initialize $\mathcal{U}^{(0)},\mathcal{S}^{(0)},\mathcal{Y}_{1,k}^{(0)}(k=1,\ldots ,K),\mathcal{Y}_{i}^{(0)} (i=2,3,4)$;
		\State $n = 0$;
		\While {A stopping criterion is not satisfied}
		
		\State $\mathcal{U}^{(n+1)}\leftarrow\mathcal{U}^{(n)} - \mathcal{G}_{\mathcal{U}}\odot(\sum_{k=1}^{K}\mathfrak{L}_{k}^{*}(\mathcal{Y}_{1,k}^{(n)})+\mathcal{Y}_{4}^{(n)})$;
		
		\State $\mathcal{S}^{\prime}\leftarrow\mathcal{S}^{(n)} - \mathcal{G}_{\mathcal{S}}\odot(\mathfrak{D}_{v}^{*}(\mathcal{Y}_{2}^{(n)})+\mathfrak{D}_{t}^{*}(\mathcal{Y}_{3}^{(n)})+\mathcal{Y}_{4}^{(n)})$;
		\State $\mathcal{S}^{(n+1)}\leftarrow \mathrm{prox}_{\mathcal{G}_{\mathcal{S}}^{-1},\lambda\|\cdot\|_{1}}(\mathcal{S}^{\prime})$ by Eq.~\eqref{eq:prox_of_l1};
		
		\For {$i=1,\cdots,K$} 
		\State $\mathcal{Y}_{1,k}^{(n)}\leftarrow\mathcal{Y}_{1,k}^{(n)}+\mathcal{G}_{\mathcal{Y}_{1,k}}\odot\mathfrak{L}_{k}(2\mathcal{U}^{(n+1)}-\mathcal{U}^{(n)})$;
		\State $\mathcal{Y}_{1,k}^{(n+1)}\leftarrow\mathcal{Y}_{1,k}^{(n)} - \mathcal{G}_{\mathcal{Y}_{1,k}}\odot\mathrm{prox}_{\mathcal{G}_{\mathcal{Y}_{1,k}},R_{k}}(\mathcal{G}_{\mathcal{Y}_{1,k}}^{-1}\odot\mathcal{Y}_{1,k}^{(n)})$;
		\EndFor

		\State $\mathcal{Y}_{2}^{(n+1)}\leftarrow\mathcal{Y}_{2}^{(n)}+\mathcal{G}_{\mathcal{Y}_{2}}\odot\mathfrak{D}_{v}(2\mathcal{S}^{(n+1)}-\mathcal{S}^{(n)})$;
		\State $\mathcal{Y}_{3}^{(n+1)}\leftarrow\mathcal{Y}_{3}^{(n)}+\mathcal{G}_{\mathcal{Y}_{3}}\odot\mathfrak{D}_{t}(2\mathcal{S}^{(n+1)}-\mathcal{S}^{(n)})$;
		\State $\mathcal{Y}_{4}^{(n)}\leftarrow\mathcal{Y}_{4}^{(n)}+\mathcal{G}_{4}\odot(2(\mathcal{L}^{(n+1)} + \mathcal{S}^{(n+1)}) - (\mathcal{L}^{(n)} + \mathcal{S}^{(n)}))$;
		\State $\mathcal{Y}_{4}^{(n+1)}\leftarrow\mathcal{Y}_{4}^{(n)} - \mathcal{G}_{4}\odot\mathcal{P}_{B_{(\mathcal{V},\varepsilon)}}(\mathcal{G}_{4}^{-1}\odot\mathcal{Y}_{4}^{(n)})$ by Eq.~\eqref{eq:prox_of_Fnorm};
		\State $n\leftarrow n+1$;
		\EndWhile
	\end{algorithmic}
\end{algorithm}

The primal variables are updated as follows:
\begin{equation}
	\mathcal{U}^{(n+1)}\leftarrow \mathcal{U}^{(n)}-\mathcal{G}_{\mathcal{U}}\odot\left(\sum_{k=1}^{K}\mathfrak{L}^{*}_{k}(\mathcal{Y}_{1,k})+\mathcal{Y}_{4}\right),
\end{equation}
\begin{align}
	& \mathcal{S}^{(n+1)}\leftarrow\mathrm{prox}_{\mathcal{G}_{\mathcal{S}}^{-1},\lambda\|\cdot\|_{1}} & \nonumber \\
	& \left(\mathcal{S}^{(n)} - \mathcal{G}_{\mathcal{S}}\odot\left(\mathfrak{D}_{v}^{*}\left(\mathcal{Y}_{2}^{(n)}\right)+\mathfrak{D}_{t}^{*}\left(\mathcal{Y}_{3}^{(n)}\right)+\mathcal{Y}_{4}^{(n)}\right)\right), &
	\label{eq:update_of_S}
\end{align}
where $\mathfrak{L}^{*}_{1},\ldots,\mathfrak{L}^{*}_{K}$, $\mathfrak{D}_{v}^{*}$, and $\mathfrak{D}_{t}^{*}$ are the adjoint operators\footnote{
	Let $\mathfrak{L}:\prod_{i=1}^{N_{0}}\mathbb{R}^{n_{i,1}\times \cdots \times n_{i,N_{i}}}\rightarrow \prod_{i=1}^{M_{0}}\mathbb{R}^{m_{i,1}\times \cdots \times m_{i,M_{i}}}$. A linear operator $\mathfrak{L}^{*}:\prod_{i=1}^{M_{0}}\mathbb{R}^{m_{i,1}\times \cdots \times m_{i,M_{i}}}\rightarrow \prod_{i=1}^{N_{0}}\mathbb{R}^{n_{i,1}\times \cdots \times n_{i,N_{i}}}$ is called \textit{adjoint operator} of $\mathfrak{L}$ if for all $\mathcal{X}\in\mathbb{R}^{n_{i,1}\times \cdots \times n_{i,N_{i}}}, \mathcal{Y}\in\prod_{i=1}^{M_{0}}\mathbb{R}^{m_{i,1}\times \cdots \times m_{i,M_{i}}}$, $\langle \mathfrak{L}(\mathcal{X}),\mathcal{Y}\rangle=\langle \mathcal{X},\mathfrak{L}^{*}(\mathcal{Y})\rangle$.
} of $\mathfrak{L}_{1},\ldots,\mathfrak{L}_{K}$, $\mathfrak{D}_{v}$, and $\mathfrak{D}_{t}$, respectively. The constants $\mathcal{G}_{\mathcal{U}}$ and $\mathcal{G}_{\mathcal{S}}$ are stepsize parameters that are called \textit{preconditioners} and $\mathcal{G}_{\mathcal{S}}^{-1}$ is the inverse tensor of $\mathcal{G}_{\mathcal{S}}$ (see line 10 of Tab.~\ref{tab:notations}). The preconditioners are given by the coefficients of the linear operations $\mathfrak{L}$ and $\mathfrak{D}_{v}$ (see Eq.~\eqref{eq:def_preconditioner1} in Appendix for the detailed definitions). 
The skewed proximity operator (see line 12 of Tab.~\ref{tab:notations} for the definition) of $\|\cdot\|_{1}$ in Eq.~\eqref{eq:update_of_S} is given by
\begin{equation}
	\label{eq:prox_of_l1}
	\mathrm{prox}_{\mathcal{G}_{\mathcal{S}}^{-1},\lambda \|\cdot\|_{1}}(\mathcal{X})=\mathrm{sgn}(\mathcal{X})\odot\max\{|\mathcal{X}|-\lambda\mathcal{G}_{\mathcal{S}}, 0\},
\end{equation}
where $\mathrm{sgn}(\mathcal{X})$, $\max\{\mathcal{X},0\}$, and $|\mathcal{X}|$ respectively denote the sign, positive part, and magnitude of $\mathcal{X}$. Their definitions are as follows:
%\begin{equation}
%	[\mathrm{sgn}(\mathcal{X})]_{i,j,k}=\left\{\begin{matrix}
%		1 & \mathrm{if}\quad\mathcal{X}(i,j,k)\geq 0, \\ -1 & \mathrm{if}\quad\mathcal{X}(i,j,k)<0,
%	\end{matrix}\right.
%\end{equation}
\begin{equation}
[\mathrm{sgn}(\mathcal{X})]_{i,j,k}=\begin{cases}
	1, & \mathrm{if} \: \mathcal{X}(i,j,k)\geq 0, \\ -1, & \mathrm{if} \: \mathcal{X}(i,j,k)<0,
\end{cases}
\end{equation}
%\begin{equation}
%	[\max\{\mathcal{X}, 0\}]_{i,j,k}=\left\{\begin{matrix}\mathcal{X}(i,j,k) & \mathrm{if}\quad \mathcal{X}(i,j,k)\geq 0,\\ 0 & \mathrm{if}\quad \mathcal{X}(i,j,k)<0,\end{matrix}\right.
%\end{equation} 
\begin{equation}
	[\max\{\mathcal{X}, 0\}]_{i,j,k}=\begin{cases}\mathcal{X}(i,j,k), & \mathrm{if} \: \mathcal{X}(i,j,k)\geq 0,\\ 0, & \mathrm{if} \: \mathcal{X}(i,j,k)<0,\end{cases}
\end{equation} 
\begin{equation}
	[|\mathcal{X}|]_{i,j,k}=|\mathcal{X}(i,j,k)|, \left\{\begin{matrix}\forall i\in\{1,\cdots,n_{1}\}, \\ \forall j\in\{1,\cdots,n_{2}\}, \\ \forall k\in\{1,\cdots,n_{3}\}.\end{matrix}\right.
\end{equation}

Then, the dual variables are updated as follows:
\begin{align}
	\mathcal{Y}_{1,k}^{(n+1)} \leftarrow &  \mathrm{prox}_{\mathcal{G}_{\mathcal{Y}_{1,k}}^{-1},R^{*}_{k}} \nonumber \\ & \left(\mathcal{Y}^{(n)}_{1}+\mathcal{G}_{\mathcal{Y}_{1,k}}\odot\left(\mathfrak{L}_{k}\left(2\mathcal{U}^{(n+1)}-\mathcal{U}^{(n)}\right)\right)\right), \nonumber \\
	& (\forall k = 1,\cdots,K)
	\label{eq:Y1_update}
\end{align}
\begin{align}
	\mathcal{Y}_{2}^{(n+1)}\leftarrow &  \mathrm{prox}_{\mathcal{G}_{\mathcal{Y}_{2}}^{-1},\iota_{\{\mathcal{O}\}}^{*}} \nonumber \\ & \left(\mathcal{Y}^{(n)}_{2}+\mathcal{G}_{\mathcal{Y}_{2}}\odot\left(\mathfrak{D}_{v}\left(2\mathcal{S}^{(n+1)}-\mathcal{S}^{(n)}\right)\right)\right),
	\label{eq:Y2_update}
\end{align}
\begin{align}
	\mathcal{Y}_{3}^{(n+1)}\leftarrow &  \mathrm{prox}_{\mathcal{G}_{\mathcal{Y}_{3}}^{-1},\iota_{\{\mathcal{O}\}}^{*}} \nonumber \\ & \left(\mathcal{Y}^{(n)}_{3}+\mathcal{G}_{\mathcal{Y}_{3}}\odot\left(\mathfrak{D}_{v}\left(2\mathcal{S}^{(n+1)}-\mathcal{S}^{(n)}\right)\right)\right),
	\label{eq:Y3_update}
\end{align}
\begin{equation}
	\begin{array}{l}\mathcal{Y}_{4}^{(n+1)}\leftarrow \mathrm{prox}_{\mathcal{G}_{\mathcal{Y}_{4}}^{-1},\iota_{B_{(\mathcal{V},\varepsilon)}}^{*}} \\  \left(\mathcal{Y}^{(n)}_{4}+\mathcal{G}_{\mathcal{Y}_{4}}\odot\left(2\left(\mathcal{U}^{(n+1)}+\mathcal{S}^{(n+1)}\right)-\left(\mathcal{U}^{(n)}+\mathcal{S}^{(n)}\right)\right)\right), \end{array}
	\label{eq:Y4_update}
\end{equation}
where the constants $\mathcal{G}_{\mathcal{Y}_{1,1}}, \ldots, \mathcal{G}_{\mathcal{Y}_{1,K}},\mathcal{G}_{\mathcal{Y}_{2}},\mathcal{G}_{\mathcal{Y}_{3}}$, and $\mathcal{G}_{\mathcal{Y}_{4}}$ are preconditioners that can be also determined automatically (see Eq.~\eqref{eq:def_preconditioner2} in Appendix). The functions $R_{k}^{*}$, $\iota_{\{\mathcal{O}\}}^{*}$, and $\iota_{B_{(\mathcal{V},\varepsilon)}}^{*}$ are the \textit{Fenchel--Rockafellar conjugate functions}\footnote{The Fenchel--Rockafellar conjugate function of $f$ is defined as
\begin{equation*}
	f^{*}(\mathcal{X}) := \max_{\mathcal{Y}}\langle \mathcal{X},\mathcal{Y}\rangle + f(\mathcal{Y}).
\end{equation*}} of $R_{k}$, $\iota_{\{\mathcal{O}\}}$, and $\iota_{B_{(\mathcal{V},\varepsilon)}}$. The skewed proximity operator has the following useful property~\cite[Corollary 6]{S_Becker}:
\begin{equation}
	\label{eq:skewed_proximity_operator_of_Fenchel_Rockafellar}
	\mathrm{prox}_{\mathcal{G}^{-1},f^{*}}(\mathcal{X}) = \mathcal{X} -  \mathcal{G}\odot\mathrm{prox}_{\mathcal{G},f}(\mathcal{G}^{-1}\odot\mathcal{X}),
\end{equation}
so that the skewed proximity operator of a Fenchel-Rockafellar conjugate function $f^{*}$ can be easily calculated if $f$ is skew proximable. The skewed proximity operators in Eq.~\eqref{eq:Y1_update} are efficiently computed because $R_{k}$ is a skew proximable function. The skewed proximity operator of $\iota_{\{\mathcal{O}\}}$ in Eqs.~\eqref{eq:Y2_update} and~\eqref{eq:Y3_update} are calculated as $\mathrm{prox}_{\mathcal{G},\iota_{\{\mathcal{O}\}}}(\mathcal{X})=\mathcal{O}$ for any $\mathcal{X}\in\mathbb{R}^{(n_{1}-1)\times n_{2}\times n_{3}}$ and $\mathcal{G}\in\mathbb{R}_{++}^{(n_{1}-1)\times n_{2}\times n_{3}}$. The skewed proximity operator of $\iota_{B_{(\mathcal{V},\varepsilon)}}$ in Eq.~\eqref{eq:Y4_update} is not proximable in general.
In our method, all entries of the preconditioner $\mathcal{G}_{4}$ are $\frac{1}{2}$. Hence, the operator $\mathrm{prox}_{\mathcal{G}_{4},\iota_{B_{(\mathcal{V},\varepsilon)}}}$ is easily calculated as
%\begin{align}
%	& \mathrm{prox}_{\mathcal{G}_{4},\iota_{B_{(\mathcal{V},\varepsilon)}}}(\mathcal{X}) \nonumber \\
%	& =\mathrm{prox}_{\mathcal{I},2\iota_{B_{(\mathcal{V},\varepsilon)}}}(\mathcal{X}) \nonumber \\
%	& =\mathcal{P}_{B_{(\mathcal{V},\varepsilon)}}(\mathcal{X})=\begin{cases} \mathcal{X}, & \mathrm{if}\quad \mathcal{X}\in B_{(\mathcal{V},\varepsilon)}, \\ \mathcal{V} + \frac{\varepsilon(\mathcal{X}-\mathcal{V})}{\|\mathcal{X}-\mathcal{V}\|_{F}}, & \mathrm{otherwise}. \end{cases} 
%	\label{eq:prox_of_Fnorm}
%\end{align}
\begin{align}
	& \mathrm{prox}_{\mathcal{G}_{4},\iota_{B_{(\mathcal{V},\varepsilon)}}}(\mathcal{X}) \nonumber \\
	& =\mathrm{prox}_{\mathcal{I},2\iota_{B_{(\mathcal{V},\varepsilon)}}}(\mathcal{X}) \nonumber \\
	& =\mathcal{P}_{B_{(\mathcal{V},\varepsilon)}}(\mathcal{X})=\begin{cases} \mathcal{X}, & \mathrm{if}\: \mathcal{X}\in B_{(\mathcal{V},\varepsilon)}, \\ \mathcal{V} + \frac{\varepsilon(\mathcal{X}-\mathcal{V})}{\|\mathcal{X}-\mathcal{V}\|_{F}}, & \mathrm{otherwise}. \end{cases} 
	\label{eq:prox_of_Fnorm}
\end{align}

Through these update steps, we obtain the solution of Prob.~\eqref{eq:flatness_constraint_vt}. We show the detailed algorithms in Alg.~\ref{algo:DP_PDS_for_zero_gradient_constraint_vt}. We note that this algorithm can handle a nonconvex optimization problem that contains the proximable nonconvex function such as the $\ell_{0}$-norm and the rank function. However, its convergence, in this case, is not guaranteed.

In temporally variant stripe noise cases, such as an HSI, the temporal constraint is removed. Following the change, the update step in~\eqref{eq:update_of_S} will be as follows:
\begin{equation}
	\mathcal{S}^{(n+1)}\leftarrow \mathrm{prox}_{\mathcal{G}_{\mathcal{S}}^{-1},\lambda\|\cdot\|}\left(\mathcal{S}^{(n)}-\mathcal{G}_{\mathcal{S}}\odot\left(\mathfrak{D}_{v}^{*}(\mathcal{Y}_{2}^{(n)})+\mathcal{Y}_{4}^{(n)}\right)\right). 
\end{equation}
Then, we remove the update step of $\mathcal{Y}_{3}$ (line 9 of Alg.~\ref{algo:DP_PDS_for_zero_gradient_constraint_vt}).

\subsection{Examples of Image Regularizations}

We give some examples of image regularization $\sum_{k=1}^{K} R_{k}(\mathfrak{L}_{k}(\mathcal{U}))$ in~\eqref{eq:flatness_constraint_vt}. First, let us consider HTV~\cite{HTV}. Since the HTV is an image regularization for HSIs, we adopt the formulation that does not involve the temporal flatness constraint. The definition of HTV is 
\begin{equation}
	\|\mathcal{U}\|_{\mbox{HTV}}:=\sum_{i,j}\sqrt{\sum_{k}\mathcal{D}_{1}(i,j,k)^{2}+\mathcal{D}_{2}(i,j,k)^{2}},
	\label{eq:HTV}
\end{equation}
where $\mathcal{D}_{1}=\mathfrak{D}_{v}(\mathcal{U})$ and $\mathcal{D}_{2}=\mathfrak{D}_{h}(\mathcal{U})$. Therefore, by letting $K=1$, $\mathfrak{L}_{1}(\mathcal{U})=\{\mathfrak{D}_{v}(\mathcal{U}),\mathfrak{D}_{h}(\mathcal{U})\}$, and $R_{1}=\|\{\mathcal{U}_{1},\mathcal{U}_{2}\}\|_{1,2}=\sum_{i,j,k}\sqrt{\mathcal{U}_{1}(i,j,k)^{2}+\mathcal{U}_{2}(i,j,k)^{2}}$, we can apply HTV to Prob.~\eqref{eq:flatness_constraint_vt}. The update of $\mathcal{U}$ is as follows:
\begin{equation}
	\mathcal{U}^{(n+1)}\leftarrow\mathcal{U}^{(n)}-\mathcal{G}_{\mathcal{U}}\odot\left(\mathfrak{D}_{v}^{*}(\mathcal{Y}_{1,1,1}^{(n)})+\mathfrak{D}_{h}^{*}(\mathcal{Y}_{1,1,2}^{(n)})+\mathcal{Y}_{3}^{(n)}\right),
\end{equation}
where $\mathcal{Y}_{1,1}^{(n)}=\{\mathcal{Y}_{1,1,1}^{(n)},\mathcal{Y}_{1,1,2}^{(n)}\}$. The proximity operator \\ of $\|\cdot\|_{1,2}$ is calculated as follows:
\begin{align}
	& \mathcal{Z}_{l}(i,j,k) = \nonumber \\
	& \max\left\{1-\frac{\mathcal{G}_{\mathcal{Y}_{1,1,l}}(i,j,k)}{\sqrt{\sum_{k^{\prime}}\mathcal{Y}_{1,1,1}(i,j,k^{\prime})^{2}+\mathcal{Y}_{1,1,2}(i,j,k^{\prime})^{2}}},0\right\} \nonumber \\
	& *\mathcal{Y}_{1,1,l}(i,j,k),
\end{align} 
where $\{\mathcal{Z}_{1},\mathcal{Z}_{2}\}=\mathrm{prox}_{\mathcal{G}_{\mathcal{Y}_{1,k}},\|\cdot\|_{1,2}}(\mathcal{Y}_{1,1})$. Preconditioners are determined as $\mathcal{G}_{\mathcal{U}}(i,j,k) = 1/(\mathcal{G}_{\mathfrak{D}_{v}^{*}}^{-1}(i,j,k) + \mathcal{G}_{\mathfrak{D}_{h}^{*}}^{-1}(i,j,k) + 1)$, and $\mathcal{G}_{\mathcal{S}}(i,j,k)=1/(\mathcal{G}_{\mathfrak{D}_{v}^{*}}^{-1}(i,j,k) + 1)$, where 
\begin{equation}
	\mathcal{G}_{\mathfrak{D}_{v}^{*}}^{-1}(i,j,k) = \begin{cases} 1, & \mathrm{if} \: i = 1,n_{1}, \\ 2, & \mathrm{otherwise}, \end{cases} 
\end{equation}
\begin{equation}
	\mathcal{G}_{\mathfrak{D}_{h}^{*}}^{-1}(i,j,k) = \begin{cases} 1, & \mathrm{if} \: j = 1,n_{2}, \\ 2, & \mathrm{otherwise}, \end{cases}
\end{equation}
$\mathcal{G}_{\mathcal{Y}_{1,1,1}}(i,j,k) = 1/2$, $\mathcal{G}_{\mathcal{Y}_{1,1,2}}(i,j,k) = 1/2$, $\mathcal{G}_{\mathcal{Y}_{2}}(i,j,k)=1/2$, $\forall i\in\{1,\cdots,n_{1}\}$, $\forall j\in\{1,\cdots,n_{2}\}$, and $\forall k\in\{1,\cdots,n_{3}\}$.
Finally, we obtain a solver for Prob.~\eqref{eq:flatness_constraint_vt} with HTV.

As another example for an IR video case, we consider ATV~\cite{video_TV}. ATV is defined as 
\begin{equation}
	\|\mathcal{U}\|_{\mbox{ATV}}:=\|\mathfrak{D}_{v}(\mathcal{U})\|_{1}+\|\mathfrak{D}_{h}(\mathcal{U})\|_{1}+\|\mathfrak{D}_{t}(\mathcal{U})\|_{1}.
	\label{eq:ATV}
\end{equation}
Therefore, we set $K=3$, $\mathfrak{L}_{1}=\mathfrak{D}_{v}$, $\mathfrak{L}_{2}=\mathfrak{D}_{h}$, $\mathfrak{L}_{3}=\mathfrak{D}_{t}$, and $R_{1}=R_{2}=R_{3}=\|\cdot\|_{1}$ to apply ATV to Eq.~\eqref{eq:DP_PDS_form_of_zero_gradient_optimization_vt}. Then, we update $\mathcal{U}$ as
\begin{align}
	 & \mathcal{U}^{(n+1)}\leftarrow\mathcal{U}^{(n)} \nonumber \\
	 & -\mathcal{G}_{\mathcal{U}}\odot\left(\mathfrak{D}_{v}^{*}(\mathcal{Y}_{1,1}^{(n)})+\mathfrak{D}_{h}^{*}(\mathcal{Y}_{1,2}^{(n)}) + \mathfrak{D}_{t}^{*}(\mathcal{Y}_{1,3}^{(n)})+\mathcal{Y}_{3}^{(n)}\right).
\end{align}
The proximity operator in line 11 of Alg.~\ref{algo:DP_PDS_for_zero_gradient_constraint_vt} is calculated by Eq.~\eqref{eq:prox_of_l1}. Preconditioners are set as $\mathcal{G}_{\mathcal{U}}(i,j,k) = 1/(\mathcal{G}_{\mathfrak{D}_{v}^{*}}^{-1}(i,j,k) + \mathcal{G}_{\mathfrak{D}_{h}^{*}}^{-1}(i,j,k) + \mathcal{G}_{\mathfrak{D}_{t}^{*}}^{-1}(i,j,k) + 1)$ and $\mathcal{G}_{\mathcal{S}}(i,j,k)=1/(\mathcal{G}_{\mathfrak{D}_{v}^{*}}^{-1}(i,j,k)  + \mathcal{G}_{\mathfrak{D}_{t}^{*}}^{-1}(i,j,k)+ 1)$, where $\mathcal{G}_{\mathfrak{D}_{v}^{*}}^{-1}$ and $\mathcal{G}_{\mathfrak{D}_{h}^{*}}^{-1}$ are already defined in the HTV example and 
\begin{equation}
	\mathcal{G}_{\mathfrak{D}_{t}^{*}}^{-1}(i,j,k) = \begin{cases} 1, & \mathrm{if} \: k = 1,n_{3}, \\ 2, & \mathrm{otherwise}, \end{cases}
\end{equation}
$\mathcal{G}_{\mathcal{Y}_{1,1}}(i,j,k) = 1/2$, $\mathcal{G}_{\mathcal{Y}_{1,2}}(i,j,k) = 1/2$, $\mathcal{G}_{\mathcal{Y}_{1,3}}(i,j,k) = 1/2$, $\mathcal{G}_{\mathcal{Y}_{2}}(i,j,k)=1/2$, $\mathcal{G}_{\mathcal{Y}_{3}}(i,j,k)=1/2$, $\forall i\in\{1,\cdots,n_{1}\}$, $\forall j\in\{1,\cdots,n_{2}\}$, and $\forall k\in\{1,\cdots,n_{3}\}$.

\begin{figure}[!t]
	\begin{center}
		\begin{minipage}{0.49\hsize}
			\centerline{\includegraphics[width=\hsize]{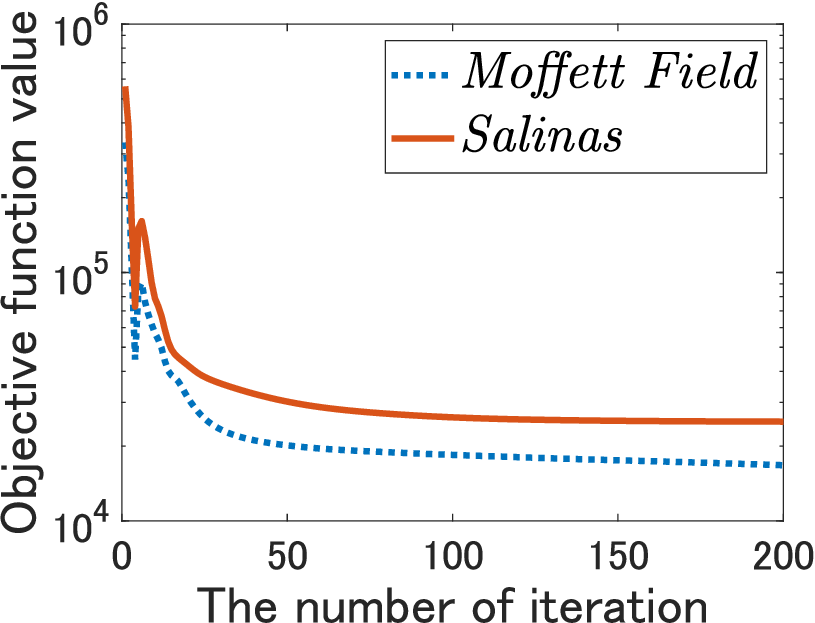}} %画像
		\end{minipage}
		\begin{minipage}{0.49\hsize}
			\centerline{\includegraphics[width=\hsize]{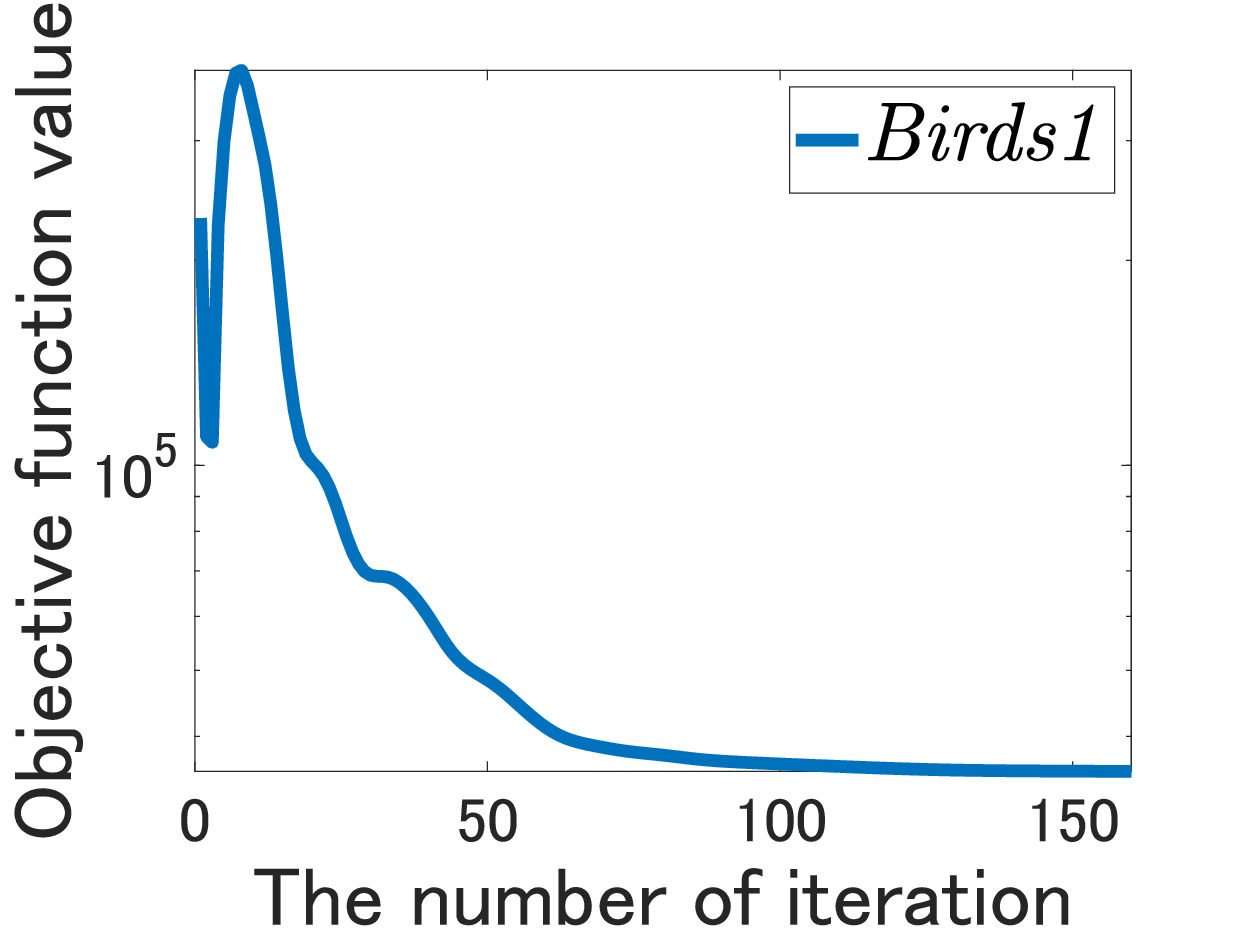}} %画像
		\end{minipage}
		
		\vspace{2mm}
		
		\begin{minipage}{0.49\hsize}
			\centerline{(a)}
		\end{minipage}
		\begin{minipage}{0.49\hsize}
			\centerline{(b)}
		\end{minipage}
	\end{center}

	\vspace{-5mm}
	
	\caption{Convergence analysis of DP-PDS for that are experimentally performed using two image regulraizations. (a) HSI destriping using HTV (Eq.~\eqref{eq:HTV}). (b) IR video destriping using ATV (Eq.~\eqref{eq:ATV}).}
	\label{fig:converge_rate_analysis}
\end{figure}

\subsection{Computational Cost and Running Time}
The complexities of lines 4, 8, and 9 of Alg.~\ref{algo:DP_PDS_for_zero_gradient_constraint_vt} depend on what image regularization is adopted. When a specific image regularization is not given, we cannot have explicit complexities. All operations of lines 5, 6, 11, 12, 13, and 14 of Alg.~\ref{algo:DP_PDS_for_zero_gradient_constraint_vt} have the complexity of $O(n_{1}n_{2}n_{3})$. Thus, the complexity for each iteration of the algorithm is the larger of $O(n_{1}n_{2}n_{3})$ or the one for the image regularization term.

We measured the actual running times using MATLAB (R2021a) on a Windows 10 computer with an Intel Core i9-10900 3.7GHz processor, 32GB of RAM, and NVIDIA GeForce RTX 3090. The actual running times [s] and total iteration numbers were 13.47 and 932, 5.123 and 317, and 1.064 and 191 for \textit{Moffett Field} destriping using HTV, \textit{Salinas} destriping using HTV, and \textit{Bats1} destriping using ATV, respectively. For the experimental settings, see Sec.~\ref{ssec:Simulated_experiments}.

\begin{figure}[t]
	\centering
	\begin{minipage}{0.25\hsize}
		\centerline{\includegraphics[width=\hsize]{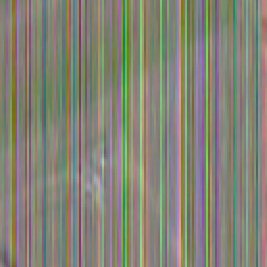}}
	\end{minipage}
	\begin{minipage}{0.06\hsize}
		~
	\end{minipage}
	\begin{minipage}{0.25\hsize}
		\centerline{\includegraphics[width=\hsize]{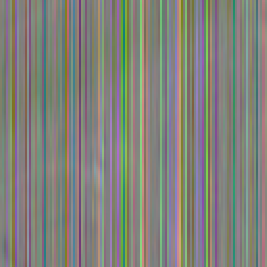}}
	\end{minipage}
	\begin{minipage}{0.06\hsize}
		~
	\end{minipage}
	\begin{minipage}{0.25\hsize}
		\centerline{\includegraphics[width=\hsize]{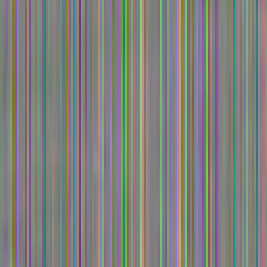}}
	\end{minipage}

\vspace{1mm}
	
	\begin{minipage}{0.32\hsize}
		\centerline{(a) Iteration = 20}
	\end{minipage}
	\begin{minipage}{0.32\hsize}
		\centerline{(b) Iteration = 40}
	\end{minipage}
	\begin{minipage}{0.32\hsize}
		\centerline{(c) Iteration = 60}
	\end{minipage}
	
	\caption{\textit{Salinas} destriping result of $\mathcal{S}^{(n)}$ in each iteration with HTV (R: 140, G: 101, B: 30).}
	\label{fig:transition_S}
\end{figure}

\begin{figure*}[t]

	%1st line
	\begin{minipage}{0.02\hsize}
		\rotatebox[origin=c]{90}{\scriptsize{Estimated HSI}}
	\end{minipage}
	\begin{minipage}{0.13\hsize}
		\centerline{\includegraphics[width=\hsize]{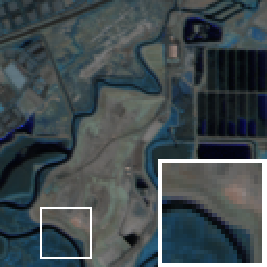}}
	\end{minipage}
	\begin{minipage}{0.13\hsize}
		\centerline{\includegraphics[width=\hsize]{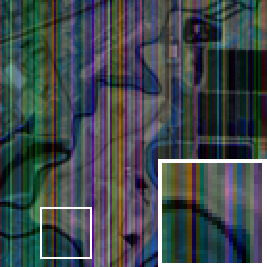}}
	\end{minipage}
	\begin{minipage}{0.13\hsize}
		\centerline{\includegraphics[width=\hsize]{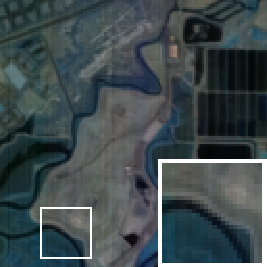}}
	\end{minipage}
	\begin{minipage}{0.13\hsize}
		\centerline{\includegraphics[width=\hsize]{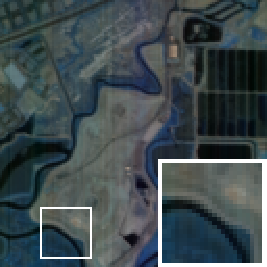}}
	\end{minipage}
	\begin{minipage}{0.13\hsize}
		\centerline{\includegraphics[width=\hsize]{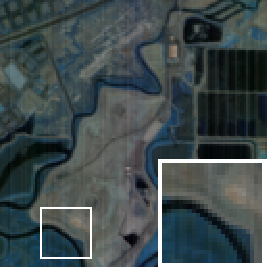}}
	\end{minipage}
	\begin{minipage}{0.13\hsize}
		\centerline{\includegraphics[width=\hsize]{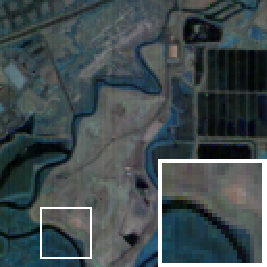}}
	\end{minipage}
	\begin{minipage}{0.13\hsize}
		\centerline{\includegraphics[width=\hsize]{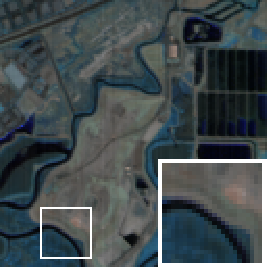}}
	\end{minipage}
	
	%1st line
	\begin{minipage}{0.02\hsize}
		\rotatebox[origin=c]{90}{\scriptsize{Stripe noise}}
	\end{minipage}
	\begin{minipage}{0.13\hsize}
		~
	\end{minipage}
	\begin{minipage}{0.13\hsize}
		\centerline{\includegraphics[width=\hsize]{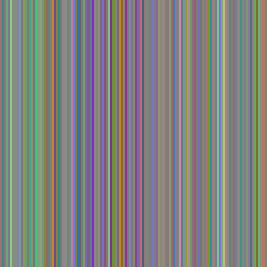}}
	\end{minipage}
	\begin{minipage}{0.13\hsize}
		\centerline{\includegraphics[width=\hsize]{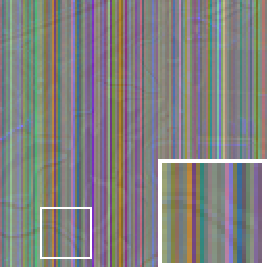}}
	\end{minipage}
	\begin{minipage}{0.13\hsize}
		\centerline{\includegraphics[width=\hsize]{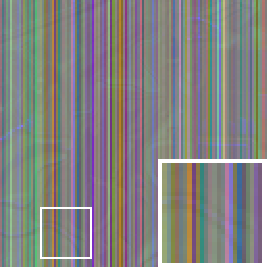}}
	\end{minipage}
	\begin{minipage}{0.13\hsize}
		\centerline{\includegraphics[width=\hsize]{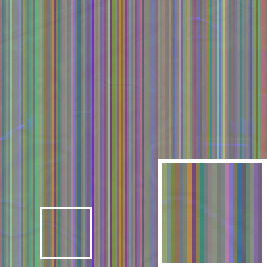}}
	\end{minipage}
	\begin{minipage}{0.13\hsize}
		\centerline{\includegraphics[width=\hsize]{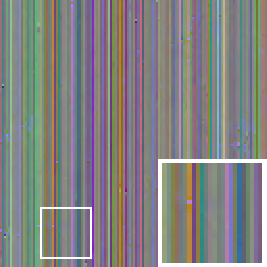}}
	\end{minipage}
	\begin{minipage}{0.13\hsize}
		\centerline{\includegraphics[width=\hsize]{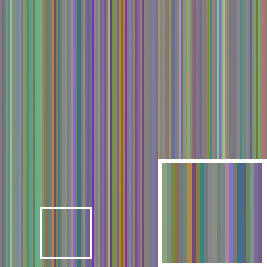}}
	\end{minipage}
	
	\vspace{1mm}

	\begin{minipage}{0.02\hsize}
		~
	\end{minipage}
	\begin{minipage}{0.13\hsize}
		\centerline{\small{(a) Ground-truth}}
	\end{minipage}
	\begin{minipage}{0.13\hsize}
		\centerline{\small{(b) Observed}}
	\end{minipage}
	\begin{minipage}{0.13\hsize}
		\centerline{\small{(c) S~\cite{LRMR}}}
	\end{minipage}
	\begin{minipage}{0.13\hsize}
		\centerline{\small{(d) GS~\cite{GLSSTV}}}
	\end{minipage}
	\begin{minipage}{0.13\hsize}
		\centerline{\small{(e) LR~\cite{NN_char}}}
	\end{minipage}
	\begin{minipage}{0.13\hsize}
		\centerline{\small{(f) TV~\cite{gradient_constraint}}}
	\end{minipage}
	\begin{minipage}{0.13\hsize}
		\centerline{\small{(g) FC}}
	\end{minipage}

	\begin{minipage}{0.02\hsize}
		~
	\end{minipage}
	\begin{minipage}{0.13\hsize}
		\centerline{\small{(MPSNR, MSSIM)}}
	\end{minipage}
	\begin{minipage}{0.13\hsize}
		\centerline{\small{~}}
	\end{minipage}
	\begin{minipage}{0.13\hsize}
		\centerline{\small{(35.10, 0.8871)}}
	\end{minipage}
	\begin{minipage}{0.13\hsize}
		\centerline{\small{(35.79, 0.8919)}}
	\end{minipage}
	\begin{minipage}{0.13\hsize}
		\centerline{\small{(36.93, 0.9005)}}
	\end{minipage}
	\begin{minipage}{0.13\hsize}
		\centerline{\small{(35.85, 0.8711)}}
	\end{minipage}
	\begin{minipage}{0.13\hsize}
		\centerline{\small{\textbf{(40.84, 0.9548)}}}
	\end{minipage}

	\caption{\textit{Moffett field} destriping results in Case (i) with SSTV (R: 126, G: 95, B: 74). The MPSNR and MSSIM of our FC are highlighted in bold.}
	\label{fig:SSTV_results_on_MoffettField_stripe_only}
\end{figure*}

\begin{figure*}[!t]
	%1st line
	\begin{minipage}{0.02\hsize}
		\rotatebox[origin=c]{90}{\scriptsize{Estimated IR video}}
	\end{minipage}
	\begin{minipage}{0.13\hsize}
		\centerline{\includegraphics[width=\hsize]{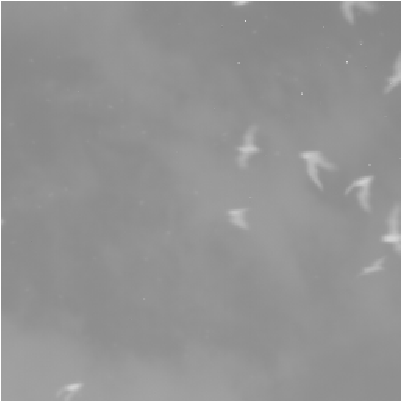}}
	\end{minipage}
	\begin{minipage}{0.13\hsize}
		\centerline{\includegraphics[width=\hsize]{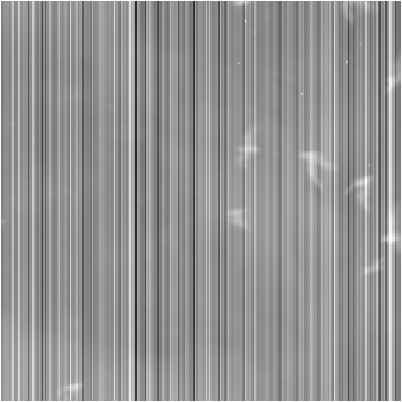}}
	\end{minipage}
	\begin{minipage}{0.13\hsize}
		\centerline{\includegraphics[width=\hsize]{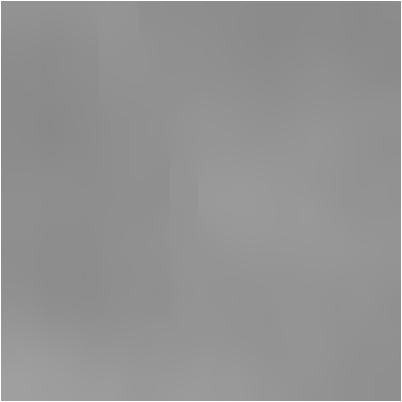}}
	\end{minipage}
	\begin{minipage}{0.13\hsize}
		\centerline{\includegraphics[width=\hsize]{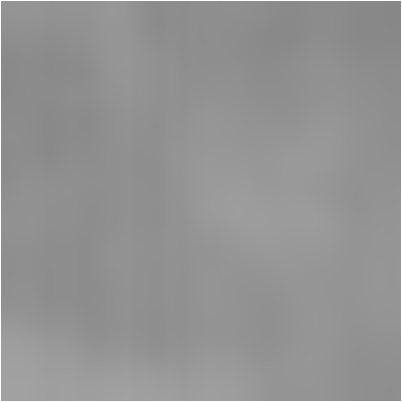}}
	\end{minipage}
	\begin{minipage}{0.13\hsize}
		\centerline{\includegraphics[width=\hsize]{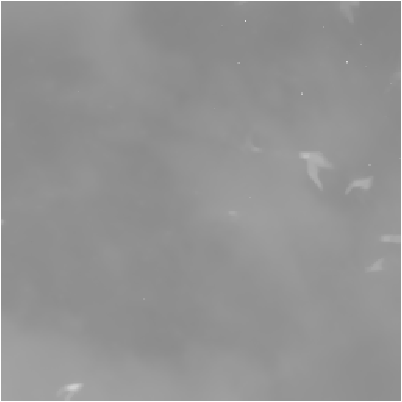}}
	\end{minipage}
	\begin{minipage}{0.13\hsize}
		\centerline{\includegraphics[width=\hsize]{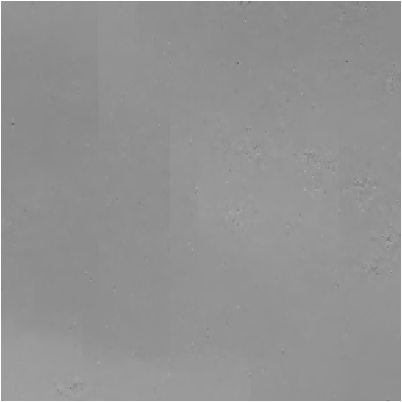}}
	\end{minipage}
	\begin{minipage}{0.13\hsize}
		\centerline{\includegraphics[width=\hsize]{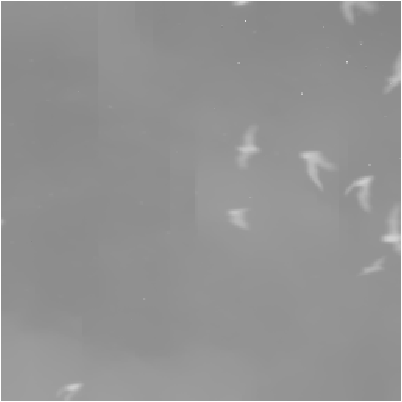}}
	\end{minipage}
	
	%1st line
	\begin{minipage}{0.02\hsize}
		\rotatebox[origin=c]{90}{\scriptsize{Stripe noise}}
	\end{minipage}
	\begin{minipage}{0.13\hsize}
		~
	\end{minipage}
	\begin{minipage}{0.13\hsize}
		\centerline{\includegraphics[width=\hsize]{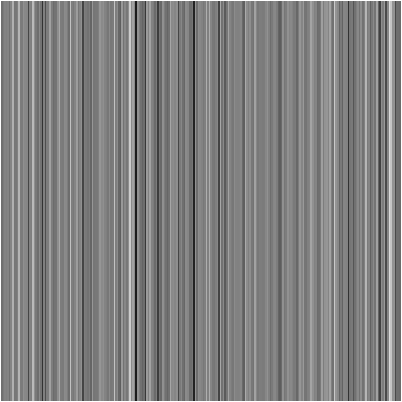}}
	\end{minipage}
	\begin{minipage}{0.13\hsize}
		\centerline{\includegraphics[width=\hsize]{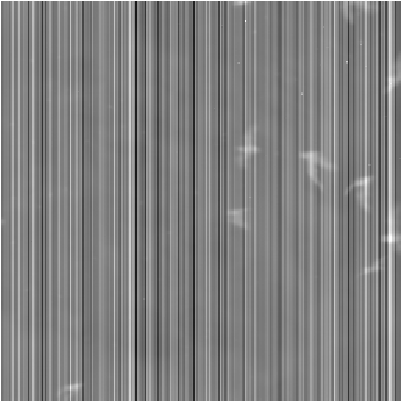}}
	\end{minipage}
	\begin{minipage}{0.13\hsize}
		\centerline{\includegraphics[width=\hsize]{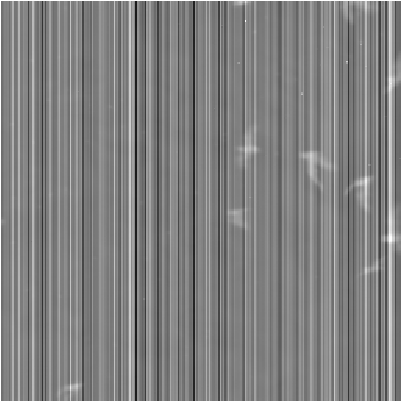}}
	\end{minipage}
	\begin{minipage}{0.13\hsize}
		\centerline{\includegraphics[width=\hsize]{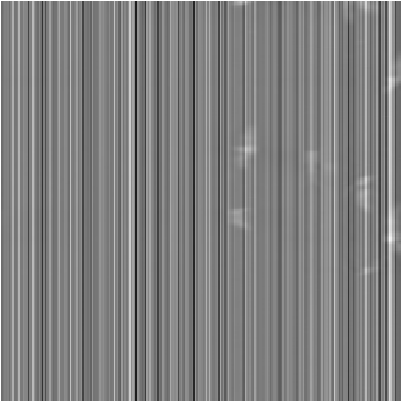}}
	\end{minipage}
	\begin{minipage}{0.13\hsize}
		\centerline{\includegraphics[width=\hsize]{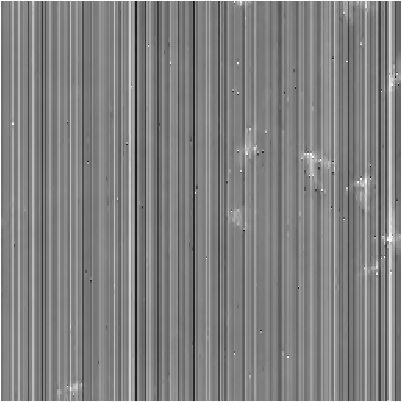}}
	\end{minipage}
	\begin{minipage}{0.13\hsize}
		\centerline{\includegraphics[width=\hsize]{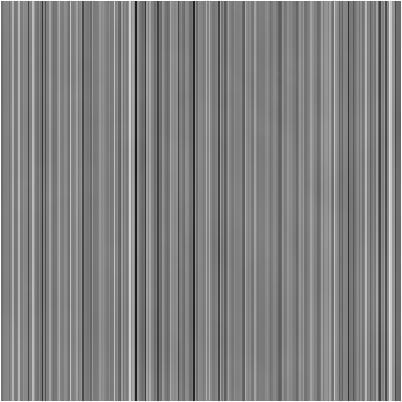}}
	\end{minipage}
	
	\vspace{1mm}
	
	\begin{minipage}{0.02\hsize}
		~
	\end{minipage}
	\begin{minipage}{0.13\hsize}
		\centerline{\small{(a) Ground-truth}}
	\end{minipage}
	\begin{minipage}{0.13\hsize}
		\centerline{\small{(b) Observed}}
	\end{minipage}
	\begin{minipage}{0.13\hsize}
		\centerline{\small{(c) S~\cite{LRMR}}}
	\end{minipage}
	\begin{minipage}{0.13\hsize}
		\centerline{\small{(d) GS~\cite{GLSSTV}}}
	\end{minipage}
	\begin{minipage}{0.13\hsize}
		\centerline{\small{(e) LR~\cite{NN_char}}}
	\end{minipage}
	\begin{minipage}{0.13\hsize}
		\centerline{\small{(f) TV~\cite{gradient_constraint}}}
	\end{minipage}
	\begin{minipage}{0.13\hsize}
		\centerline{\small{(g) FC}}
	\end{minipage}

	\begin{minipage}{0.02\hsize}
		~
	\end{minipage}
	\begin{minipage}{0.13\hsize}
		\centerline{\small{(MPSNR, MSSIM)}}
	\end{minipage}
	\begin{minipage}{0.13\hsize}
		\centerline{\small{~}}
	\end{minipage}
	\begin{minipage}{0.13\hsize}
		\centerline{\small{(31.89, 0.9541)}}
	\end{minipage}
	\begin{minipage}{0.13\hsize}
		\centerline{\small{(32.36, 0.9552)}}
	\end{minipage}
	\begin{minipage}{0.13\hsize}
		\centerline{\small{(38.67, 0.9945)}}
	\end{minipage}
	\begin{minipage}{0.13\hsize}
		\centerline{\small{(31.76, 0.9420)}}
	\end{minipage}
	\begin{minipage}{0.13\hsize}
		\centerline{\small{\textbf{(41.94, 0.9953)}}}
	\end{minipage}

	\caption{\textit{Bats1} destriping results in Case (ii) with ATV. The MPSNR and MSSIM of our FC are highlighted in bold.}
	\label{fig:TV1_results_on_test1_stripe_only}
\end{figure*}

\begin{table}
	\begin{center}
		\caption{All Methods (Stripe Noise Characterization+Image Regularization Examined in Our Experiments)}
		\label{tab:Reg_and_strcha}
		\scalebox{0.9}{
		\begin{tabular}{cccccc}
			\toprule
			\diagbox[height=15mm,width=32mm]{\begin{tabular}{l}Image \\ regularization\end{tabular}}{\begin{tabular}{r}Characterization of \\ stripe noise\end{tabular}}& S~\cite{LRMR} & GS~\cite{GLSSTV} & LR~\cite{NN_char} & TV~\cite{gradient_constraint} & FC \\
			\cmidrule(lr){1-6}
			HTV (HSI)& \cite{HTV}& \cite{GLSSTV}& \cite{NN_char}& \cite{gradient_constraint}& Ours\\
			SSTV (HSI)& \cite{SSTV}& None& None& None& Ours\\
			ASSTV (HSI)& \cite{ASSTV}& None& None& None& Ours\\
			TNN (HSI)& \cite{LRTR}& None& None& None& Ours\\
			SSTV+TNN (HSI)& \cite{SSTV_LRTD}& None& None& None& Ours\\
			$l_{0}$-$l_{1}$HTV (HSI)& \cite{l0l1HTV}& None& None& None& Ours\\
			\cmidrule(lr){1-6}
			ATV (IR video)& \cite{video_TV}& None& None& None& Ours\\
			ITV (IR video)& \cite{video_TV}& None& None& None& Ours\\
			ATV+NN (IR video)& \cite{video_TV_NN}& None& None& None& Ours\\
			\bottomrule
		\end{tabular}}
	\end{center}
\end{table}

\begin{table*}[t]
	\begin{center}
		\caption{MPSNRs and MSSIMs of the HSI Destriping Results in Case (i)}
		\label{tab:PSNR_and_SSIM_so}
		%		\begin{tabular}{|c|c|c|cc|c|c|cc|c|c|c|c|}
		\scriptsize
		%		\tiny
		\vspace{-2mm}
		\begin{tabular}{ccccccccccccc}
			\toprule
			\multirow{2}{*}{Image data} & Range of & Regularization & \multicolumn{5}{c}{MPSNR} & \multicolumn{5}{c}{MSSIM} \\ \addlinespace[-1pt]
			\cmidrule(lr){4-8} 
			\cmidrule(lr){9-13} \addlinespace[-1pt]
			& stripe noise & function & S~\cite{LRMR} & GS~\cite{GLSSTV} & LR~\cite{NN_char} & TV~\cite{gradient_constraint} & FC & S~\cite{LRMR} & GS~\cite{GLSSTV} & LR~\cite{NN_char} & TV~\cite{gradient_constraint} & FC \\ 
			\addlinespace[-1pt]\midrule
			& \multirow{5}{*}{$[-0.2,0.2]$} 
			& HTV
			& 28.70 & 32.00 & \textbf{38.29} & \underline{37.65} & 37.35
			& 0.7601 & 0.8980 & 0.9835 & \textbf{0.9937} & \underline{0.9929}\\
			& & SSTV
			& 36.09 & 36.34 & \underline{38.61} & 36.69 & \textbf{41.00}
			& 0.9344 & 0.9411 & \underline{0.9628} & 0.9266 & \textbf{0.9751}\\
			& & ASSTV
			& 36.92 & 38.92 & \textbf{41.64} & 37.29 & \underline{39.29}
			& 0.9656 & 0.9756 & \textbf{0.9922} & 0.9646 & \underline{0.9900}\\
			& & TNN
			& 21.78 & 25.48 & 25.47 & \textbf{28.36} & \textbf{28.36}
			& 0.3230 & 0.5227 & 0.9404 & \textbf{0.9589} & \textbf{0.9589}\\
			& & SSTV+TNN
			& 32.95 & 34.54 & \textbf{39.04} & 37.58 & \underline{37.61}
			& 0.9285 & 0.9276 & \textbf{0.9889} & 0.9735 & \underline{0.9799}\\
			& & $l_{0}$-$l_{1}$HTV
			& 36.66 & 35.96 & \textbf{41.53} & 38.52 & \underline{39.92} &
			0.9515 & 0.9429 & \textbf{0.9877} & 0.9655 & \underline{0.9837} \\
			\addlinespace[-1pt]\cmidrule(lr){2-13}\addlinespace[-1pt]
			& \multirow{5}{*}{$[-0.25,0.25]$} 
			& HTV
			& 28.51 & 31.63 & \textbf{37.17} & \underline{37.21} & 36.92 &
			0.8309 & 0.8949 & 0.9825 & \textbf{0.9930} & \underline{0.9925} \\
			& & SSTV
			& 35.84 & 36.10 & \underline{37.75} & 36.55 & \textbf{40.78} &
			0.9340 & 0.9407 & \underline{0.9587} & 0.9264 & \textbf{0.9775} \\
			& & ASSTV
			& 36.79 & 38.69 & \textbf{40.87} & 37.18 & \underline{38.82} &
			0.9652 & 0.9753 & \textbf{0.9918} & 0.9644 & \underline{0.9791} \\
			& & TNN
			& 22.75 & 24.79 & 24.86 & \textbf{28.25} & \textbf{28.25} &
			0.4181 & 0.4739 & 0.9335 & \textbf{0.9480} & \textbf{0.9480} \\
			& & SSTV+TNN
			& 32.66 & 34.26 & \textbf{37.89} & 37.21 & \underline{37.25} &
			0.9263 & 0.9260 & \textbf{0.9852} & 0.9737 & \underline{0.9798} \\
			& & $l_{0}$-$l_{1}$HTV
			& 36.38 & 35.70 & \textbf{40.87} & 38.39 & \underline{39.69} &
			0.9502 & 0.9412 & \textbf{0.9873} & 0.9651 & \underline{0.9833} \\
			\addlinespace[-1pt]\cmidrule(lr){2-13}\addlinespace[-1pt]
			& \multirow{5}{*}{$[-0.3,0.3]$} 
			& HTV 
			& 28.41 & 31.46 & 36.37 & \textbf{37.07} & \underline{36.78} 
			& 0.8292 & 0.8929 & 0.9817 & \textbf{0.9928} & \underline{0.9924} \\
			& & SSTV 
			& 35.73 & 35.96 & \underline{36.90} & 36.39 & \textbf{40.55} 
			& 0.9328 & 0.9393 & \underline{0.9509} & 0.9248 & \textbf{0.9764} \\
			\textit{Salinas} & & ASSTV 
			& 36.70 & 38.62 & \textbf{40.35} & 37.11 & \underline{38.76} 
			& 0.9648 & 0.9797 & \textbf{0.9914} & 0.9640 & \underline{0.9789} \\
			& & TNN 
			& 22.30 & 24.06 & 24.19 & \textbf{28.04} & \textbf{28.04} 
			& 0.6844 & 0.7069 & \textbf{0.9224} & \underline{0.9179} & \underline{0.9179} \\
			& & SSTV+TNN
			& 32.53 & 34.14 & \underline{37.13} & \underline{37.13} & \textbf{37.14} 
			& 0.9252 & 0.9250 & \textbf{0.9839} & 0.9736 & \underline{0.9796} \\
			& & $l_{0}$-$l_{1}$HTV
			& 36.25 & 35.57 & \textbf{40.16} & 38.24 & \underline{39.52} &
			0.9492 & 0.9402 & \textbf{0.9868} & 0.9647 & \underline{0.9831} \\
			\addlinespace[-1pt]\cmidrule(lr){2-13}\addlinespace[-1pt]
			& \multirow{5}{*}{$[-0.35,0.35]$} 
			& HTV 
			& 28.34 & 31.38 & 36.07 & \textbf{36.84} & \underline{36.62} &
			0.8281 & 0.8916 & 0.9815 & \textbf{0.9925} & \underline{0.9922} \\
			& & SSTV 
			& 35.60 & 35.86 & 35.86 & \underline{36.10} & \textbf{40.07} &
			0.9301 & 0.9374 & \underline{0.9312} & 0.9216 & \textbf{0.9743} \\
			& & ASSTV 
			& 36.52 & 38.38 & \textbf{39.68} & 36.95 & \underline{38.54} &
			0.9638 & 0.9736 & \textbf{0.9905} & 0.9632 & \underline{0.9781} \\
			& & TNN 
			& 21.83 & 23.36 & 23.59 & \textbf{27.36} & \textbf{27.36} &
			0.3349 & 0.3871 & \textbf{0.9051} & \underline{0.8355} & \underline{0.8355} \\
			& & SSTV+TNN
			& 32.47 & 34.17 & 36.71 & \underline{37.04} & \textbf{37.14} &
			0.9252 & 0.9257 & \textbf{0.9824} & 0.9727 & \underline{0.9794} \\
			& & $l_{0}$-$l_{1}$HTV
			& 36.10 & 35.45 & \textbf{39.31} & 38.02 & \underline{39.23} &
			0.9485 & 0.9393 & \textbf{0.9858} & 0.9641 & \underline{0.9825} \\
			\addlinespace[-1pt]\cmidrule(lr){2-13}\addlinespace[-1pt]
			& \multirow{5}{*}{$[-0.4,0.4]$} 
			& HTV
			& 28.28 & 31.27 & 35.02 & \textbf{36.82} & \underline{36.65}
			& 0.8268 & 0.8898 & 0.9796 & \underline{0.9922} & \textbf{0.9931}\\
			& & SSTV
			& 35.68 & 35.96 & 35.49 & \underline{36.14} & \textbf{40.16}
			& 0.9287 & 0.9360 & \underline{0.9353} & 0.9218 & \textbf{0.9726}\\
			& & ASSTV
			& 36.57 & 38.52 & \textbf{39.68} & 37.02 & \underline{39.09}
			& 0.9638 & 0.9742 & \textbf{0.9950} & 0.9633 & \underline{0.9898}\\
			& & TNN
			& 21.36 & 22.68 & 23.07 & \textbf{26.35} & \textbf{26.35}
			& 0.3015 & 0.3521 & \textbf{0.9003} & \underline{0.7340} & 0.7339\\
			& & SSTV+TNN
			& 32.44 & 34.14 & 36.41 & \underline{37.13} & \textbf{37.25}
			& 0.9238 & 0.9227 & \textbf{0.9865} & 0.9731 & \underline{0.9799}\\
			& & $l_{0}$-$l_{1}$HTV
			& 36.26 & 35.53 & \underline{38.56} & 38.30 & \textbf{39.60} &
			0.9488 & 0.9387 & \textbf{0.9857} & 0.9648 & \underline{0.9833} \\
			\addlinespace[-1pt]\midrule
			& \multirow{5}{*}{$[-0.2,0.2]$} 
			& HTV
			& 27.95 & 29.32 & \textbf{36.88} & 36.07 & \underline{36.18} 
			& 0.6351 & 0.7237 & \textbf{0.9199} & \underline{0.9165} & 0.9139\\
			& & SSTV
			& 35.33 & 35.97 & \underline{38.69} & 36.17 & \textbf{40.91} 
			& 0.8926 & 0.8952 & \underline{0.9285} & 0.8825 & \textbf{0.9535}\\
			& & ASSTV
			& 30.45 & 32.68 & \textbf{44.31} & 35.05 & \underline{38.99} 
			& 0.8418 & 0.8898 & \textbf{0.9847} & 0.9222 & \underline{0.9691}\\
			& & TNN
			& 24.51 & 26.26 & 32.67 & \textbf{35.54} & \textbf{35.54} 
			& 0.4283 & 0.5301 & 0.7779 & \textbf{0.9390} & \textbf{0.9390}\\
			& & SSTV+TNN
			& 32.63 & 35.14 & \textbf{39.94} & 37.38 & \underline{38.00} 
			& 0.8682 & 0.8857 & \underline{0.9479} & 0.9465 & \textbf{0.9481}\\
			& & $l_{0}$-$l_{1}$HTV
			& 35.51 & 35.14 & \textbf{41.24} & 37.77 & \underline{39.17} &
			0.8984 & 0.8834 & \textbf{0.9551} & 0.9269 & \underline{0.9429} \\
			\addlinespace[-1pt]\cmidrule(lr){2-13}\addlinespace[-1pt]
			& \multirow{5}{*}{$[-0.25,0.25]$} 
			& HTV
			& 27.62 & 29.19 & \textbf{36.36} & 36.06 & \underline{36.19} &
			0.6216 & 0.7201 & \underline{0.9151} & \textbf{0.9165} & 0.9144 \\
			& & SSTV
			& 35.39 & 36.05 & \underline{37.86} & 37.32 & \textbf{41.07} &
			0.8960 & 0.8998 & \underline{0.9218} & 0.9092 & \textbf{0.9580} \\
			& & ASSTV
			& 30.37 & 32.59 & \textbf{44.07} & 33.76 & \underline{38.97} &
			0.8411 & 0.8886 & \textbf{0.9840} & 0.9111 & \underline{0.9703} \\
			& & TNN
			& 23.9 & 25.44 & 31.71 & \textbf{35.74} & \textbf{35.74} &
			0.3846 & 0.4781 & 0.7496 & \textbf{0.9342} & \textbf{0.9342} \\
			& & SSTV+TNN
			& 32.57 & 35.29 & \textbf{39.50} & 37.34 & \underline{38.02} &
			0.8701 & 0.8909 & 0.9450 & \underline{0.9477} & \textbf{0.9498} \\
			& & $l_{0}$-$l_{1}$HTV
			& 35.56 & 35.24 & \textbf{40.36} & 37.75 & \underline{39.25} &
			0.9025 & 0.8878 & \textbf{0.9526} & 0.9303 & \underline{0.9466} \\
			\addlinespace[-1pt]\cmidrule(lr){2-13}\addlinespace[-1pt]
			& \multirow{5}{*}{$[-0.3,0.3]$} 
			& HTV
			& 27.18 & 29.05 & 35.72 & \underline{35.96} & \textbf{36.07} 
			& 0.6100 & 0.7135 & 0.9014 & \underline{0.9133} & \textbf{0.9107} \\
			& & SSTV 
			& 35.10 & 35.79 & \underline{36.93} & 35.85 & \textbf{40.84} 
			& 0.8871 & 0.8919 & \underline{0.9005} & 0.8711 & \textbf{0.9548} \\
			\textit{Moffett Field} & & ASSTV 
			& 38.31 & 39.59 & \textbf{43.53} & 34.81 & \underline{38.85} 
			& 0.9672 & 0.9689 & \textbf{0.9819} & 0.9201 & \underline{0.9691} \\
			& & TNN 
			& 23.33 & 24.65 & 30.66 & \textbf{36.59} & \underline{36.41} 
			& 0.3546 & 0.4308 & 0.7108 & \textbf{0.8873} & \underline{0.8786} \\
			& & SSTV+TNN
			& 32.33 & 35.04 & \textbf{38.72} & 37.28 & \underline{37.95 }
			& 0.8634 & 0.8851 & 0.9342 & \underline{0.9464} & \textbf{0.9487} \\
			& & $l_{0}$-$l_{1}$HTV
			& 35.28 & 34.93 & \textbf{39.85} & 37.69 & \underline{39.02} &
			0.8936 & 0.8795 & \textbf{0.9410} & 0.9243 & \underline{0.9399} \\
			\addlinespace[-1pt]\cmidrule(lr){2-13}\addlinespace[-1pt]
			& \multirow{5}{*}{$[-0.35,0.35]$} 
			& HTV
			& 27.20 & 28.96 & 35.27 & \underline{35.83} & \textbf{35.93} &
			0.6100 & 0.7135 & 0.9014 & \textbf{0.9133} & \underline{0.9107} \\
			& & SSTV 
			& 35.10 & 35.80 & 36.42 & \underline{37.01} & \textbf{40.47} &
			0.8871 & 0.8919 & 0.9005 & \underline{0.9045} & \textbf{0.9548} \\
			& & ASSTV 
			& 30.27 & 32.46 & \textbf{43.10} & 33.67 & \underline{38.78} &
			0.8399 & 0.8875 & \textbf{0.9819} & 0.9087 & \underline{0.9691} \\
			& & TNN 
			& 22.80 & 23.94 & 30.32 & \underline{36.73} & \textbf{36.75} &
			0.3450 & 0.4308 & 0.7108 & \underline{0.8873} & \textbf{0.8786} \\
			& & SSTV+TNN
			& 32.31 & 34.95 & \textbf{38.36} & 37.13 & \underline{37.71} &
			0.8634 & 0.8851 & 0.9342 & \underline{0.9464} & \textbf{0.9487} \\
			& & $l_{0}$-$l_{1}$HTV
			& 35.34 & 34.91 & \textbf{39.33} & 37.65 & \underline{38.99} &
			0.8972 & 0.8801 & \textbf{0.9421} & 0.9268 & \underline{0.9420} \\
			\addlinespace[-1pt]\cmidrule(lr){2-13}\addlinespace[-1pt]
			& \multirow{5}{*}{$[-0.4,0.4]$} 
			& HTV
			& 26.98 & 28.80 & 34.47 & \underline{35.57} & \textbf{35.67}
			& 0.5959 & 0.7015 & 0.8820 & \textbf{0.9039} & \underline{0.9014}\\
			& & SSTV
			& 34.96 & \underline{35.61} & 35.57 & \underline{35.61} & \textbf{40.27}
			& 0.8816 & 0.8871 & \underline{0.8789} & 0.8634 & \textbf{0.9498}\\
			& & ASSTV
			& 30.29 & 32.47 & \textbf{42.75} & 34.83 & \underline{38.96}
			& 0.8408 & 0.8878 & \textbf{0.9763} & 0.9187 & \underline{0.9691}\\
			& & TNN
			& 22.27 & 23.26 & 30.00 & \underline{35.91} & \textbf{36.75}
			& 0.2867 & 0.3602 & 0.6974 & \underline{0.8848} & \textbf{0.8935}\\
			& & SSTV+TNN
			& 32.14 & 34.74 & \underline{37.24} & 37.09 & \textbf{37.66}
			& 0.8563 & 0.8801 & 0.9179 & \underline{0.9415} & \textbf{0.9425}\\
			& & $l_{0}$-$l_{1}$HTV
			& 35.13 & 34.74 & \underline{38.24} & 37.36 & \textbf{38.60} &
			0.8886 & 0.8749 & \underline{0.9254} & 0.9169 & \textbf{0.9321} \\
			\bottomrule
		\end{tabular}
	\end{center}
\end{table*}

%%%%%%%%%%%%%%%%%%%%%%%%%%%%%%%%%%%%%%%%%%%%%%%%%%%%%%%%%%%%%%%%%%%%
%% PSNR, SSIM and time of simulated destriping results
%%%%%%%%%%%%%%%%%%%%%%%%%%%%%%%%%%%%%%%%%%%%%%%%%%%%%%%%%%%%%%%%%%%%
\begin{table*}[t]
	\begin{center}
		\caption{MPSNRs and MSSIMs of the IR Destriping Results in Case (ii)}
		\label{tab:PSNR_and_SSIM_so_IR_video}
		%		\begin{tabular}{|c|c|c|cc|c|c|cc|c|c|c|c|}
		\scriptsize
		\begin{tabular}{ccccccccccccc}
			\toprule
			\multirow{2}{*}{IR video data} & Range of & Regularization & \multicolumn{5}{c}{MPSNR} & \multicolumn{5}{c}{MSSIM} \\
			\cmidrule(lr){4-8} 
			\cmidrule(lr){9-13}
			& stripe noise & function & S~\cite{LRMR} & GS~\cite{GLSSTV} & LR~\cite{NN_char} & TV~\cite{gradient_constraint} & FC & S~\cite{LRMR} & GS~\cite{GLSSTV} & LR~\cite{NN_char} & TV~\cite{gradient_constraint} & FC \\ 
			\midrule
			
			& \multirow{3}{*}{$[-0.2,0.2]$} 
			& ATV
			& 30.15 & 30.48 & \underline{34.85} & 29.97 & \textbf{36.53}
			& 0.9532 & 0.9540 & \underline{0.9955} & 0.9400 & \textbf{0.9956}\\
			& & ITV
			& 30.15 & 30.53 & \underline{34.26} & 29.98 & \textbf{36.53}
			& 0.9532 & 0.9524 & \underline{0.9935} & 0.9414 & \textbf{0.9957} \\
			& & ATV+NN
			& 30.18 & 30.50 & \underline{34.76} & 29.98 & \textbf{35.28}
			& 0.9531 & 0.9530 & \textbf{0.9951} & 0.9486 & \textbf{0.9951}\\
			\cmidrule(lr){2-13}
			& \multirow{3}{*}{$[-0.25,0.25]$} 
			& ATV
			& 30.02 & 30.44 & \underline{32.09} & 29.98 & \textbf{36.82} &
			0.9526 & 0.9540 & \underline{0.9771} & 0.9399 & \textbf{0.9956} \\
			& & ITV
			& 30.02 & 30.44 & \underline{33.45} & 29.99 & \textbf{36.87} &
			0.9525 & 0.9540 & \underline{0.9936} & 0.9410 & \textbf{0.9959} \\
			& & ATV+NN
			& 30.06 & 30.44 & \underline{33.87} & 29.99 & \textbf{35.36} &
			0.9528 & 0.9538 & \underline{0.9950} & 0.9485 & \textbf{0.9951} \\
			\cmidrule(lr){2-13}
			& \multirow{3}{*}{$[-0.3,0.3]$} 
			& ATV
			& 31.89 & 32.36 & \underline{38.67} & 31.76 & \textbf{41.94}
			& 0.9541 & 0.9552 & \underline{0.9945} & 0.9420 & \textbf{0.9953} \\
			\textit{Bats1} & & ITV 
			& 31.90 & 32.37 & \underline{37.85} & 31.82 & \textbf{42.13}
			& 0.9539 & 0.9551 & \underline{0.9939} & 0.9430 & \textbf{0.9954} \\
			& & ATV+NN 
			& 31.90 & 32.37 & \underline{34.96} & 31.91 & \textbf{41.73} 
			& 0.9539 & 0.9551 & 0.9775 & 0.9504 & \textbf{0.9953} \\
			\cmidrule(lr){2-13}
			& \multirow{3}{*}{$[-0.35,0.35]$} 
			& ATV
			& 31.78 & 32.23 & \underline{35.84} & 31.63 & \textbf{40.70} &
			0.9539 & 0.9548 & \underline{0.9914} & 0.9422 & \textbf{0.9955} \\
			& & ITV
			& 31.78 & 32.37 & \underline{34.88} & 31.68 & \textbf{40.54} &
			0.9539 & 0.9553 & \underline{0.9844} & 0.9424 & \textbf{0.9958} \\
			& & ATV+NN
			& 31.82 & 32.39 & \underline{35.69} & 31.98 & \textbf{40.69} &
			0.9540 & 0.9554 & \underline{0.9909} & 0.9503 & \textbf{0.9956} \\
			\cmidrule(lr){2-13}
			& \multirow{3}{*}{$[-0.4,0.4]$} 
			& ATV
			& 31.43 & 31.91 & \underline{34.58} & 31.19 & \textbf{39.10}
			& 0.9537 & 0.9541 & 0.9899 & 0.9378 & \textbf{0.9954} \\
			& & ITV
			& 31.43 & 31.91 & \underline{34.99} & 31.27 & \textbf{39.38}
			& 0.9537 & 0.9541 & \underline{0.9840} & 0.9393 & \textbf{0.9956} \\
			& & ATV+NN
			& 31.37 & 31.85 & \underline{34.48} & 31.36 & \textbf{38.98}
			& 0.9536 & 0.9540 & \underline{0.9900} & 0.9467 & \textbf{0.9953} \\
			\bottomrule
		\end{tabular}
	\end{center}
\end{table*}

\begin{figure*}[t]
	
	%1st line
	\begin{minipage}{0.02\hsize}
		\rotatebox[origin=c]{90}{\scriptsize{Estimated HSI}}
	\end{minipage}
	\begin{minipage}{0.13\hsize}
		\centerline{\includegraphics[width=\hsize]{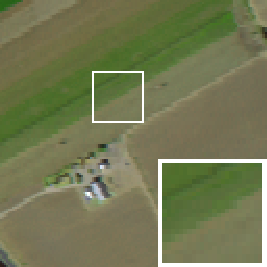}}
	\end{minipage}
	\begin{minipage}{0.13\hsize}
		\centerline{\includegraphics[width=\hsize]{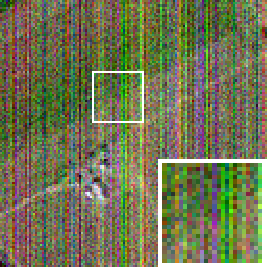}}
	\end{minipage}
	\begin{minipage}{0.13\hsize}
		\centerline{\includegraphics[width=\hsize]{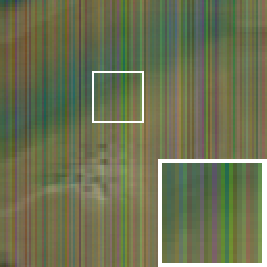}}
	\end{minipage}
	\begin{minipage}{0.13\hsize}
		\centerline{\includegraphics[width=\hsize]{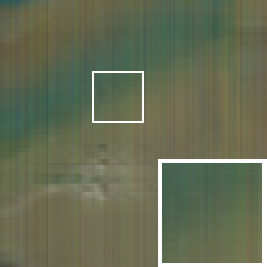}}
	\end{minipage}
	\begin{minipage}{0.13\hsize}
		\centerline{\includegraphics[width=\hsize]{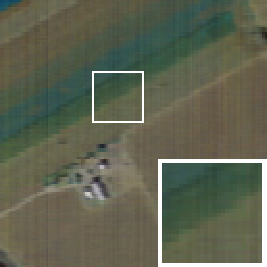}}
	\end{minipage}
	\begin{minipage}{0.13\hsize}
		\centerline{\includegraphics[width=\hsize]{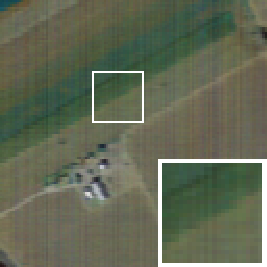}}
	\end{minipage}
	\begin{minipage}{0.13\hsize}
		\centerline{\includegraphics[width=\hsize]{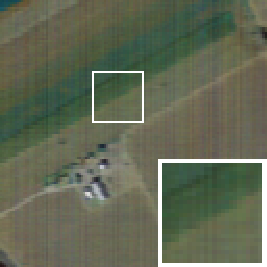}}
	\end{minipage}
	
	%2st line
	\begin{minipage}{0.02\hsize}
		\rotatebox[origin=c]{90}{\scriptsize{Stripe noise}}
	\end{minipage}
	\begin{minipage}{0.13\hsize}
		~
	\end{minipage}
	\begin{minipage}{0.13\hsize}
		\centerline{\includegraphics[width=\hsize]{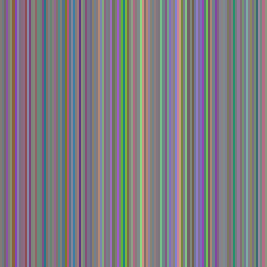}}
	\end{minipage}
	\begin{minipage}{0.13\hsize}
		\centerline{\includegraphics[width=\hsize]{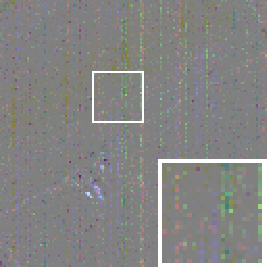}}
	\end{minipage}
	\begin{minipage}{0.13\hsize}
		\centerline{\includegraphics[width=\hsize]{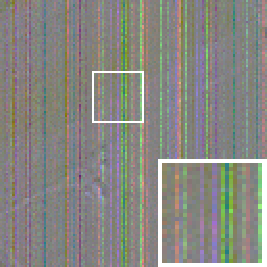}}
	\end{minipage}
	\begin{minipage}{0.13\hsize}
		\centerline{\includegraphics[width=\hsize]{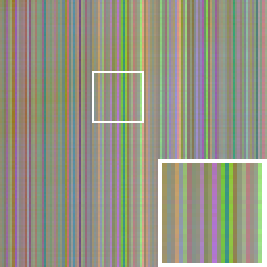}}
	\end{minipage}
	\begin{minipage}{0.13\hsize}
		\centerline{\includegraphics[width=\hsize]{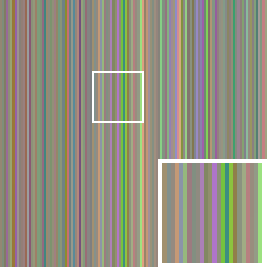}}
	\end{minipage}
	\begin{minipage}{0.13\hsize}
		\centerline{\includegraphics[width=\hsize]{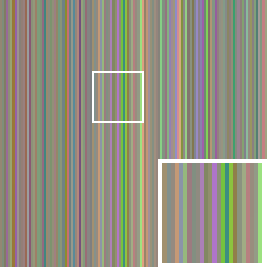}}
	\end{minipage}
	
	\vspace{1mm}
	
	\begin{minipage}{0.02\hsize}
		~
	\end{minipage}
	\begin{minipage}{0.13\hsize}
		\centerline{\small{(a) Ground-truth}}
	\end{minipage}
	\begin{minipage}{0.13\hsize}
		\centerline{\small{(b) Observed}}
	\end{minipage}
	\begin{minipage}{0.13\hsize}
		\centerline{\small{(c) S~\cite{LRMR}}}
	\end{minipage}
	\begin{minipage}{0.13\hsize}
		\centerline{\small{(d) GS~\cite{GLSSTV}}}
	\end{minipage}
	\begin{minipage}{0.13\hsize}
		\centerline{\small{(e) LR~\cite{NN_char}}}
	\end{minipage}
	\begin{minipage}{0.13\hsize}
		\centerline{\small{(f) TV~\cite{gradient_constraint}}}
	\end{minipage}
	\begin{minipage}{0.13\hsize}
		\centerline{\small{(g) FC}}
	\end{minipage}

	\begin{minipage}{0.02\hsize}
		~
	\end{minipage}
	\begin{minipage}{0.13\hsize}
		\centerline{\small{(MPSNR, MSSIM)}}
	\end{minipage}
	\begin{minipage}{0.13\hsize}
		\centerline{\small{~}}
	\end{minipage}
	\begin{minipage}{0.13\hsize}
		\centerline{\small{(22.43, 0.3678)}}
	\end{minipage}
	\begin{minipage}{0.13\hsize}
		\centerline{\small{(23.78, 0.7004)}}
	\end{minipage}
	\begin{minipage}{0.13\hsize}
		\centerline{\small{(23.15, 0.8421)}}
	\end{minipage}
	\begin{minipage}{0.13\hsize}
		\centerline{\small{(28.30, 0.8311)}}
	\end{minipage}
	\begin{minipage}{0.13\hsize}
		\centerline{\small{\textbf{(28.30, 0.8311)}}}
	\end{minipage}

	\caption{\textit{Salinas} destriping results in Case (iii) with TNN (R: 140, G: 101, B: 30). The MPSNR and MSSIM of our FC are highlighted in bold.}
	\label{fig:TNN_results_on_Salinas}
\end{figure*}

\subsection{Convergence Analysis}
The convergence property of Alg.~\ref{algo:DP_PDS_for_zero_gradient_constraint_vt} is given in Appendix~\ref{app:convergence_of_our_algorithm}. Moreover, we experimentally confirm the convergence properties. We plotted the objective function values $\sum_{k=1}^{K}R_{k}(\mathfrak{L}_{k}(\mathcal{U}^{(n)})) + \lambda\|\mathcal{S}^{(n)}\|_{1}$ versus iterations $n$ on the experiments using HTV and ATV in Fig.~\ref{fig:converge_rate_analysis}, where our algorithm minimizes the objective function. Figure~\ref{fig:transition_S} shows \textit{Salinas} destriping results of $\mathcal{S}^{(n)}$ in each iteration. From these results, we can see that the stripe noise becomes flat along the vertical direction as the number of iterations is large. The convergence speed of the stripe noise component depends on what image regularization is adopted.

\section{Experiments}
\label{sec:experiments}
In this section, we illustrate the effectiveness of our framework through comprehensive experiments. Specifically, these experiments aim to show that
\begin{itemize}
	\item Our flatness constraint accurately separates stripe noise from striped images,
	\item Our framework achieves good destriping performance on average, whatever image regularizations are used.
\end{itemize}

The specific experimental procedure is as follows.
\begin{enumerate}
	\item Select image regularizations to be used.
	\item Develop DP-PDS-based solvers for all optimization problems that include all combinations of the image regularizations and the stripe noise characterizations summarized in Table~\ref{tab:Reg_and_strcha}.
	\item Set some parameters such as the weight of image regularization, the gradient regularization weight $\mu$ of the TV-based model, the data-fidelity parameter $\varepsilon$, and the parameter of the sparse term $\lambda$. (Their detailed settings are given in each experimental section).
	\item Conduct destriping experiments using these solvers and parameters.
\end{enumerate}

\subsection{Image Regularizations and Stripe Noise Characterizations}
In HSI experiments, we adopted Hyperspectral Total Variation (HTV)~\cite{HTV}, Spatio-Spectral Total Variation (SSTV)~\cite{SSTV}, Anisotropic Spectral-Spatial Total Variation (ASSTV)~\cite{ASSTV}, Tensor Nuclear Norm (TNN)~\cite{LRTR}, Spatial-Spectral Total Variation with Tensor Nuclear Norm (SSTV+TNN)~\cite{SSTV_LRTD}, and $l_{0}$-$l_{1}$ Hybrid Total Variation ($l_{0}$-$l_{1}$HTV)~\cite{l0l1HTV}, which are often used for HSI regularization. The parameters of ASSTV were experimentally determined as the values that can achieve the best performance. The parameter of SSTV+TNN was set to the values recommended in~\cite{SSTV_LRTD}. In IR video experiments, we adopted Anisotropic Total Variation (ATV), Isotropic Total Variation (ITV)~\cite{video_TV}, and Anisotropic Total Variation with Nuclear Norm (ATV+NN)~\cite{video_TV_NN}, which are known as video regularization. 
We compared the proposed flatness constraint (FC) with the sparsity-based model (S), the group-sparsity-based model (GS), the low-rank-based model (LR), and the TV-based model (TV). For convenience, we denote each method that combines a particular stripe noise characterization and a particular image regularization shortly by connecting each name with a hyphen. For example, the destriping method using the sparsity-based model and HTV is denoted as S-HTV.

\begin{table*}[t]
	\begin{center}
		\caption{MPSNRs and MSSIMs of the HSI Destriping Results in Case (iii)}
		\label{tab:PSNR_and_SSIM_sg}
		\scriptsize
		\vspace{-2mm}
		\begin{tabular}{ccccccccccccc}
			\midrule 
			\multirow{2}{*}{HSI} & Range of & Regularization & \multicolumn{5}{c}{MPSNR} & \multicolumn{5}{c}{MSSIM} \\
			\addlinespace[-1pt]\cmidrule(lr){4-8}\cmidrule(lr){9-13}\addlinespace[-1pt]
			& stripe noise & function & S~\cite{LRMR} & GS~\cite{GLSSTV} & LR~\cite{NN_char} & TV~\cite{gradient_constraint} & FC & S~\cite{LRMR} & GS~\cite{GLSSTV} & LR~\cite{NN_char} & TV~\cite{gradient_constraint} & FC \\ 
			\addlinespace[-1pt]\midrule
			& \multirow{5}{*}{$[-0.2,0.2]$} 
			& HTV
			& 29.16 & 29.30 & 31.05 & \textbf{31.08} & \textbf{31.08}
			& 0.8371 & 0.8371 & \textbf{0.8717} & \underline{0.8698} & \underline{0.8698}\\
			& & SSTV
			& 33.55 & 33.91 & \textbf{34.60} & \underline{34.89} & 34.49
			& 0.8643 & 0.8770 & \underline{0.8910} & 0.8772 & \textbf{0.8913}\\
			& & ASSTV
			& 28.93 & 28.98 & \textbf{29.11} & 28.83 & \underline{28.96}
			& 0.6669 & 0.6268 & 0.6473 & \underline{0.6613} & \textbf{0.6648}\\
			& & TNN
			& 24.10 & 24.35 & 24.15 & \textbf{26.52} & \textbf{26.52}
			& 0.5456 & 0.4218 & 0.8565 & \textbf{0.8662} & \textbf{0.8662}\\
			& & SSTV+TNN
			& 32.38 & 32.95 & \underline{34.63} & \textbf{34.64} & 33.84
			& 0.8893 & 0.9153 & \textbf{0.9382} & \underline{0.9209} & 0.9097\\
			& & $l_{0}$-$l_{1}$HTV
			& 35.58 & 35.90 & \underline{37.10} & 37.09 & \textbf{37.17} &
			0.9384 & 0.9401 & \textbf{0.9523} & 0.9480 & \underline{0.9483} \\
			\addlinespace[-1pt]\cmidrule(lr){2-13}\addlinespace[-1pt]
			& \multirow{5}{*}{$[-0.25,0.25]$} 
			& HTV
			& 28.74 & 29.10 & 30.68 & \textbf{30.98} & \textbf{30.98} &
			0.8317 & 0.8354 & \textbf{0.8702} & \underline{0.8696} & \underline{0.8696} \\
			& & SSTV
			& 33.01 & 33.59 & 34.16 & \textbf{34.39} & \textbf{34.39} &
			0.8575 & 0.8757 & 0.8882 & \textbf{0.8917} & \textbf{0.8917} \\
			& & ASSTV
			& 28.86 & 28.92 & \textbf{29.03} & 28.79 & \underline{28.94} &
			\underline{0.6643} & 0.6260 & 0.6471 & 0.6600 & \textbf{0.6638} \\
			& & TNN
			& 23.34 & 23.89 & 23.67 & \textbf{26.40} & \textbf{26.40} &
			0.4505 & 0.7146 & 0.8511 & \textbf{0.8549} & \textbf{0.8549} \\
			& & SSTV+TNN
			& 31.63 & 32.49 & \underline{34.14} & \textbf{34.39} & 33.51 &
			0.8738 & 0.9103 & \textbf{0.9370} & \underline{0.9194} & 0.9072 \\
			& & $l_{0}$-$l_{1}$HTV
			& 35.12 & 35.28 & 36.71 & \underline{36.91} & \textbf{36.99} &
			0.9360 & 0.9357 & \textbf{0.9518} & 0.9474 & \underline{0.9476} \\
			\addlinespace[-1pt]\cmidrule(lr){2-13}\addlinespace[-1pt]
			& \multirow{5}{*}{$[-0.3,0.3]$} 
			& HTV 
			& 28.35 & 28.97 & 30.36 & \textbf{30.94} & \textbf{30.94} 
			& 0.8247 & 0.8338 & 0.8688 & \textbf{0.8695} & \textbf{0.8695} \\
			& & SSTV 
			& 32.63 & 33.39 & 33.73 & \textbf{36.14} & \underline{34.34}  
			& 0.8495 & 0.8727 & 0.8817 & \textbf{0.9090} & \underline{0.8910} \\
			\textit{Salinas} & & ASSTV 
			& 28.80 & 28.89 & \textbf{28.98} & 28.77 & \underline{28.92}  
			& 0.6624 & 0.6256 & 0.6468 & \underline{0.6591} & \textbf{0.6632} \\
			& & TNN 
			& 22.43 & 23.78 & 23.15 & \underline{26.30} & \textbf{28.30} 
			& 0.3678 & 0.7004 & \textbf{0.8421} & \underline{0.8311} & \underline{0.8311} \\
			& & SSTV+TNN
			& 31.14 & 32.24 & \underline{33.79} & \textbf{34.29} & 33.41 
			& 0.8611 & 0.9067 & \textbf{0.9362} & \underline{0.9185} & 0.9058 \\
			& & $l_{0}$-$l_{1}$HTV
			& 34.80 & 34.85 & 36.35 & \underline{36.84} & \textbf{36.92} &
			0.9337 & 0.9318 & \textbf{0.9513} & 0.9472 & \underline{0.9474} \\
			\addlinespace[-1pt]\cmidrule(lr){2-13}\addlinespace[-1pt]
			& \multirow{5}{*}{$[-0.35,0.35]$} 
			& HTV 
			& 28.03 & 28.86 & 30.17 & \textbf{30.90} & \textbf{30.90} &
			0.8166 & 0.8323 & 0.8680 & \textbf{0.8693} & \textbf{0.8693} \\
			& & SSTV 
			& 32.30 & 33.17 & 33.22 & \textbf{34.20} & \textbf{34.20} &
			0.8392 & 0.8668 & 0.8649 & \textbf{0.8881} & \textbf{0.888} \\
			& & ASSTV 
			& 28.74 & 28.85 & \textbf{28.91} & 28.75 & \textbf{28.91} &
			\underline{0.6608} & 0.6251 & 0.6463 & 0.6586 & \textbf{0.6628} \\
			& & TNN 
			& 21.50 & 23.69 & 22.69 & \textbf{26.10} & \textbf{26.10} &
			0.3066 & 0.6758 & \textbf{0.8276} & \underline{0.7774} & \underline{0.7774} \\
			& & SSTV+TNN
			& 30.84 & 32.10 & \underline{33.58} & \textbf{34.22} & 33.37 &
			0.8527 & 0.9047 & \textbf{0.9349} & \underline{0.9179} & 0.9052 \\
			& & $l_{0}$-$l_{1}$HTV
			& 34.49 & 34.50 & 36.00 & \underline{36.72} & \textbf{36.80} &
			0.9306 & 0.9281 & \textbf{0.9501} & 0.9468 & \underline{0.9471} \\
			\addlinespace[-1pt]\cmidrule(lr){2-13}\addlinespace[-1pt]
			& \multirow{5}{*}{$[-0.4,0.4]$} 
			& HTV
			& 27.69 & 28.73 & 30.12 & \textbf{30.88} & \textbf{30.88}
			& 0.8098 & 0.8309 & 0.8599 & \textbf{0.8688} & \textbf{0.8688}\\
			& & SSTV
			& 32.17 & 33.17 & 32.99 & \textbf{34.60} & \underline{34.31}
			& 0.8370 & 0.8671 & 0.8686 & \underline{0.8730} & \textbf{0.8898}\\
			& & ASSTV
			& 28.70 & 28.84 & \underline{28.86} & 28.73 & \textbf{28.90}
			& 0.6598 & 0.6251 & 0.6460 & \underline{0.6581} & \textbf{0.6623}\\
			& & TNN
			& 21.55 & 23.47 & 22.26 & \textbf{25.54} & \textbf{25.54}
			& 0.6607 & 0.6198 & \textbf{0.8237} & \underline{0.6978} & \underline{0.6978}\\
			& & SSTV+TNN
			& 30.58 & 31.99 & 33.25 & \textbf{34.22} & \underline{33.38}
			& 0.8462 & 0.9029 & \textbf{0.9347} & \underline{0.9173} & 0.9046\\
			& & $l_{0}$-$l_{1}$HTV
			& 34.39 & 34.31 & \underline{35.58} & 36.86 & \textbf{36.94} &
			0.9304 & 0.9266 & \textbf{0.9500} & 0.9471 & \underline{0.9474} \\
			\addlinespace[-1pt]\midrule
			& \multirow{5}{*}{$[-0.2,0.2]$} 
			&HTV
			& 27.44 & 28.04 & 29.01 & \underline{29.10} & \textbf{29.15}
			& 0.6387 & 0.6467 & 0.7043 & \underline{0.7153} & \textbf{0.7275}\\
			& & SSTV
			& 33.51 & 33.90 & \textbf{34.35} & 33.78 & \textbf{34.35}
			& 0.8421 & 0.8466 & \textbf{0.8574} & 0.8317 & \underline{0.8548}\\
			& & ASSTV
			& 27.08 & 28.00 & \underline{28.02} & 28.00 & \textbf{28.18}
			& 0.6118 & 0.6327 & \textbf{0.6353} & 0.6234 & \underline{0.6290}\\
			& & TNN
			& 23.91 & 25.50 & 29.13 & \textbf{31.29} & \textbf{31.29}
			& 0.5512 & 0.6240 & 0.7505 & \textbf{0.8581} & \textbf{0.8581}\\
			& & SSTV+TNN
			& 32.40 & 33.19 & \textbf{34.78} & 33.85 & \underline{33.92}
			& 0.8413 & 0.8406 & \textbf{0.8933} & \underline{0.8783} & 0.8698\\
			& & $l_{0}$-$l_{1}$HTV
			&  33.98 &  34.60 & \underline{35.90} & 35.76 & \textbf{35.92}
			&  0.8778 &  0.8812 & \textbf{0.9060} & 0.9008 & \underline{0.9022} \\
			\addlinespace[-1pt]\cmidrule(lr){2-13}\addlinespace[-1pt]
			& \multirow{5}{*}{$[-0.25,0.25]$} 
			& HTV
			& 27.22 & 27.69 & 28.82 & \textbf{29.19} & \underline{29.14} &
			0.6269 & 0.6294 & 0.6977 & \textbf{0.7291} & \underline{0.7278} \\
			& & SSTV
			& 33.16 & 33.68 & 34.05 & \underline{34.37} & \underline{34.37} &
			0.8339 & 0.8436 & 0.8519 & \textbf{0.8565} & \underline{0.8564} \\
			& & ASSTV
			& 27.05 & 27.96 & \underline{27.98} & 27.97 & \underline{28.17} &
			0.6112 & \underline{0.6320} & \textbf{0.6339} & 0.6222 & 0.6284 \\
			& & TNN
			& 23.85 & 25.42 & 28.70 & \textbf{31.23} & \textbf{31.23} &
			0.5477 & 0.6203 & 0.7184 & \textbf{0.8506} & \textbf{0.8506} \\
			& & SSTV+TNN
			& 32.03 & 32.94 & \textbf{34.61} & 33.79 & \underline{33.90} &
			0.8312 & 0.8364 & \textbf{0.8910} & \underline{0.8799} & 0.8715 \\
			& & $l_{0}$-$l_{1}$HTV
			& 33.74 & 34.20 & 35.66 & \underline{35.74} & \textbf{35.91} &
			0.8738 & 0.8747 & \textbf{0.9039} & 0.9023 & \underline{0.9037} \\
			\addlinespace[-1pt]\cmidrule(lr){2-13}\addlinespace[-1pt]
			& \multirow{5}{*}{$[-0.3,0.3]$} 
			& HTV 
			& 26.96 & 27.38 & 28.56 & \underline{29.05} & \textbf{29.12} 
			& 0.6092 & 0.6125 & 0.6834 & \underline{0.7120} & \textbf{0.7256} \\
			& & SSTV 
			& 32.71 & 33.36 & \underline{33.55} & 33.52 & \textbf{34.27} 
			& 0.8175 & 0.8317 & \underline{0.8337} & 0.8241 & \textbf{0.8497} \\
			\textit{Moffett Field} & & ASSTV 
			& 25.98 & 26.15 & 26.04 & \textbf{27.94} & \underline{26.14} 
			& 0.4805 & \underline{0.4871} & 0.4838 & \textbf{0.6195} & 0.4870 \\
			& & TNN 
			& 23.77 & 25.36 & 28.18 & \textbf{30.97} & \textbf{30.97} 
			& 0.5427 & 0.6138 & 0.6822 & \textbf{0.8387} & \textbf{0.8387} \\
			& & SSTV+TNN
			& 31.59 & 32.57 & \textbf{34.23} & 33.65 & \underline{33.79}
			& 0.8138 & 0.8235 & \textbf{0.8820} & \underline{0.8754} & 0.8669 \\
			& & $l_{0}$-$l_{1}$HTV
			& 33.39 & 33.74 & 35.35 & \underline{35.63} & \textbf{35.80} &
			0.8615 & 0.8600 & 0.8926 & \underline{0.8958} & \textbf{0.8972} \\
			\addlinespace[-1pt]\cmidrule(lr){2-13}\addlinespace[-1pt]
			& \multirow{5}{*}{$[-0.35,0.35]$} 
			& HTV 
			& 26.73 & 27.11 & 28.34 & \textbf{29.15} & \underline{29.11} &
			0.5994 & 0.6024 & 0.6790 & \textbf{0.7279} & \underline{0.7271} \\
			& & SSTV 
			& 32.60 & 33.39 & 33.30 & \textbf{34.32} & \underline{34.31} &
			0.8170 & 0.8348 & 0.8309 & \textbf{0.8548} & \underline{0.8542} \\
			& & ASSTV 
			& 27.01 & 27.92 & 27.93 & \underline{27.94} & \textbf{28.16} &
			0.6087 & \underline{0.6303} & \textbf{0.6319} & 0.6207 & 0.6277 \\
			& & TNN 
			& 23.75 & 25.39 & 27.97 & \textbf{31.38} & \textbf{31.38} &
			0.5379 & 0.6006 & 0.6643 & \textbf{0.8403} & \textbf{0.8403} \\
			& & SSTV+TNN
			& 31.34 & 32.43 & \textbf{34.18} & 33.67 & \underline{33.79} &
			0.8629 & 0.8605 & 0.8932 & \underline{0.9012} & \textbf{0.9025} \\
			& & $l_{0}$-$l_{1}$HTV
			& 33.30 & 33.62 & 35.14 & \underline{35.70} & \textbf{35.87} &
			0.8629 & 0.8605 & 0.8932 & \underline{0.9012} & \textbf{0.9025} \\
			\addlinespace[-1pt]\cmidrule(lr){2-13}\addlinespace[-1pt]
			& \multirow{5}{*}{$[-0.4,0.4]$} 
			& HTV
			& 26.40 & 26.79 & 27.94 & \underline{28.93} & \textbf{29.05}
			& 0.5807 & 0.5871 & 0.6603 & \underline{0.7071} & \textbf{0.7210}\\
			& & SSTV
			& 32.31 & 33.14 & 32.85 & \underline{33.43} & \textbf{34.19}
			& 0.8069 & 0.8263 & 0.8150 & \underline{0.8222} & \textbf{0.8472}\\
			& & ASSTV
			& 26.98 & 27.90 & 27.89 & \underline{27.93} & \textbf{28.16}
			& 0.6073 & 0.6280 & \textbf{0.6283} & 0.6184 & \underline{0.6257}\\
			& & TNN
			& 23.72 & 25.26 & 27.63 & \textbf{31.41} & \textbf{31.41}
			& 0.5359 & 0.5713 & 0.6521 & \textbf{0.8293} & \textbf{0.8293}\\
			& & SSTV+TNN
			& 30.93 & 32.09 & \textbf{33.65} & 33.52 & \underline{33.63}
			& 0.7939 & 0.8127 & \underline{0.8670} & \textbf{0.8710} & 0.8630\\
			& & $l_{0}$-$l_{1}$HTV
			& 33.01 & 33.19 & 34.64 & \underline{35.53} & \textbf{35.71} &
			0.8504 & 0.8472 & 0.8777 & \underline{0.8926} & \textbf{0.8939} \\
			\bottomrule
		\end{tabular}
	\end{center}
\end{table*}

\begin{figure}[t]
	\centering
	\begin{minipage}{0.49\hsize}
		\centerline{\includegraphics[width=\hsize]{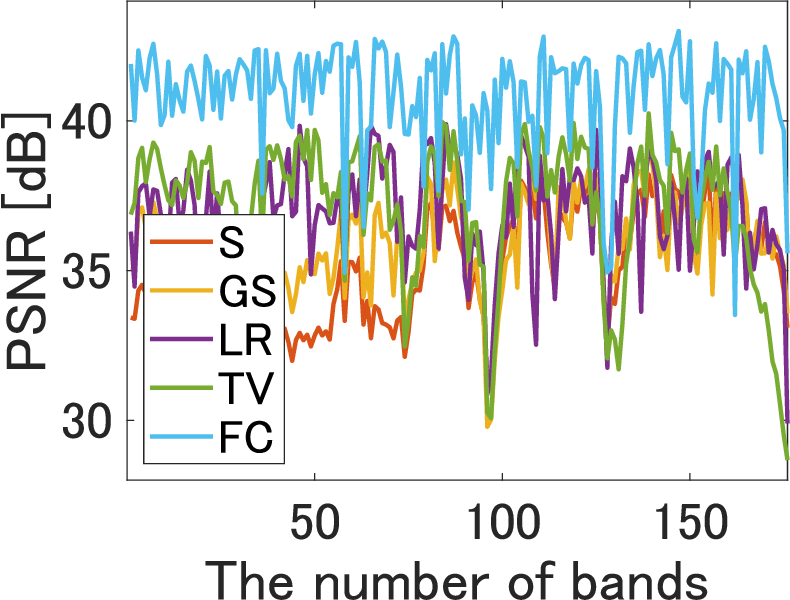}}
	\end{minipage}
	\begin{minipage}{0.49\hsize}
		\centerline{\includegraphics[width=\hsize]{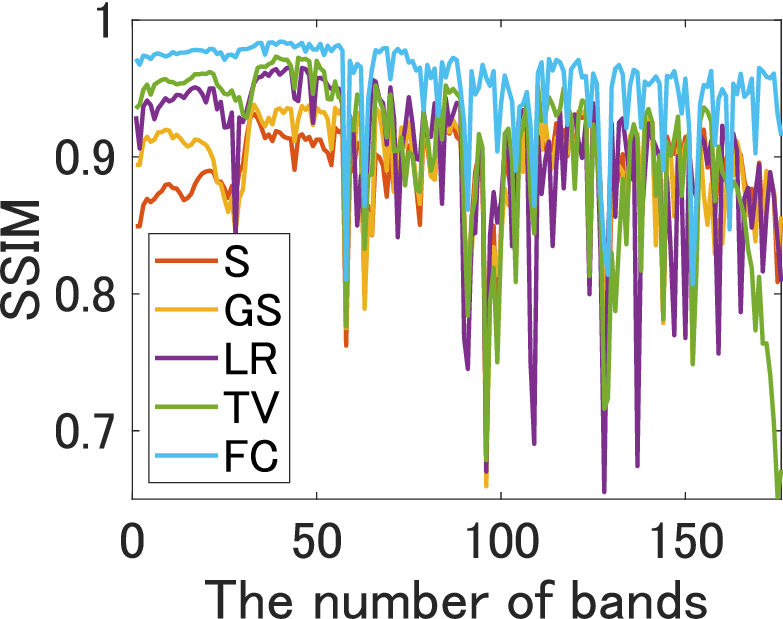}}
	\end{minipage}
	
	\vspace{1mm}
	
	\begin{minipage}{0.49\hsize}
		\centerline{(a)}
	\end{minipage}
	\begin{minipage}{0.49\hsize}
		\centerline{(b)}
	\end{minipage}

	\begin{minipage}{0.49\hsize}
		\centerline{\includegraphics[width=\hsize]{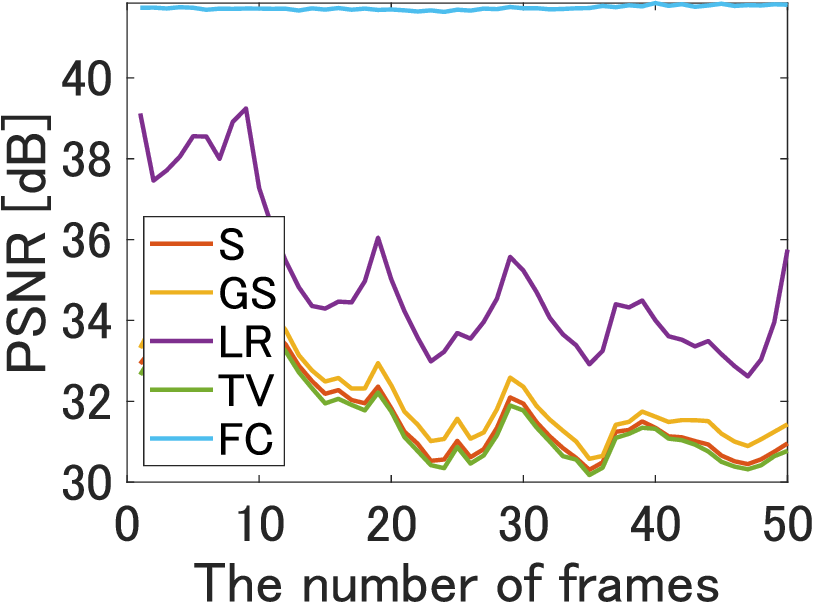}}
	\end{minipage}
	\begin{minipage}{0.49\hsize}
		\centerline{\includegraphics[width=\hsize]{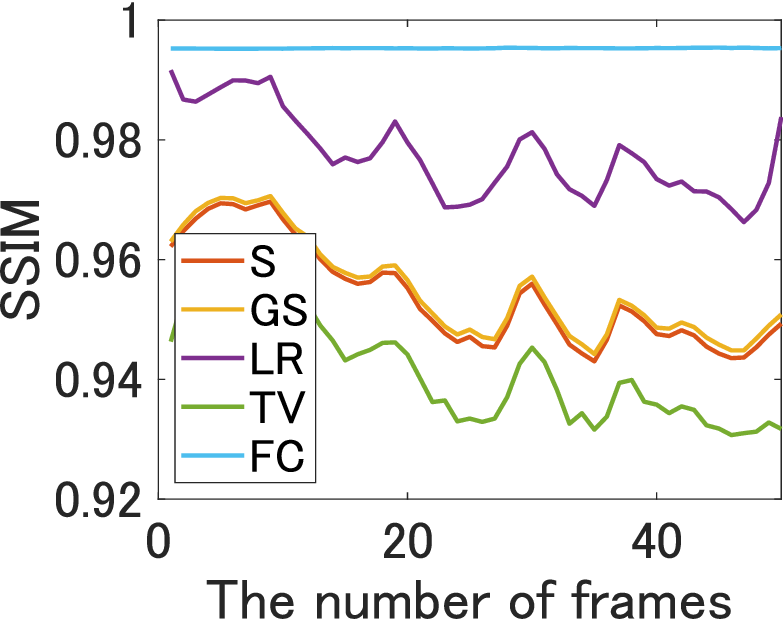}}
	\end{minipage}
	
	\vspace{1mm}
	
	\begin{minipage}{0.49\hsize}
		\centerline{(c)}
	\end{minipage}
	\begin{minipage}{0.49\hsize}
		\centerline{(d)}
	\end{minipage}
	
	\begin{minipage}{0.49\hsize}
		\centerline{\includegraphics[width=\hsize]{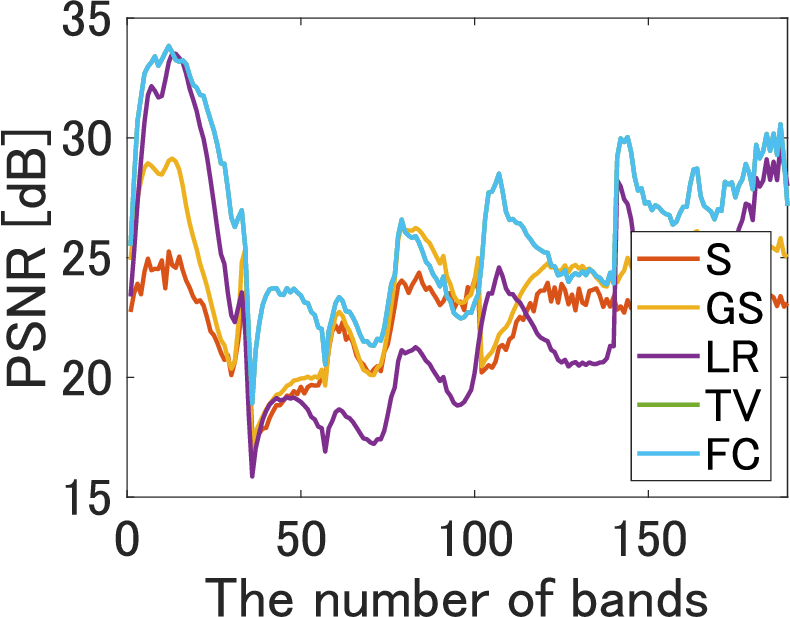}}
	\end{minipage}
	\begin{minipage}{0.49\hsize}
		\centerline{\includegraphics[width=\hsize]{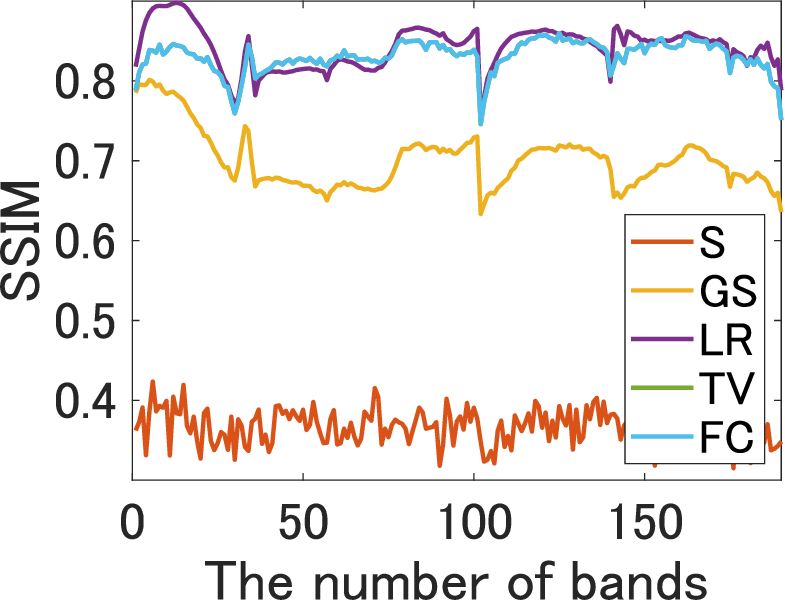}}
	\end{minipage}
	
	\vspace{1mm}
	
	\begin{minipage}{0.49\hsize}
		\centerline{(e)}
	\end{minipage}
	\begin{minipage}{0.49\hsize}
		\centerline{(f)}
	\end{minipage}

	\caption{Band-wise or frame-wise PSNRs and SSIMs. (a) and (b)  PSNRs and SSIMs of the \textit{Moffett field} destriping results in Case (i) using SSTV. (c) and (d) PSNRs and SSIMs of the \textit{Bats1} destriping results in Case (ii) using ATV. (e) and (f) PSNRs and SSIMs of the \textit{Salinas} destriping results in Case (iii) using TNN.}
	\label{fig:bw_PSNR_SSIM_TNN}
\end{figure}

%\begin{figure}[t]
%	\centering
%	\begin{minipage}{0.49\hsize}
%		\centerline{\includegraphics[width=\hsize]{./fig/all_PSNR_mean.eps}}
%	\end{minipage}
%	\begin{minipage}{0.49\hsize}
%		\centerline{\includegraphics[width=\hsize]{./fig/all_PSNR_mean_IR.eps}}
%	\end{minipage}
%	\begin{minipage}{0.49\hsize}
%		\centerline{\includegraphics[width=\hsize]{./fig/all_SSIM_mean.eps}}
%	\end{minipage}
%	\begin{minipage}{0.49\hsize}
%		\centerline{\includegraphics[width=\hsize]{./fig/all_SSIM_mean_IR.eps}}
%	\end{minipage}
%	
%	\caption{Fig. xx. }
%	\label{fig:PSNR_SSIM_mean}
%\end{figure}

\begin{figure}[t]
	\centering
	\begin{minipage}{0.49\hsize}
		\centerline{\includegraphics[width=\hsize]{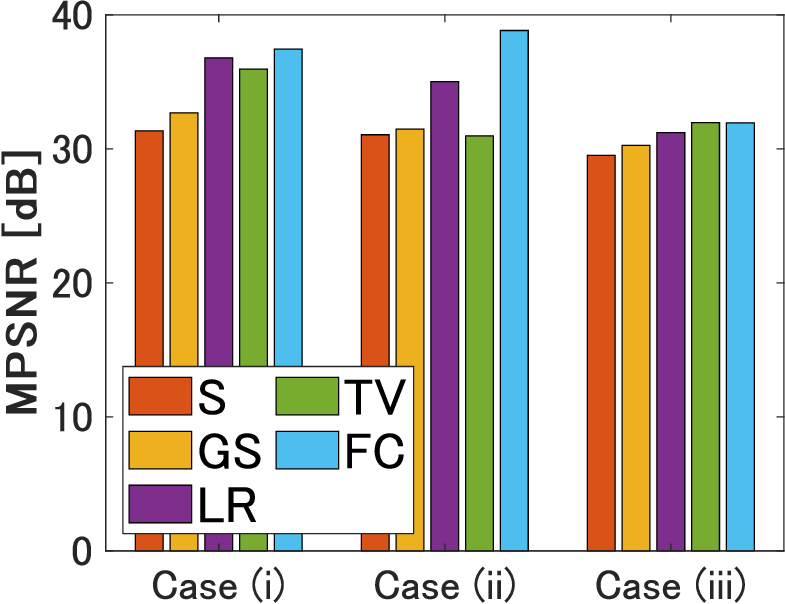}}
	\end{minipage}
	\begin{minipage}{0.49\hsize}
		\centerline{\includegraphics[width=\hsize]{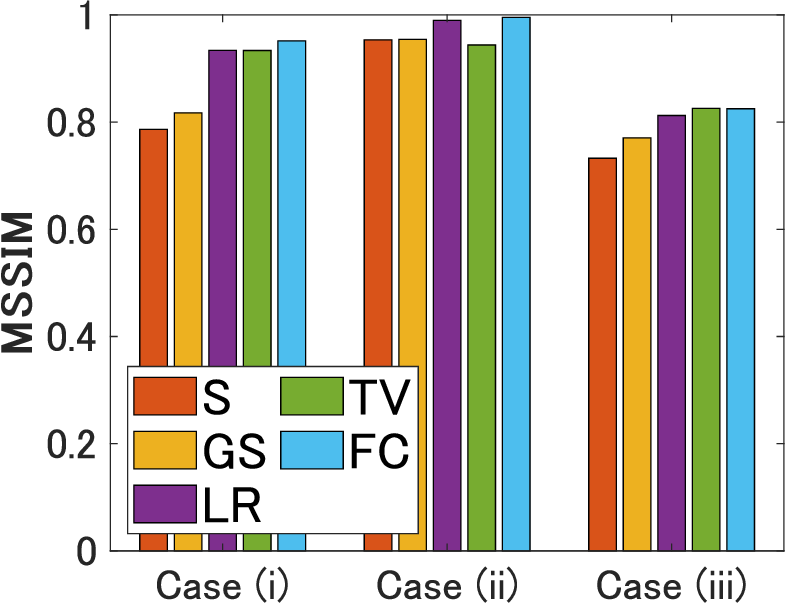}}
	\end{minipage}	
	
	\caption{Means of MPSNRs and MSSIMs in each noise case.}
	\label{fig:all_PSNR_SSIM_mean_case}
\end{figure}

%\begin{figure}[t]
%	\centering
%	\begin{minipage}{0.49\hsize}
%		\centerline{\includegraphics[width=\hsize]{./fig/all_PSNR_mean.eps}}
%	\end{minipage}
%	\begin{minipage}{0.49\hsize}
%		\centerline{\includegraphics[width=\hsize]{./fig/all_PSNR_mean_IR.eps}}
%	\end{minipage}
%	\begin{minipage}{0.49\hsize}
%		\centerline{\includegraphics[width=\hsize]{./fig/all_SSIM_mean.eps}}
%	\end{minipage}
%	\begin{minipage}{0.49\hsize}
%		\centerline{\includegraphics[width=\hsize]{./fig/all_SSIM_mean_IR.eps}}
%	\end{minipage}
%	
%	\caption{Fig. xx. }
%	\label{fig:PSNR_SSIM_mean}
%\end{figure}

\begin{figure}[t]
	\centering
	\begin{minipage}{0.49\hsize}
		\centerline{\includegraphics[width=\hsize]{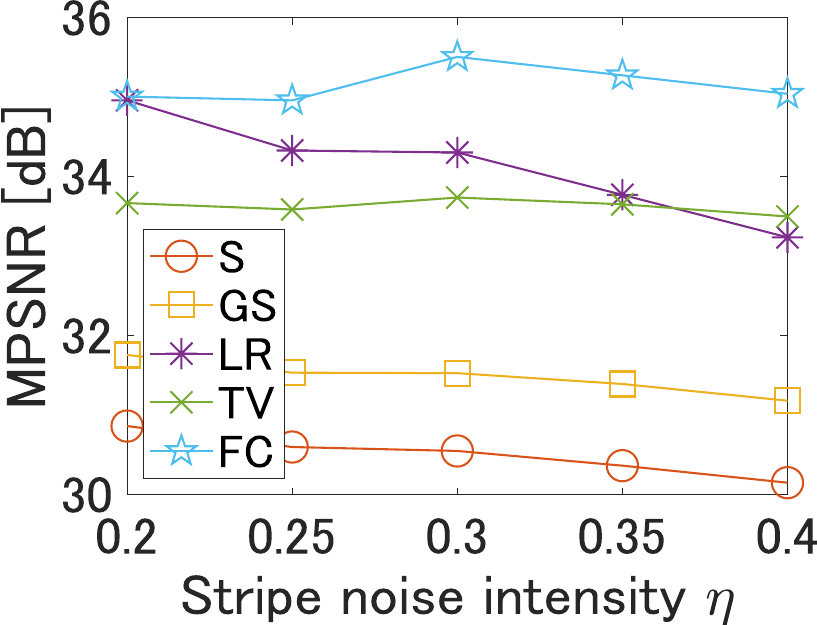}}
	\end{minipage}
	\begin{minipage}{0.49\hsize}
		\centerline{\includegraphics[width=\hsize]{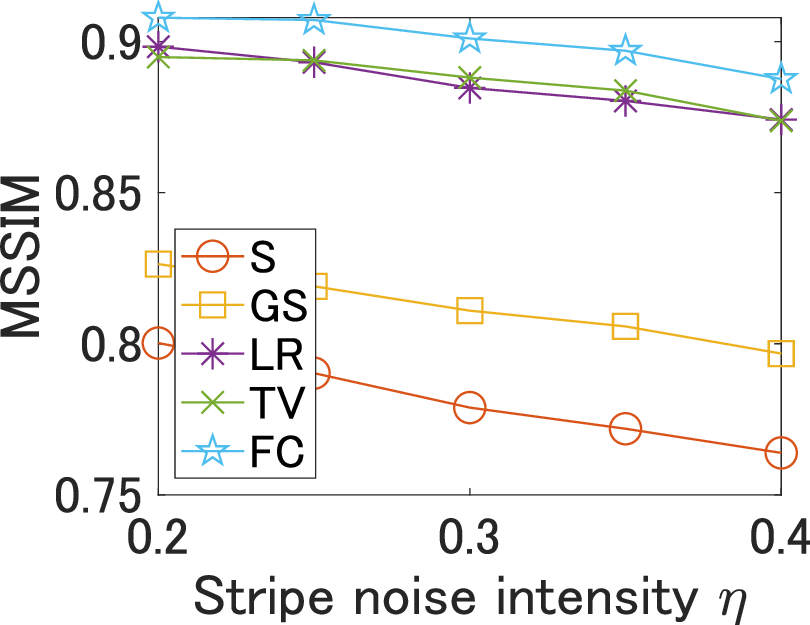}}
	\end{minipage}

	\caption{Means of MPSNRs and MSSIMs in each stripe noise intensity range $[-\eta, \eta]$.}
	\label{fig:all_PSNR_SSIM_mean}
\end{figure}

%\begin{figure}[t]
%	\centering
%	\begin{minipage}{0.49\hsize}
%		\centerline{\includegraphics[width=\hsize]{./fig/all_PSNR_mean.eps}}
%	\end{minipage}
%	\begin{minipage}{0.49\hsize}
%		\centerline{\includegraphics[width=\hsize]{./fig/all_PSNR_mean_IR.eps}}
%	\end{minipage}
%	\begin{minipage}{0.49\hsize}
%		\centerline{\includegraphics[width=\hsize]{./fig/all_SSIM_mean.eps}}
%	\end{minipage}
%	\begin{minipage}{0.49\hsize}
%		\centerline{\includegraphics[width=\hsize]{./fig/all_SSIM_mean_IR.eps}}
%	\end{minipage}
%	
%	\caption{Fig. xx. }
%	\label{fig:PSNR_SSIM_mean}
%\end{figure}

\begin{figure}[t]
	\centering
	\begin{minipage}{0.49\hsize}
		\centerline{\includegraphics[width=\hsize]{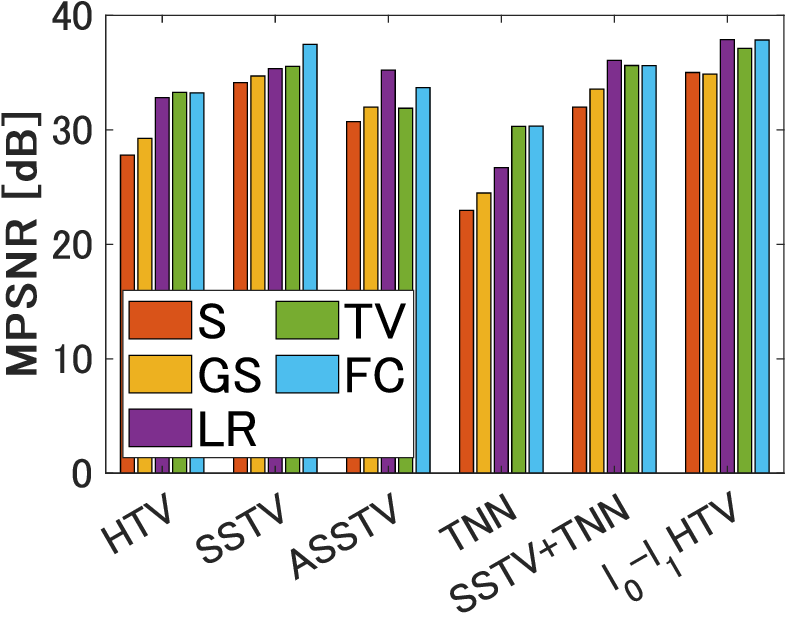}}
	\end{minipage}
	\begin{minipage}{0.49\hsize}
		\centerline{\includegraphics[width=\hsize]{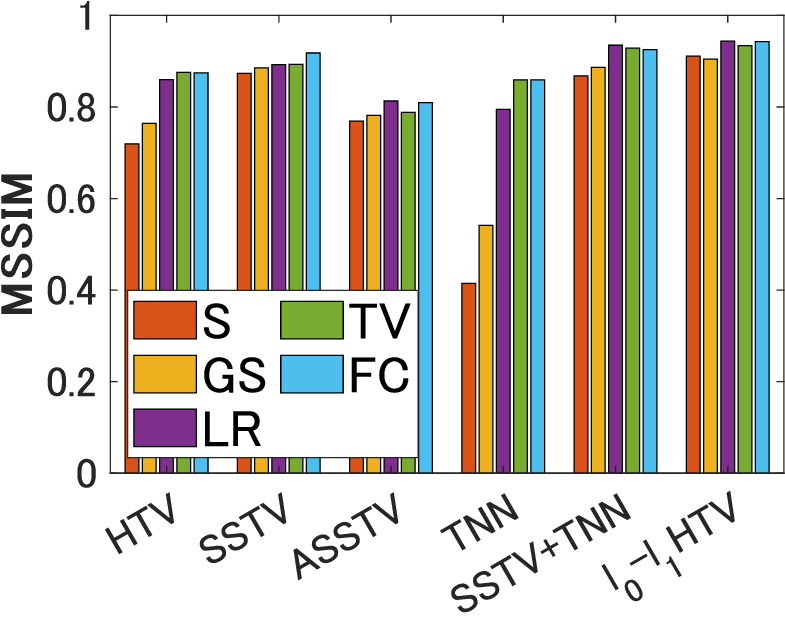}}
	\end{minipage}
	
	\begin{minipage}{0.49\hsize}
		\centerline{(a)}
	\end{minipage}
	\begin{minipage}{0.49\hsize}
		\centerline{(b)}
	\end{minipage}

	\begin{minipage}{0.49\hsize}
		\centerline{\includegraphics[width=\hsize]{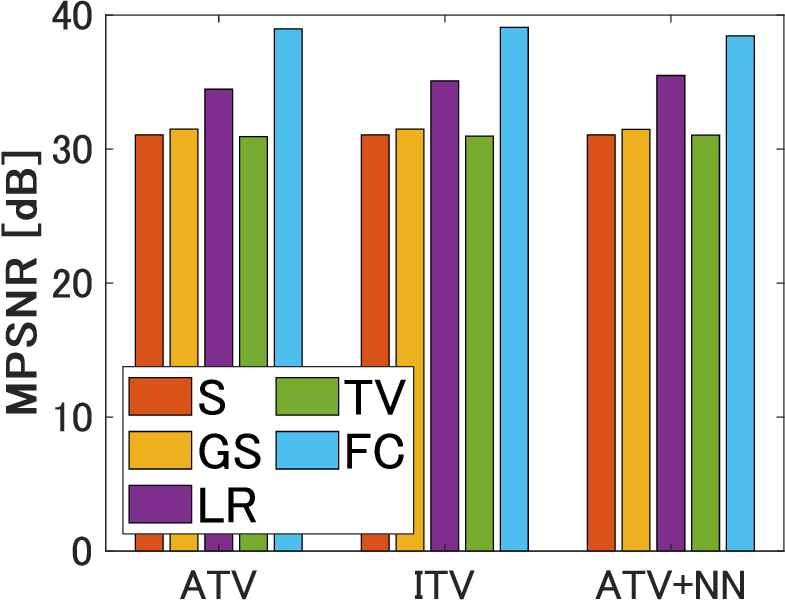}}
	\end{minipage}
	\begin{minipage}{0.49\hsize}
		\centerline{\includegraphics[width=\hsize]{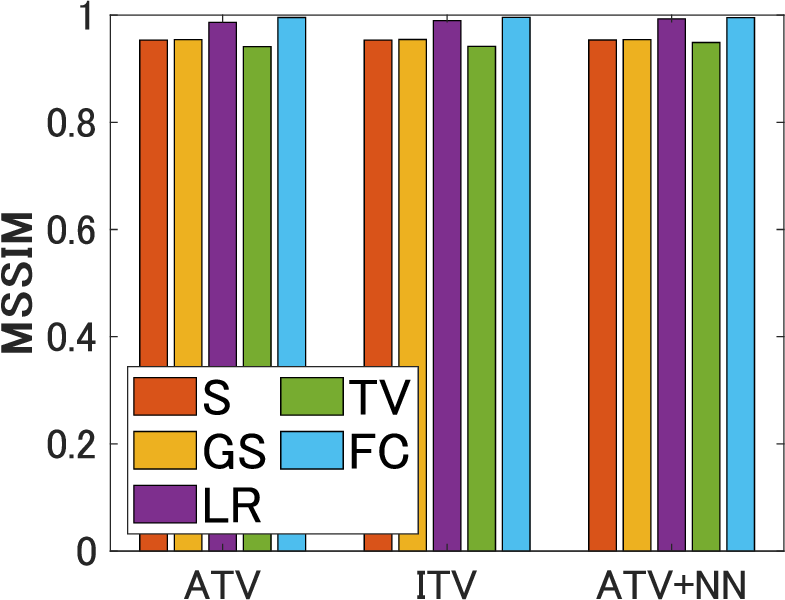}}
	\end{minipage}
	
	\begin{minipage}{0.49\hsize}
		\centerline{(c)}
	\end{minipage}
	\begin{minipage}{0.49\hsize}
		\centerline{(d)}
	\end{minipage}

	\caption{Means of MPSNRs and MSSIMs in each image regularization. (a) and (b) Means of MPSNRs and MSSIMs in the HSI experiments. (c) and (d) Means of MPSNRs and MSSIMs in the IR video experiments.}
	\label{fig:all_PSNR_SSIM_mean_reg}
\end{figure}

%%%%%%%%%%%%%%%%%%%%%%%%%%%%%%%%%%%%%%%%%%%%%%%%%%%%%%%%%%%%%%%%%%%%
%% real data
%%%%%%%%%%%%%%%%%%%%%%%%%%%%%%%%%%%%%%%%%%%%%%%%%%%%%%%%%%%%%%%%%%%%
\begin{figure}[t]
	\centering
	\begin{minipage}{0.30\hsize}
		\centerline{\includegraphics[width=\hsize]{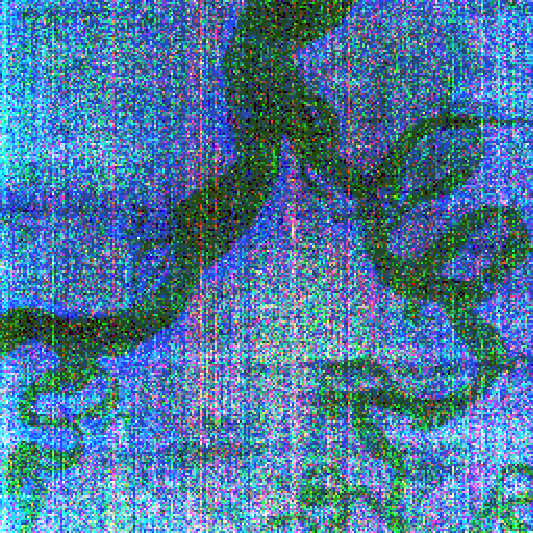}}
	\end{minipage}
	\begin{minipage}{0.30\hsize}
		\centerline{\includegraphics[width=\hsize]{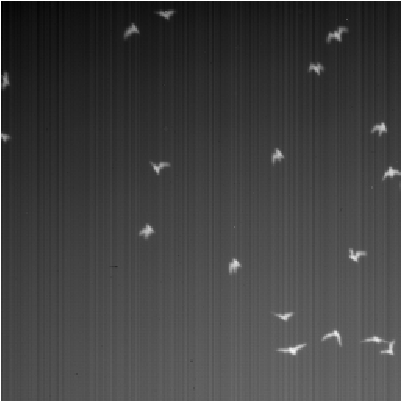}}
	\end{minipage}
	
	\vspace{1mm}
	
	\begin{minipage}{0.30\hsize} %この一つのminipageが一つの部屋って感じで，これをまず配置する
		\centerline{(a)}
	\end{minipage}
	\begin{minipage}{0.30\hsize}
		\centerline{(b)}
	\end{minipage}
	
	\vspace{-1mm}
	
	\caption{HSI and IR video data used for experiments in real noise cases. (a) \textit{Suwannee} (R: 357, G: 275, B: 120). (b) \textit{Bats2} (an IR video).}
	\label{fig:real_data}
	
	\vspace{-2mm}
	
\end{figure}

%%%%%%%%%%%%%%%%%%%%%%%%%%%%%%%%%%%%%%%%%%%%%%%%%%%%%%%
%% real HSI data destriping results with HTV~TNN
%%%%%%%%%%%%%%%%%%%%%%%%%%%%%%%%%%%%%%%%%%%%%%%%%%%%%%%
\begin{figure*}[!t]
	
	\begin{minipage}{0.48\hsize}
		\centerline{HTV}
	\end{minipage}
	\begin{minipage}{0.48\hsize}
		\centerline{SSTV}
	\end{minipage}
	
	\vspace{1mm}
	
	\begin{minipage}{0.01\hsize}
		\rotatebox[origin=c]{90}{\scriptsize{Brightened image}}
	\end{minipage}
	\begin{minipage}{0.09\hsize}
		\centerline{\includegraphics[width=\hsize]{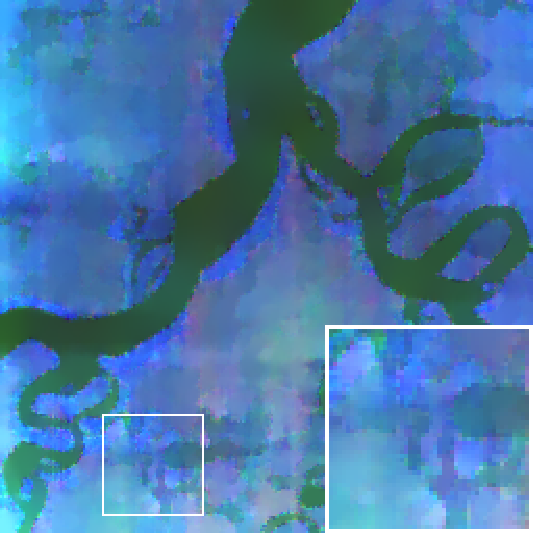}}
	\end{minipage}
	\begin{minipage}{0.09\hsize}
		\centerline{\includegraphics[width=\hsize]{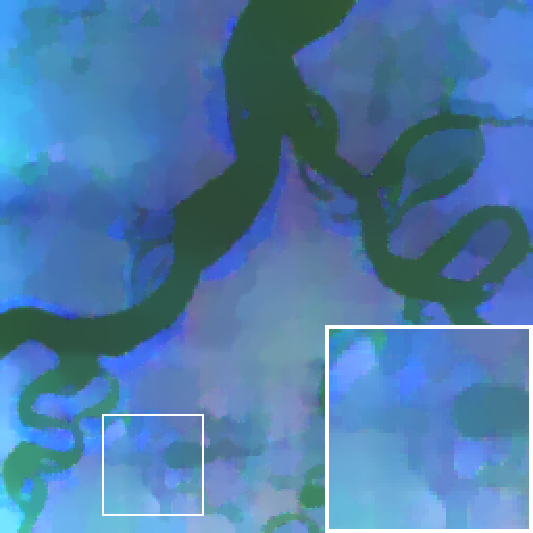}}
	\end{minipage}
	\begin{minipage}{0.09\hsize}
		\centerline{\includegraphics[width=\hsize]{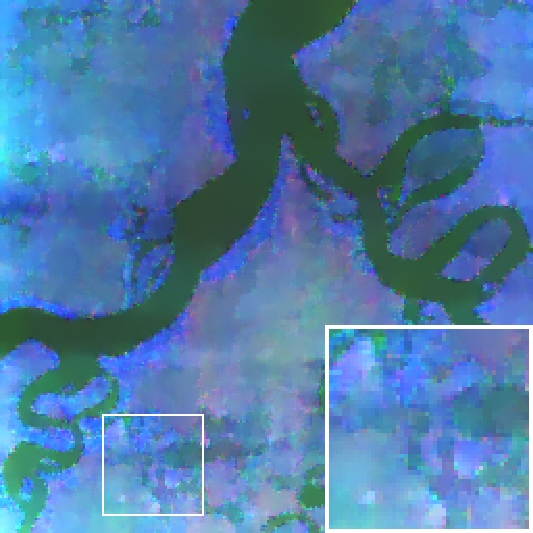}}
	\end{minipage}
	\begin{minipage}{0.09\hsize}
		\centerline{\includegraphics[width=\hsize]{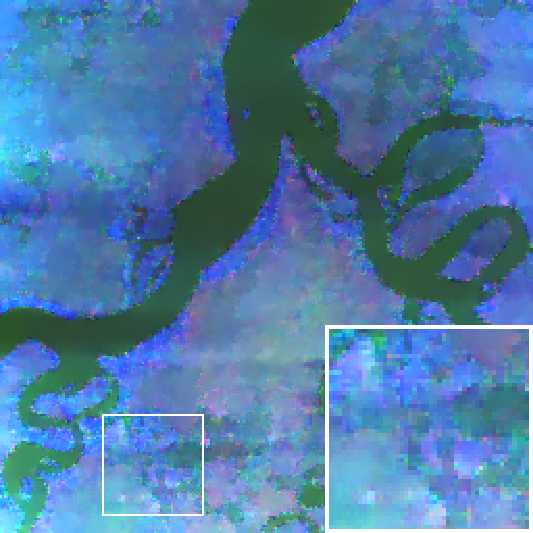}}
	\end{minipage}
	\begin{minipage}{0.09\hsize}
		\centerline{\includegraphics[width=\hsize]{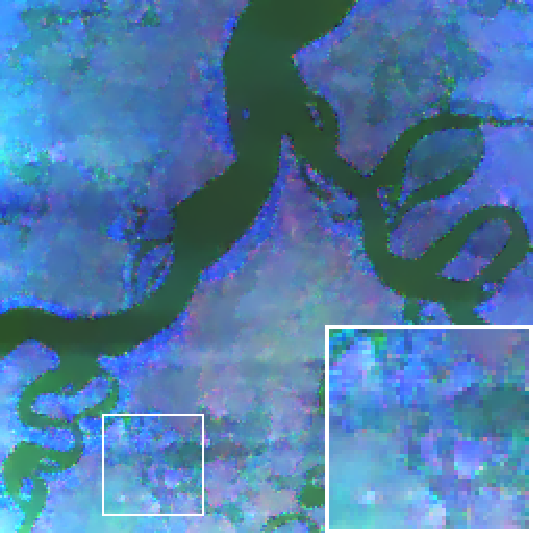}}
	\end{minipage}
	\begin{minipage}{0.01\hsize}
		\rotatebox[origin=c]{90}{\scriptsize{Brightened image}}
	\end{minipage}
	\begin{minipage}{0.09\hsize}
		\centerline{\includegraphics[width=\hsize]{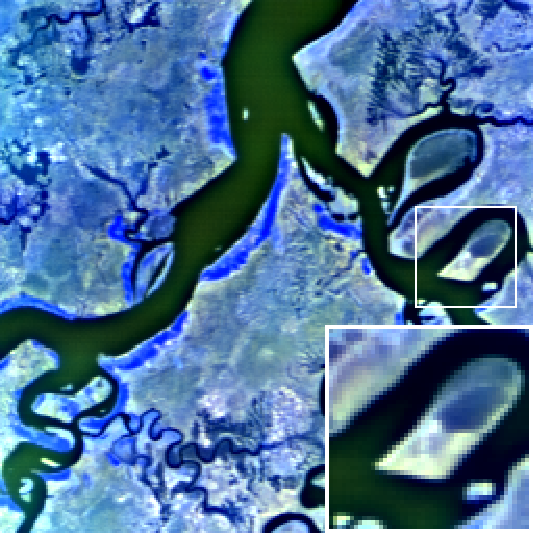}}
	\end{minipage}
	\begin{minipage}{0.09\hsize}
		\centerline{\includegraphics[width=\hsize]{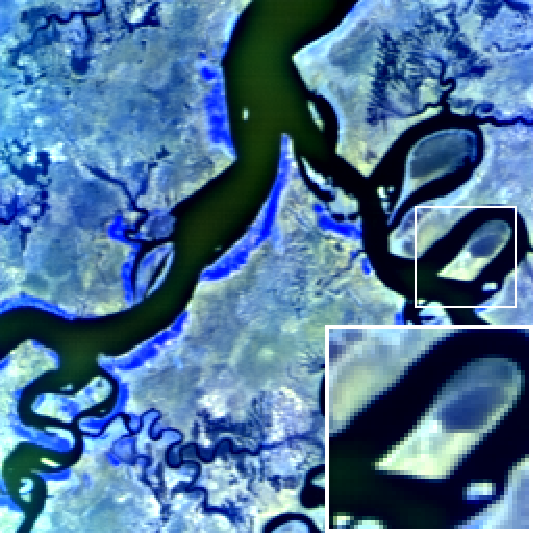}}
	\end{minipage}
	\begin{minipage}{0.09\hsize}
		\centerline{\includegraphics[width=\hsize]{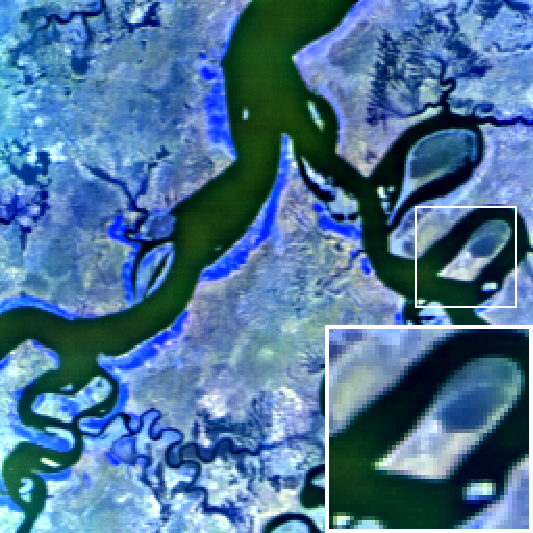}}
	\end{minipage}
	\begin{minipage}{0.09\hsize}
		\centerline{\includegraphics[width=\hsize]{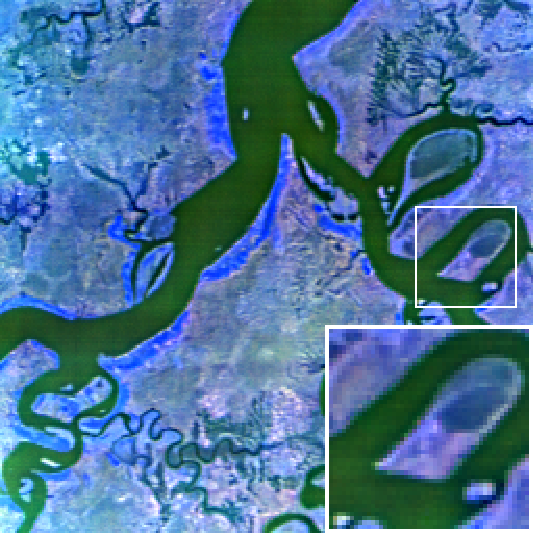}}
	\end{minipage}
	\begin{minipage}{0.09\hsize}
		\centerline{\includegraphics[width=\hsize]{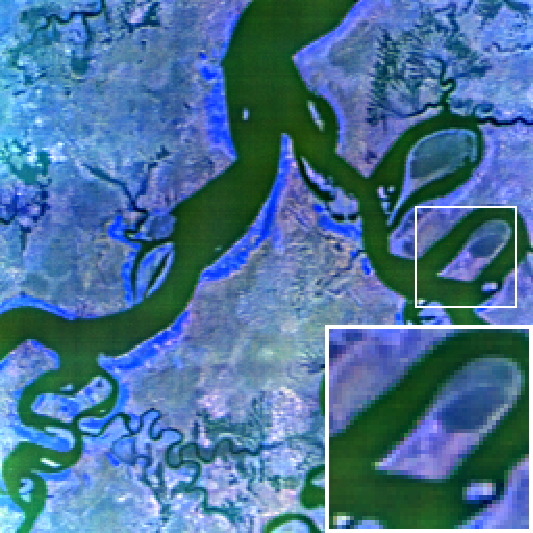}}
	\end{minipage}
	
	\begin{minipage}{0.01\hsize}
		\rotatebox[origin=c]{90}{\scriptsize{Stripe noise}}
	\end{minipage}
	\begin{minipage}{0.09\hsize}
		\centerline{\includegraphics[width=\hsize]{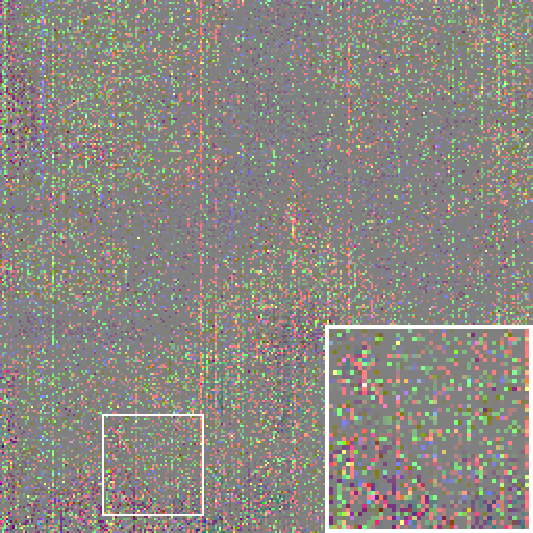}}
	\end{minipage}
	\begin{minipage}{0.09\hsize}
		\centerline{\includegraphics[width=\hsize]{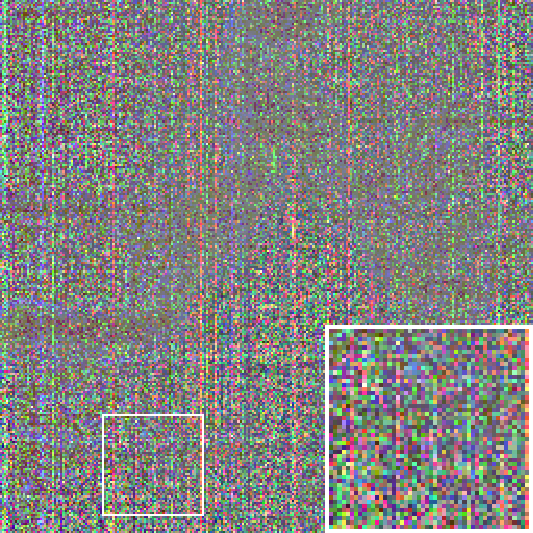}}
	\end{minipage}
	\begin{minipage}{0.09\hsize}
		\centerline{\includegraphics[width=\hsize]{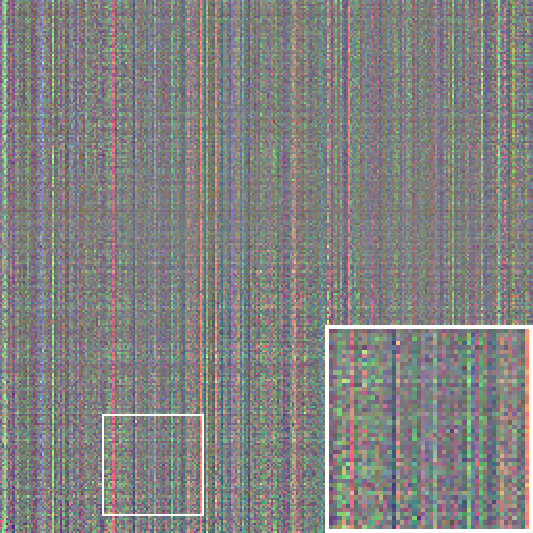}}
	\end{minipage}
	\begin{minipage}{0.09\hsize}
		\centerline{\includegraphics[width=\hsize]{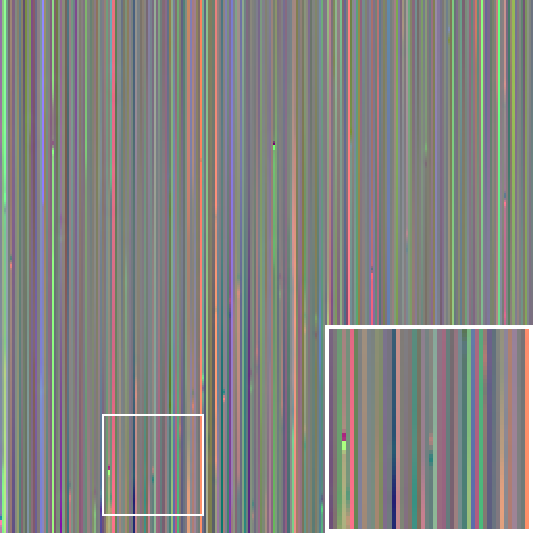}}
	\end{minipage}
	\begin{minipage}{0.09\hsize}
		\centerline{\includegraphics[width=\hsize]{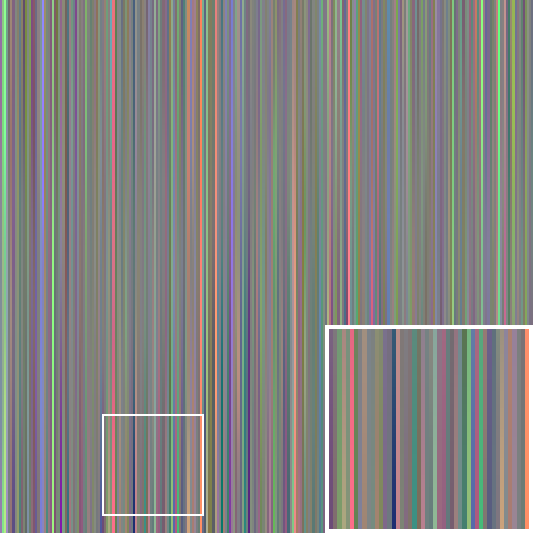}}
	\end{minipage}
	\begin{minipage}{0.01\hsize}
		\rotatebox[origin=c]{90}{\scriptsize{Stripe noise}}
	\end{minipage}
	\begin{minipage}{0.09\hsize}
		\centerline{\includegraphics[width=\hsize]{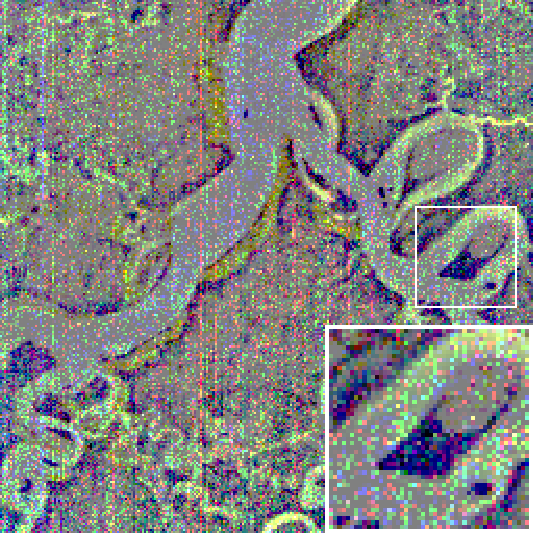}}
	\end{minipage}
	\begin{minipage}{0.09\hsize}
		\centerline{\includegraphics[width=\hsize]{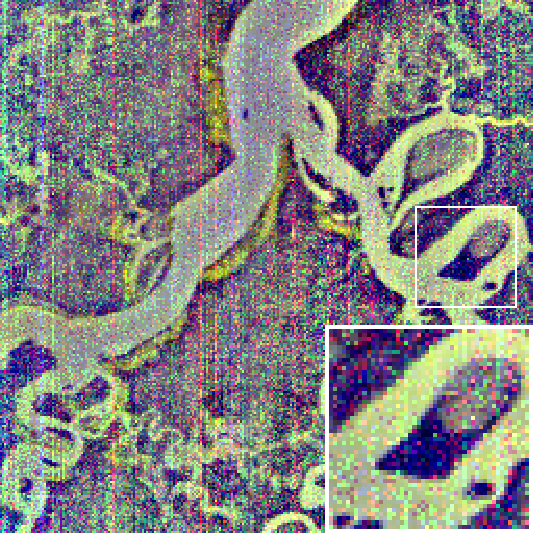}}
	\end{minipage}
	\begin{minipage}{0.09\hsize}
		\centerline{\includegraphics[width=\hsize]{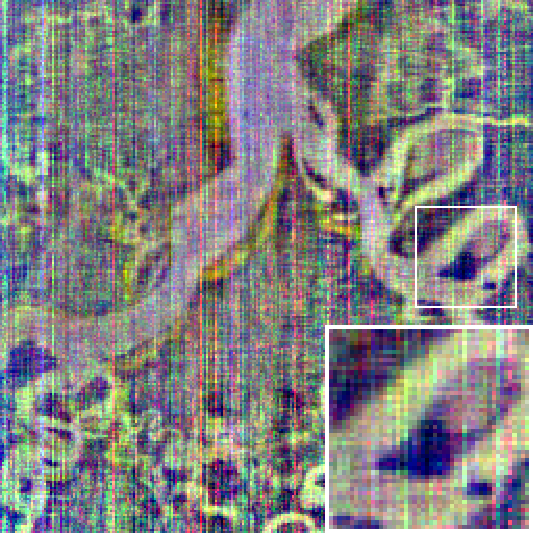}}
	\end{minipage}
	\begin{minipage}{0.09\hsize}
		\centerline{\includegraphics[width=\hsize]{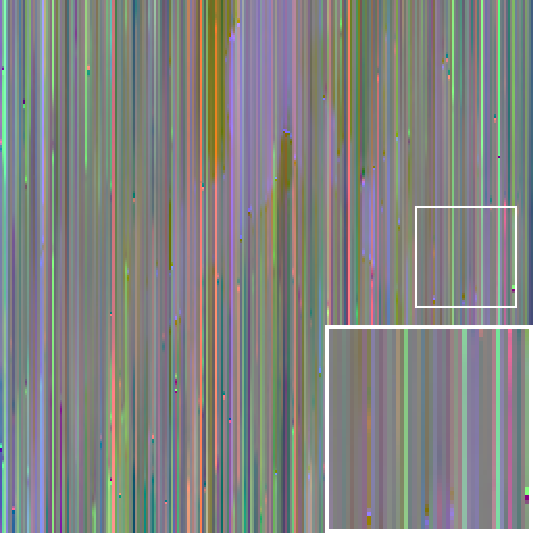}}
	\end{minipage}
	\begin{minipage}{0.09\hsize}
		\centerline{\includegraphics[width=\hsize]{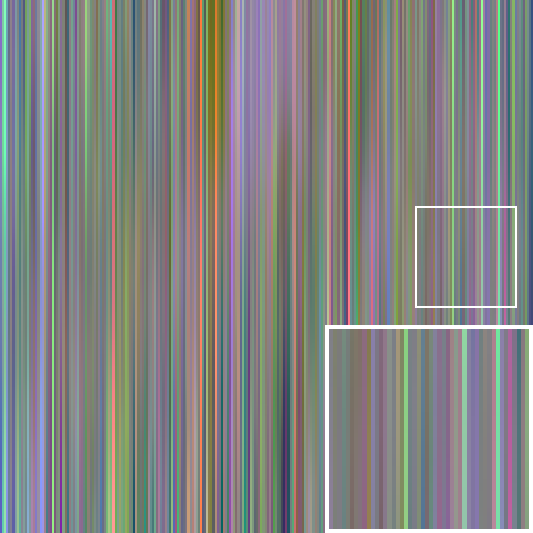}}
	\end{minipage}
	
	\vspace{1mm}
	
	\begin{minipage}{0.01\hsize}
		~
	\end{minipage}
	\begin{minipage}{0.09\hsize}
		\centerline{\small{(a1) S~\cite{LRMR}}}
	\end{minipage}
	\begin{minipage}{0.09\hsize}
		\centerline{\small{(b1) GS~\cite{GLSSTV}}}
	\end{minipage}
	\begin{minipage}{0.09\hsize}
		\centerline{\small{(c1) LR~\cite{NN_char}}}
	\end{minipage}
	\begin{minipage}{0.09\hsize}
		\centerline{\small{(d1) TV~\cite{gradient_constraint}}}
	\end{minipage}
	\begin{minipage}{0.09\hsize}
		\centerline{\small{(e1) FC}}
	\end{minipage}
	\begin{minipage}{0.01\hsize}
		~
	\end{minipage}
	\begin{minipage}{0.09\hsize}
		\centerline{\small{(a2) S~\cite{LRMR}}}
	\end{minipage}
	\begin{minipage}{0.09\hsize}
		\centerline{\small{(b2) GS~\cite{GLSSTV}}}
	\end{minipage}
	\begin{minipage}{0.09\hsize}
		\centerline{\small{(c2) LR~\cite{NN_char}}}
	\end{minipage}
	\begin{minipage}{0.09\hsize}
		\centerline{\small{(d2) TV~\cite{gradient_constraint}}}
	\end{minipage}
	\begin{minipage}{0.09\hsize}
		\centerline{\small{(e2) FC}}
	\end{minipage}

	\begin{minipage}{0.48\hsize}
		\centerline{ASSTV}
	\end{minipage}
	\begin{minipage}{0.48\hsize}
		\centerline{TNN}
	\end{minipage}
	
	\vspace{1mm}
	
	\begin{minipage}{0.01\hsize}
		\rotatebox[origin=c]{90}{\scriptsize{Brightened HSI}}
	\end{minipage}
	\begin{minipage}{0.09\hsize}
		\centerline{\includegraphics[width=\hsize]{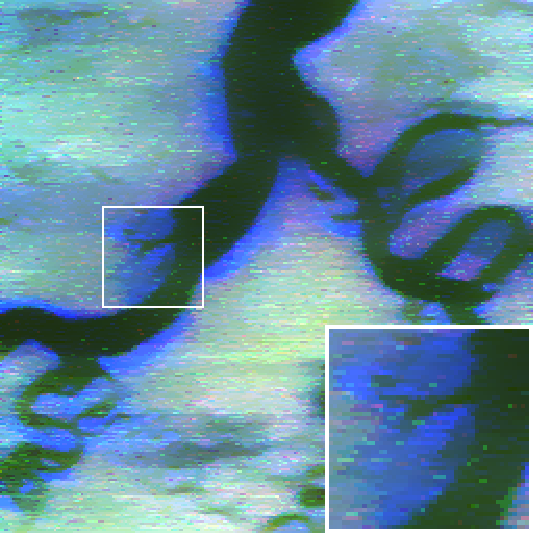}}
	\end{minipage}
	\begin{minipage}{0.09\hsize}
		\centerline{\includegraphics[width=\hsize]{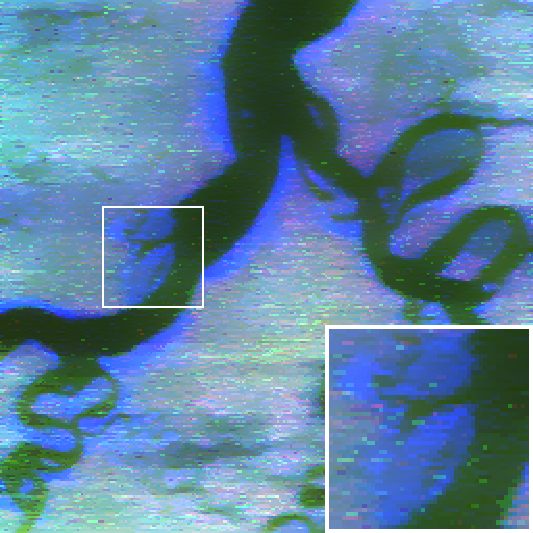}}
	\end{minipage}
	\begin{minipage}{0.09\hsize}
		\centerline{\includegraphics[width=\hsize]{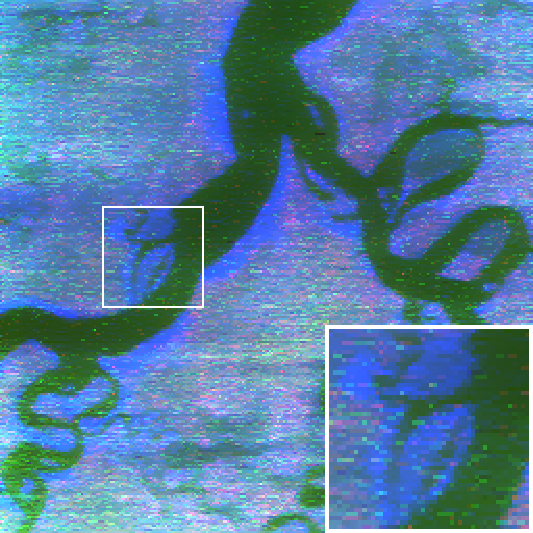}}
	\end{minipage}
	\begin{minipage}{0.09\hsize}
		\centerline{\includegraphics[width=\hsize]{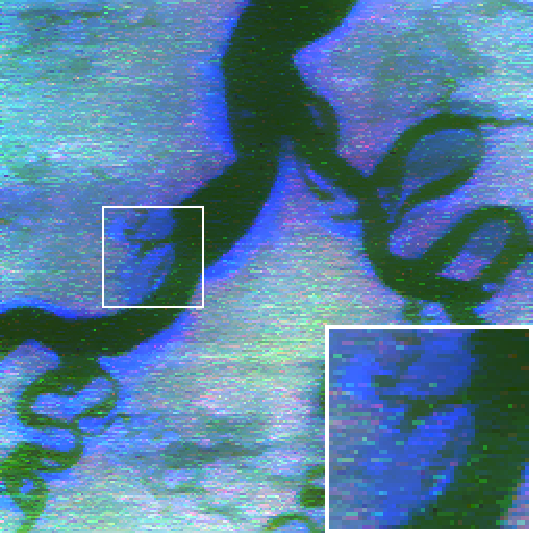}}
	\end{minipage}
	\begin{minipage}{0.09\hsize}
		\centerline{\includegraphics[width=\hsize]{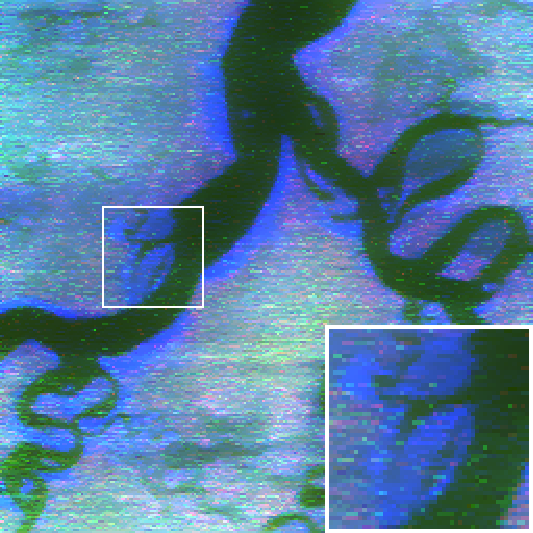}}
	\end{minipage}
	\begin{minipage}{0.01\hsize}
		\rotatebox[origin=c]{90}{\scriptsize{Brightened HSI}}
	\end{minipage}
	\begin{minipage}{0.09\hsize}
		\centerline{\includegraphics[width=\hsize]{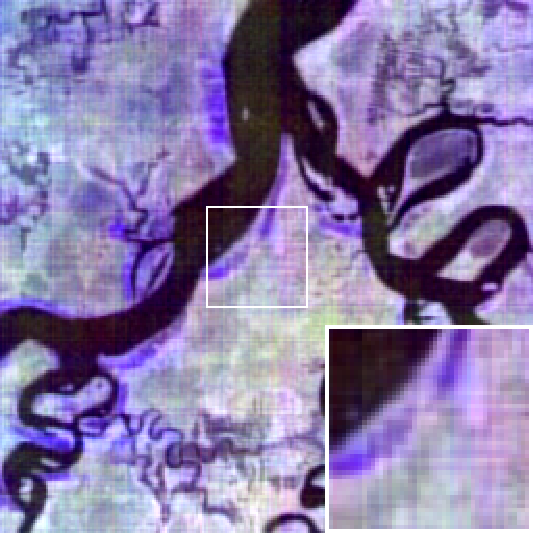}}
	\end{minipage}
	\begin{minipage}{0.09\hsize}
		\centerline{\includegraphics[width=\hsize]{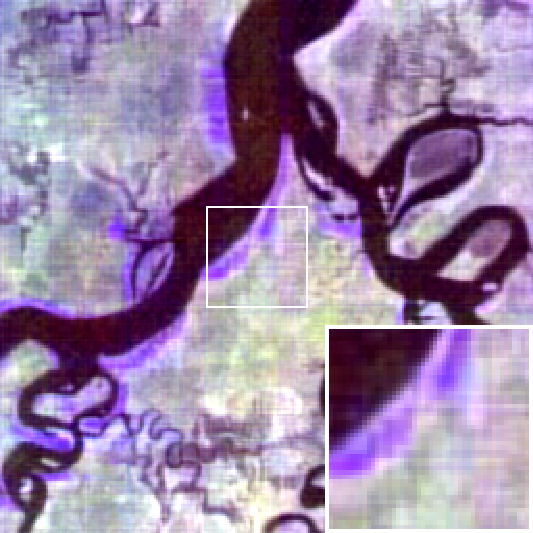}}
	\end{minipage}
	\begin{minipage}{0.09\hsize}
		\centerline{\includegraphics[width=\hsize]{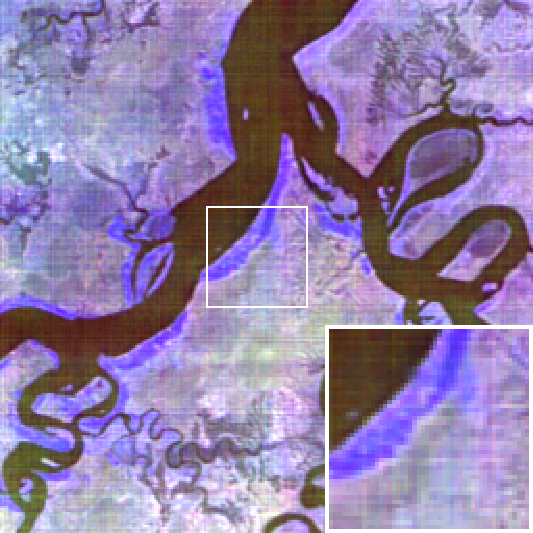}}
	\end{minipage}
	\begin{minipage}{0.09\hsize}
		\centerline{\includegraphics[width=\hsize]{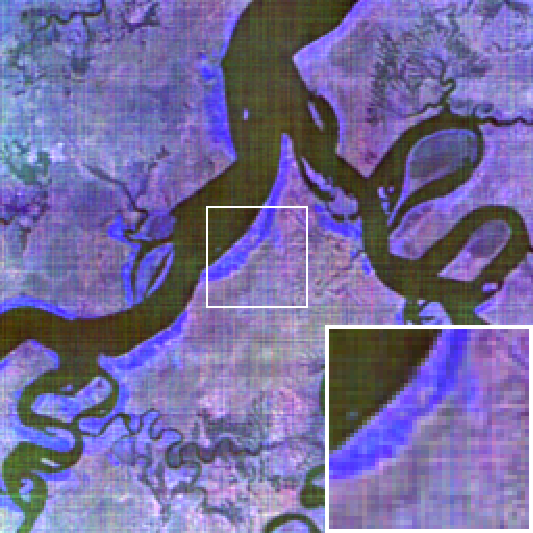}}
	\end{minipage}
	\begin{minipage}{0.09\hsize}
		\centerline{\includegraphics[width=\hsize]{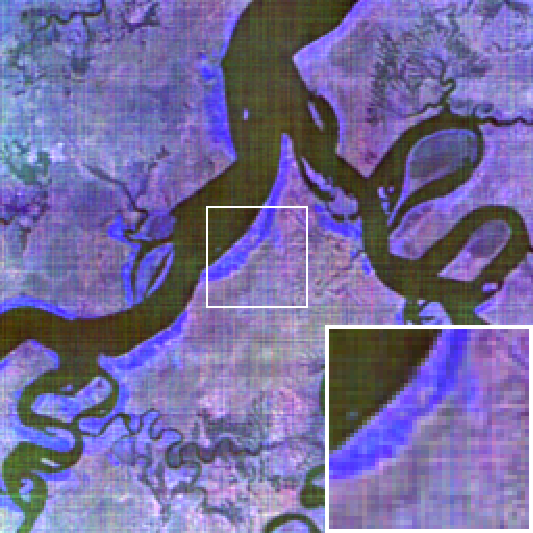}}
	\end{minipage}

	\begin{minipage}{0.01\hsize}
		\rotatebox[origin=c]{90}{\scriptsize{Stripe noise}}
	\end{minipage}
	\begin{minipage}{0.09\hsize}
		\centerline{\includegraphics[width=\hsize]{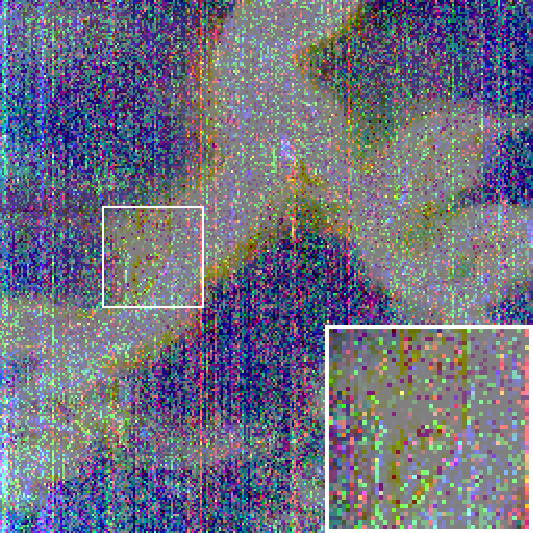}}
	\end{minipage}
	\begin{minipage}{0.09\hsize}
		\centerline{\includegraphics[width=\hsize]{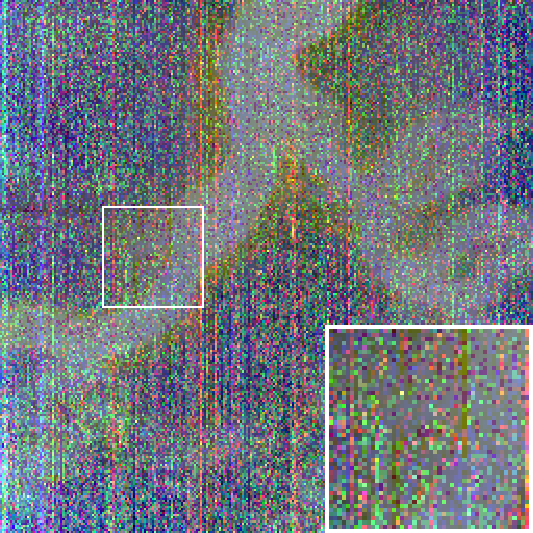}}
	\end{minipage}
	\begin{minipage}{0.09\hsize}
		\centerline{\includegraphics[width=\hsize]{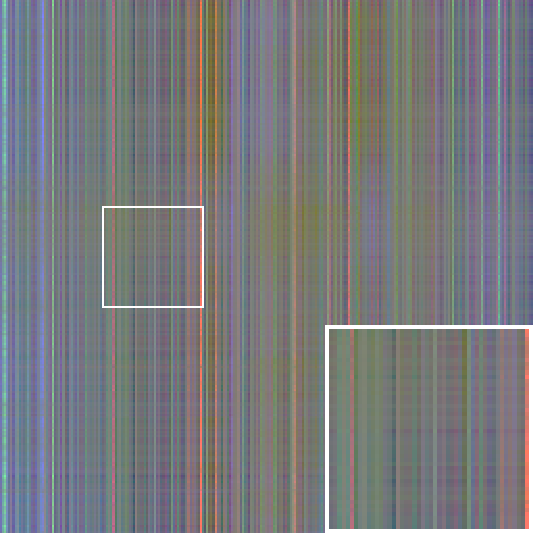}}
	\end{minipage}
	\begin{minipage}{0.09\hsize}
		\centerline{\includegraphics[width=\hsize]{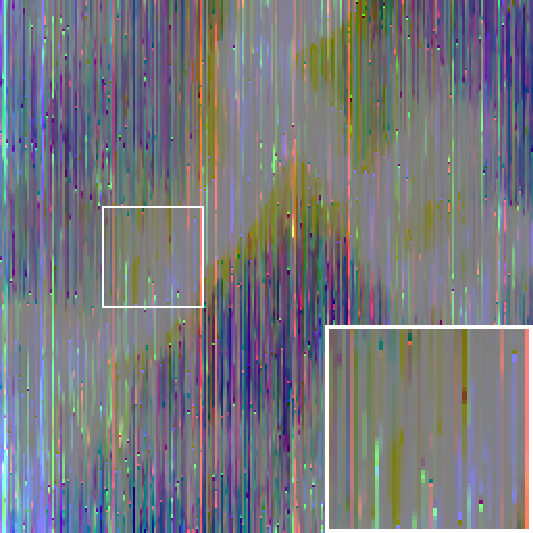}}
	\end{minipage}
	\begin{minipage}{0.09\hsize}
		\centerline{\includegraphics[width=\hsize]{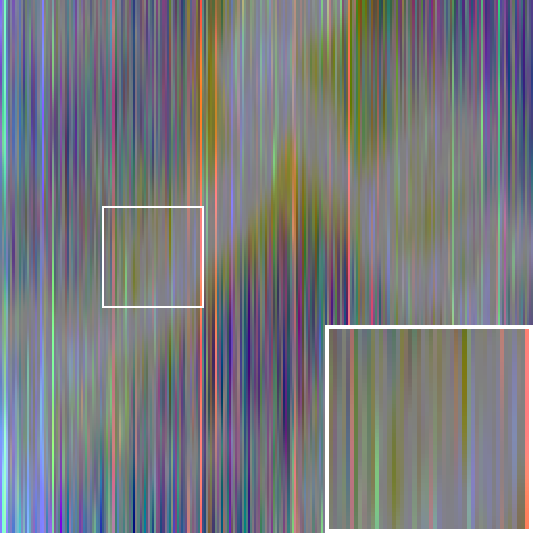}}
	\end{minipage}
	\begin{minipage}{0.01\hsize}
		\rotatebox[origin=c]{90}{\scriptsize{Stripe noise}}
	\end{minipage}
	\begin{minipage}{0.09\hsize}
		\centerline{\includegraphics[width=\hsize]{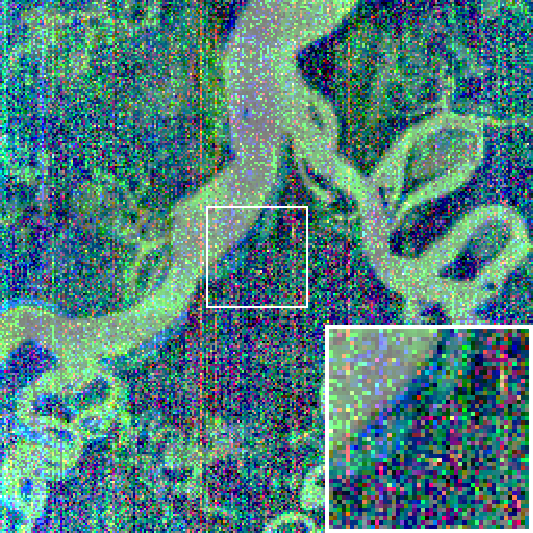}}
	\end{minipage}
	\begin{minipage}{0.09\hsize}
		\centerline{\includegraphics[width=\hsize]{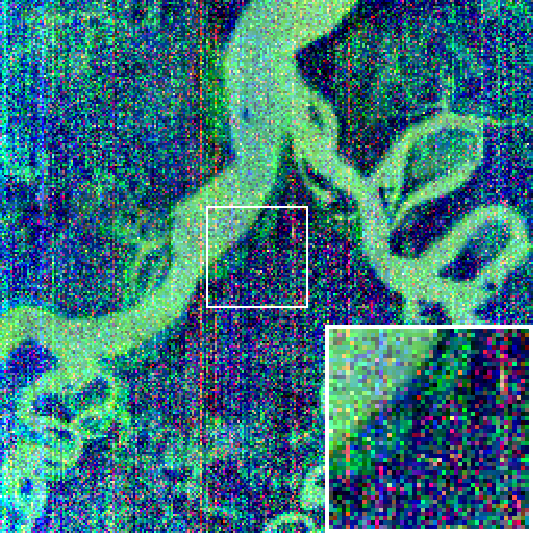}}
	\end{minipage}
	\begin{minipage}{0.09\hsize}
		\centerline{\includegraphics[width=\hsize]{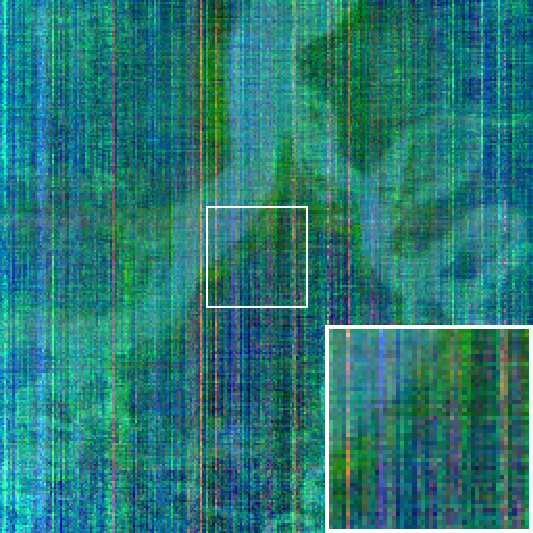}}
	\end{minipage}
	\begin{minipage}{0.09\hsize}
		\centerline{\includegraphics[width=\hsize]{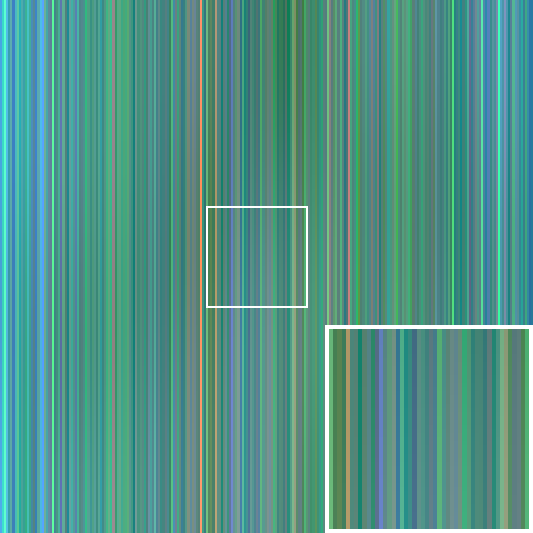}}
	\end{minipage}
	\begin{minipage}{0.09\hsize}
		\centerline{\includegraphics[width=\hsize]{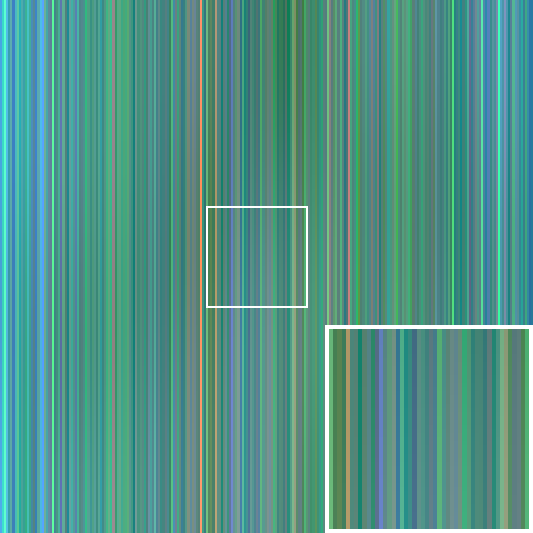}}
	\end{minipage}

	\vspace{1mm}
	
	\begin{minipage}{0.01\hsize}
		~
	\end{minipage}
	\begin{minipage}{0.09\hsize}
		\centerline{\small{(a3) S~\cite{LRMR}}}
	\end{minipage}
	\begin{minipage}{0.09\hsize}
		\centerline{\small{(b3) GS~\cite{GLSSTV}}}
	\end{minipage}
	\begin{minipage}{0.09\hsize}
		\centerline{\small{(c3) LR~\cite{NN_char}}}
	\end{minipage}
	\begin{minipage}{0.09\hsize}
		\centerline{\small{(d3) TV~\cite{gradient_constraint}}}
	\end{minipage}
	\begin{minipage}{0.09\hsize}
		\centerline{\small{(e3) FC}}
	\end{minipage}
	\begin{minipage}{0.01\hsize}
		~
	\end{minipage}
	\begin{minipage}{0.09\hsize}
		\centerline{\small{(a4) S~\cite{LRMR}}}
	\end{minipage}
	\begin{minipage}{0.09\hsize}
		\centerline{\small{(b4) GS~\cite{GLSSTV}}}
	\end{minipage}
	\begin{minipage}{0.09\hsize}
		\centerline{\small{(c4) LR~\cite{NN_char}}}
	\end{minipage}
	\begin{minipage}{0.09\hsize}
		\centerline{\small{(d4) TV~\cite{gradient_constraint}}}
	\end{minipage}
	\begin{minipage}{0.09\hsize}
		\centerline{\small{(e4) FC}}
	\end{minipage}

	\begin{minipage}{0.48\hsize}
		\centerline{SSTV+TNN}
	\end{minipage}
	\begin{minipage}{0.48\hsize}
		\centerline{$l_{0}$-$l_{1}$HTV}
	\end{minipage}
	\vspace{1mm}

	\begin{minipage}{0.01\hsize}
		\rotatebox[origin=c]{90}{\scriptsize{Brightened HSI}}
	\end{minipage}
	\begin{minipage}{0.09\hsize}
		\centerline{\includegraphics[width=\hsize]{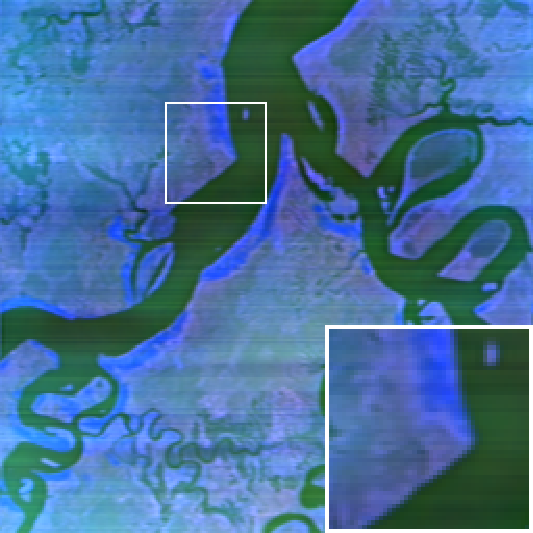}}
	\end{minipage}
	\begin{minipage}{0.09\hsize}
		\centerline{\includegraphics[width=\hsize]{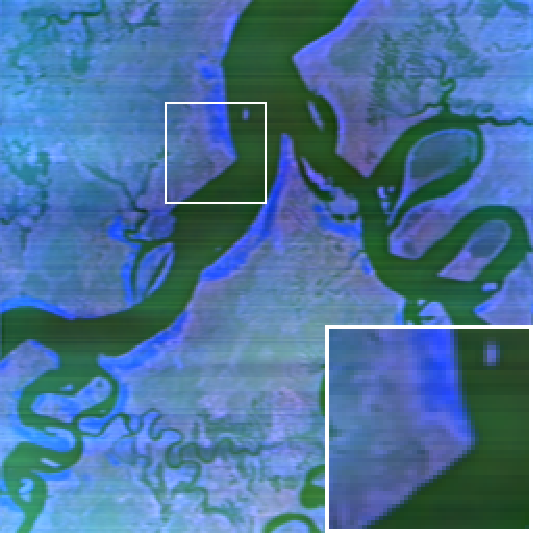}}
	\end{minipage}
	\begin{minipage}{0.09\hsize}
		\centerline{\includegraphics[width=\hsize]{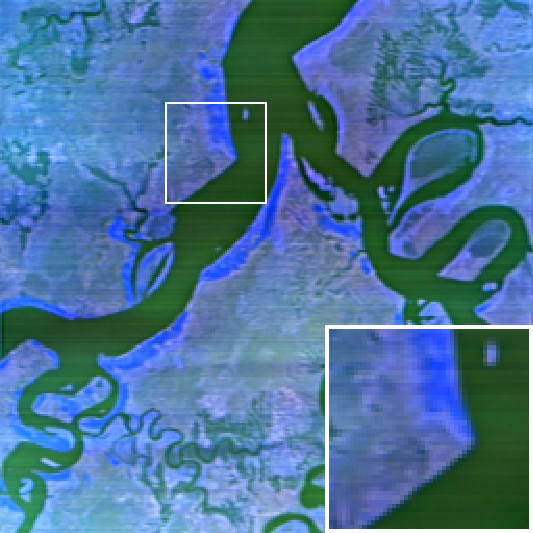}}
	\end{minipage}
	\begin{minipage}{0.09\hsize}
		\centerline{\includegraphics[width=\hsize]{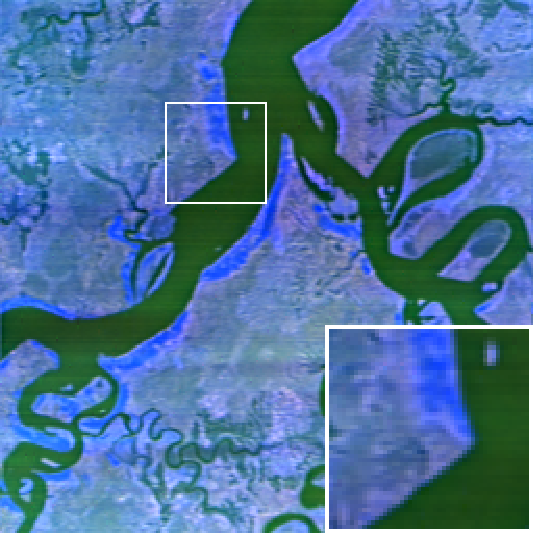}}
	\end{minipage}
	\begin{minipage}{0.09\hsize}
		\centerline{\includegraphics[width=\hsize]{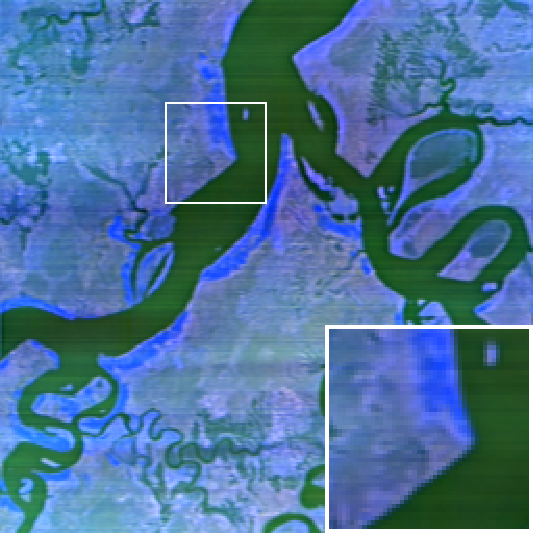}}
	\end{minipage}
	\begin{minipage}{0.01\hsize}
		\rotatebox[origin=c]{90}{\scriptsize{Brightened HSI}}
	\end{minipage}
	\begin{minipage}{0.09\hsize}
		\centerline{\includegraphics[width=\hsize]{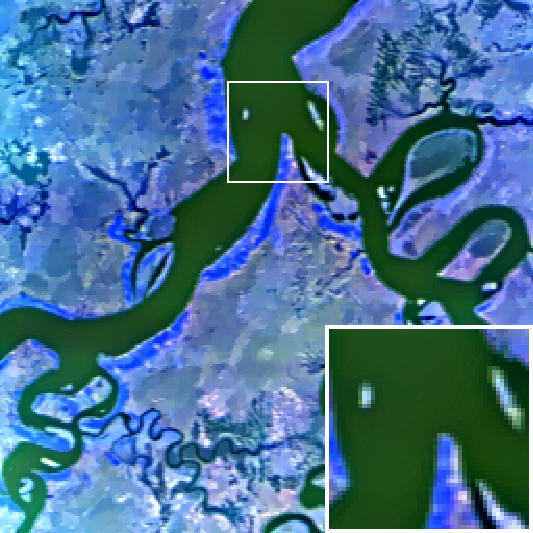}}
	\end{minipage}
	\begin{minipage}{0.09\hsize}
		\centerline{\includegraphics[width=\hsize]{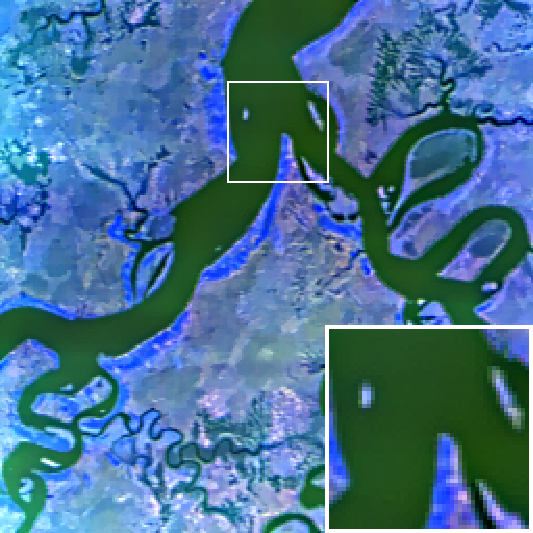}}
	\end{minipage}
	\begin{minipage}{0.09\hsize}
		\centerline{\includegraphics[width=\hsize]{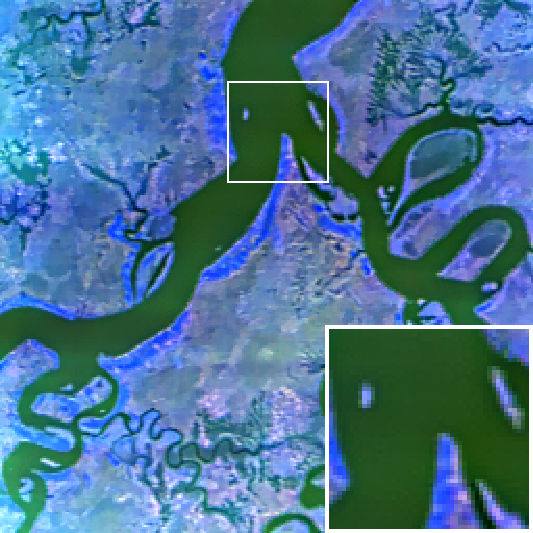}}
	\end{minipage}
	\begin{minipage}{0.09\hsize}
		\centerline{\includegraphics[width=\hsize]{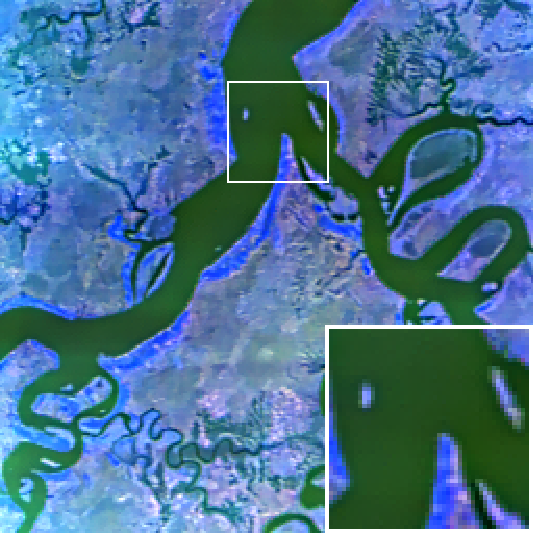}}
	\end{minipage}
	\begin{minipage}{0.09\hsize}
		\centerline{\includegraphics[width=\hsize]{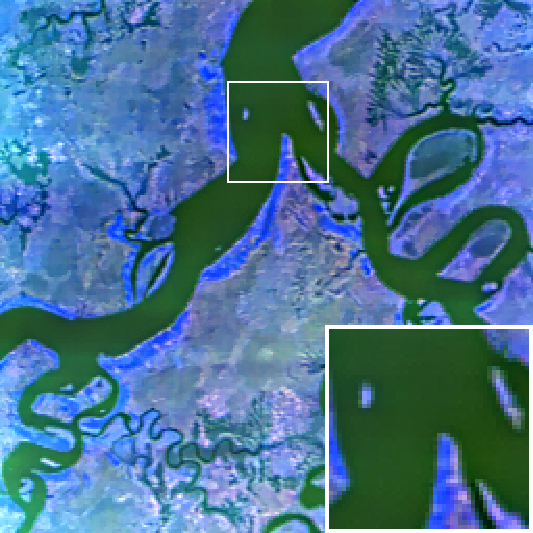}}
	\end{minipage}

	\begin{minipage}{0.01\hsize}
		\rotatebox[origin=c]{90}{\scriptsize{Stripe noise}}
	\end{minipage}
	\begin{minipage}{0.09\hsize}
			\centerline{\includegraphics[width=\hsize]{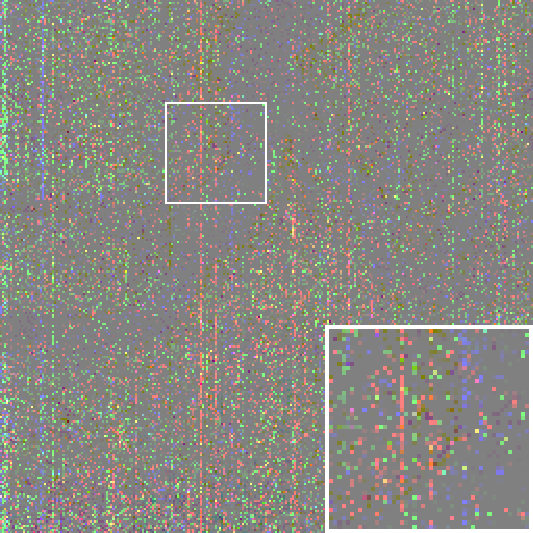}}
	\end{minipage}
	\begin{minipage}{0.09\hsize}
			\centerline{\includegraphics[width=\hsize]{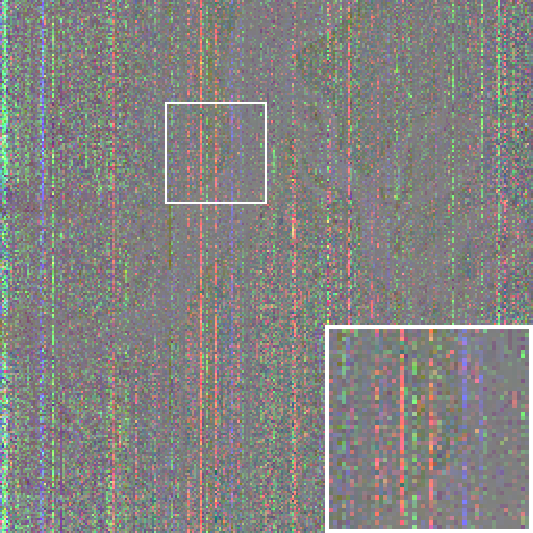}}
	\end{minipage}
	\begin{minipage}{0.09\hsize}
			\centerline{\includegraphics[width=\hsize]{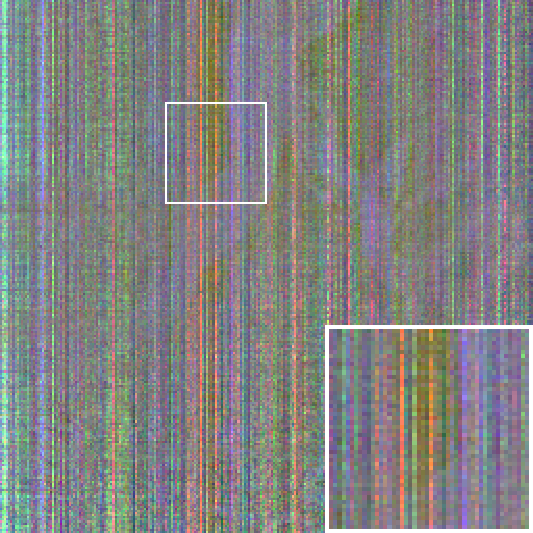}}
	\end{minipage}
	\begin{minipage}{0.09\hsize}
			\centerline{\includegraphics[width=\hsize]{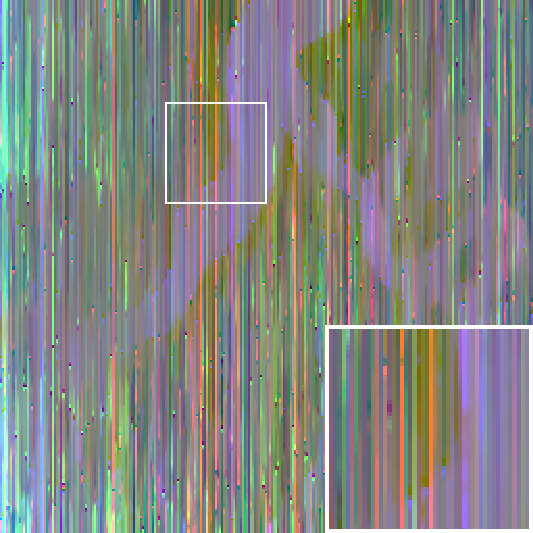}}
	\end{minipage}
	\begin{minipage}{0.09\hsize}
			\centerline{\includegraphics[width=\hsize]{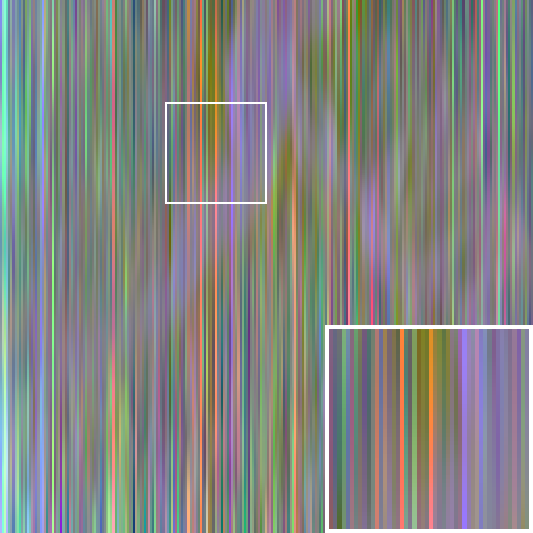}}
	\end{minipage}
	\begin{minipage}{0.01\hsize}
		\rotatebox[origin=c]{90}{\scriptsize{Stripe noise}}
	\end{minipage}
	\begin{minipage}{0.09\hsize}
		\centerline{\includegraphics[width=\hsize]{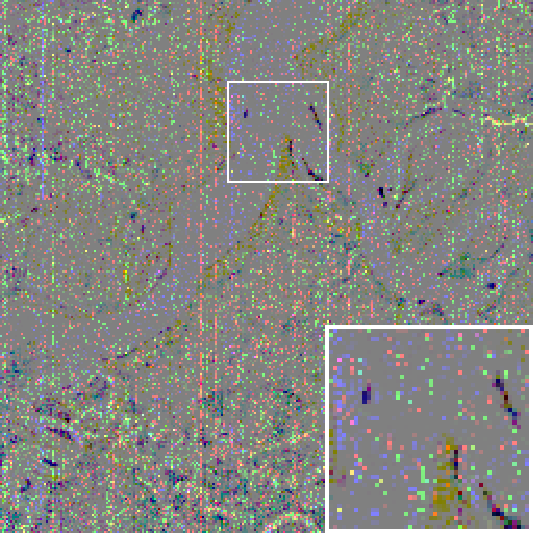}}
	\end{minipage}
	\begin{minipage}{0.09\hsize}
		\centerline{\includegraphics[width=\hsize]{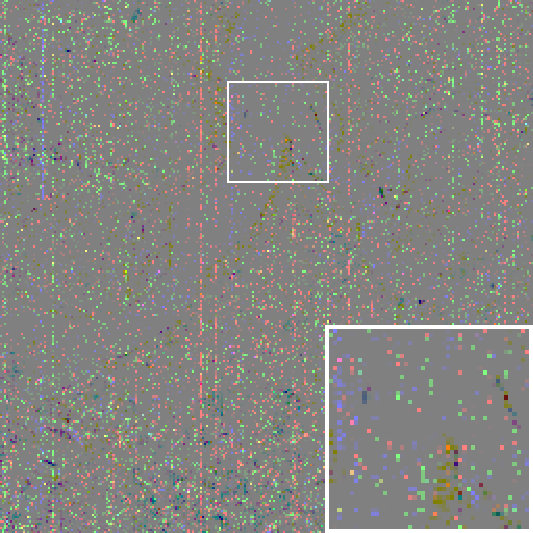}}
	\end{minipage}
	\begin{minipage}{0.09\hsize}
		\centerline{\includegraphics[width=\hsize]{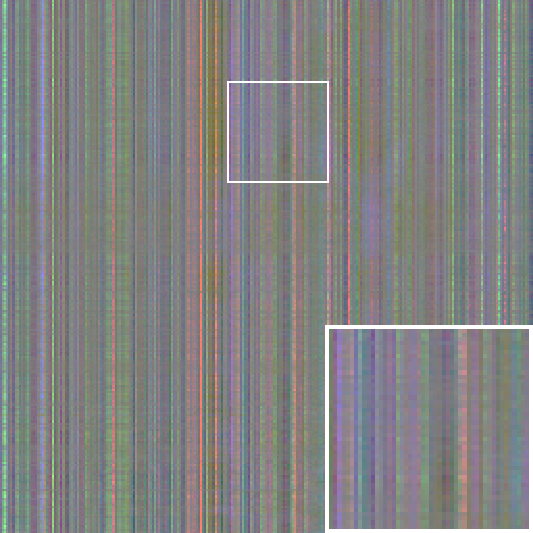}}
	\end{minipage}
	\begin{minipage}{0.09\hsize}
		\centerline{\includegraphics[width=\hsize]{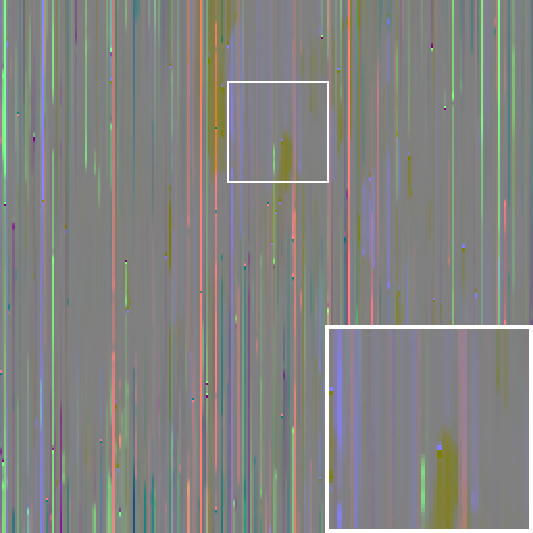}}
	\end{minipage}
	\begin{minipage}{0.09\hsize}
		\centerline{\includegraphics[width=\hsize]{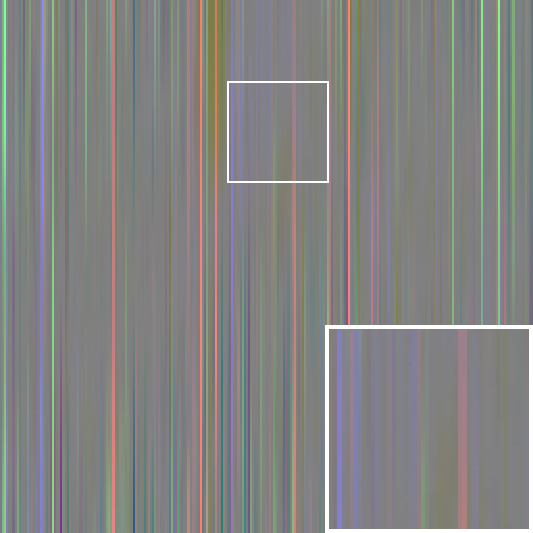}}
	\end{minipage}
	
	\vspace{1mm}
	
	\begin{minipage}{0.01\hsize} %この一つのminipageが一つの部屋って感じで，これをまず配置する
		~
	\end{minipage}
	\begin{minipage}{0.09\hsize}
		\centerline{\small{(a5) S~\cite{LRMR}}}
	\end{minipage}
	\begin{minipage}{0.09\hsize}
		\centerline{\small{(b5) GS~\cite{GLSSTV}}}
	\end{minipage}
	\begin{minipage}{0.09\hsize}
		\centerline{\small{(c5) LR~\cite{NN_char}}}
	\end{minipage}
	\begin{minipage}{0.09\hsize}
		\centerline{\small{(d5) TV~\cite{gradient_constraint}}}
	\end{minipage}
	\begin{minipage}{0.09\hsize}
		\centerline{\small{(e5) FC}}
	\end{minipage}
	\begin{minipage}{0.01\hsize} %この一つのminipageが一つの部屋って感じで，これをまず配置する
		~
	\end{minipage}
	\begin{minipage}{0.09\hsize}
		\centerline{\small{(a6) S~\cite{LRMR}}}
	\end{minipage}
	\begin{minipage}{0.09\hsize}
		\centerline{\small{(b6) GS~\cite{GLSSTV}}}
	\end{minipage}
	\begin{minipage}{0.09\hsize}
		\centerline{\small{(c6) LR~\cite{NN_char}}}
	\end{minipage}
	\begin{minipage}{0.09\hsize}
		\centerline{\small{(d6) TV~\cite{gradient_constraint}}}
	\end{minipage}
	\begin{minipage}{0.09\hsize}
		\centerline{\small{(e6) FC}}
	\end{minipage}
	
	\caption{HSI destriping results in real noise cases (R: 357, G: 275, B: 120). 
		The top rows and bottom rows are the estimated HSIs and the estimated stripe noise, respectively.}
	\label{fig:HTV_TNN_results_on_Suwannee}
\end{figure*}

Table~\ref{tab:Reg_and_strcha} summarizes all combinations of stripe noise characterizations and image regularizations examined in our experiments, where we indicate reference numbers for specific combinations that have been proposed in existing studies ("None" means that the combination has not been considered yet).

\subsection{Dataset Descriptions}

We employed three HSI datasets and two IR datasets for experiments in simulated and real noise cases. All images were normalized between $[0,1]$.

The \textit{Moffett Field}~\cite{MoffettField} was acquired by Airborne Visible/Infrared Imaging Spectrometer (AVIRIS) over the urban and rural area in Moffett Field, CA, USA, with a spatial resolution of $20$ m. This image consists of $224$ spectral bands in the range of $400-2500$ nm. After removing noisy bands, we used a sub-image of size $395 \times 185 \times 176$ (Fig.~\ref{fig:SSTV_results_on_MoffettField_stripe_only} (a)) for experiments in simulated noise cases.

The \textit{Salinas}~\cite{GIC} was collected by AVIRIS over the field area in Salinas Valley, CA, USA, with a spatial resolution of $3.7$ m. This image consists of $224$ spectral bands in the range of $400-2500$ nm. After removing noisy bands, we used a sub-image of size $360 \times 217 \times 190$ (Fig.~\ref{fig:TNN_results_on_Salinas} (a)) for experiments in simulated noise cases.

The \textit{Suwannee}~\cite{SpecTIR} acquired by AVIRIS over National Wildlife Reserves in the Gulf of Mexico with a spatial resolution of $2$ m. This image consists of $360$ spectral bands in the range of $395-2450$ nm. We used a sub-image of size $256 \times 256 \times 360$ (Fig.~\ref{fig:real_data} (a)) for experiments in real noise cases.

The \textit{Bats1} and \textit{Bats2}~\cite{TIV_data}, which include hundreds of bats, were collected with three FLIR SC6000 thermal infrared cameras at a frame rate of $125$ Hz. For more detailed descriptions, see also~\cite{TIV_visual_analysis,description_IR_2009,description_IR_2012}. We used denoised and raw sub-images of size $256 \times 256 \times 50$ (Figs.~\ref{fig:TV1_results_on_test1_stripe_only} (a) and~\ref{fig:real_data} (b)) for experiments in simulated and real noise cases, respectively.

\subsection{Experiments in Simulated Noise Cases}
\label{ssec:Simulated_experiments}

For the HSI destriping experiments, the parameter $\lambda$ of each stripe noise characterization model summarized in Tab.~\ref{tab:various_stripe_noise_characterization} was set to a hand-optimized value, so as to achieve the best MPSNR. For fair comparison, we set $\varepsilon$ to the oracle value, i.e., $\varepsilon=\|\mathcal{N}\|_{F}$. As quantitative evaluations, we employed the mean peak signal-to-noise ratio (MPSNR):
\begin{equation}
	\label{eq:MPSNR}
	\mathrm{MPSNR} = \frac{1}{n_{3}} \sum_{k=1}^{n_{3}} 10\log_{10}\frac{n_{1}n_{2}}{\|\mathcal{U}_{k} - \bar{\mathcal{U}}_{k}\|_{2}^{2}}, 
\end{equation}
and the mean structural similarity overall bands (MSSIM)~\cite{MSSIM}:
\begin{equation}
	\label{eq:MSSIM}
	\mathrm{MSSIM} = \frac{1}{n_{3}} \sum_{k=1}^{n_{3}} \mathrm{SSIM}(\mathcal{U}_{k}, \bar{\mathcal{U}_{k}}), 
\end{equation}
where $\mathcal{U}_{k}$ is the $k$th band of $\mathcal{U}$.
The larger these values are, the better the destriping results are. The stopping criterion of Alg.~\ref{algo:DP_PDS_for_zero_gradient_constraint_vt} was set as $\frac{\|\mathcal{U}^{(n+1)}-\mathcal{U}^{(n)}\|_{F}}{\|\mathcal{U}^{(n)}\|_{F}}<1.0\times 10^{-4}$.

We generated the three types of degraded images:
\begin{itemize}
	\item[(i)] HSIs with vertical stripe noise,
	\item[(ii)] IR videos with time-invariant vertical stripe noise,
	\item[(iii)] HSIs with vertical stripe noise and white Gaussian noise.
\end{itemize}
In the IR video experiments, we only consider stripe noise because Gaussian-like random noise does not appear in raw IR video data~\cite{Filtering_based_destriping_2016,Filtering_based_destriping_2020}. For the variety of experiments, we considered the following five types of the intensity range of stripe noise: $[-0.2,0.2]$, $[-0.25,0.25]$, $[-0.3,0.3]$, $[-0.35,0.35]$, and $[-0.4,0.4]$. The standard deviation of white Gaussian noise was set to $0.05$.

Tables~\ref{tab:PSNR_and_SSIM_so},~\ref{tab:PSNR_and_SSIM_so_IR_video}, and~\ref{tab:PSNR_and_SSIM_sg} list the resulting MPSNR and MSSIM values in Case (i), Case (ii), and Case (iii), respectively. The best and second-best values are highlighted in bold and underline, respectively. The proposed FC achieved the best/second-best MPSNR and MSSIM values in most cases. S and GS performed worse overall. LR and TV performed better than S and GS. However, the performance of LR and TV is significantly degraded in the cases where they are combined with a low-rank image regularization (LR-TNN) and TV image regularizations (TV-SSTV and TV-ASSTV), respectively. 

Figures~\ref{fig:SSTV_results_on_MoffettField_stripe_only},~\ref{fig:TV1_results_on_test1_stripe_only}, and~\ref{fig:TNN_results_on_Salinas} depict the \textit{Moffett field} destriping results in Case (i) using SSTV, the \textit{Bats1} destriping results in Case (ii) using ATV, and the \textit{Salinas} destriping results in Case (iii) using TNN, respectively. Figure~\ref{fig:bw_PSNR_SSIM_TNN} plots their band-wise or frame-wise PSNRs and SSIMs. In the 95th-band results of Figs.~\ref{fig:bw_PSNR_SSIM_TNN} (a) and (b), the PSNRs and SSIMs of S-SSTV, GS-SSTV, and LR-SSTV dropped to 30 [dB] and 0.7, respectively. This is because S-SSTV, GS-SSTV, and LR-SSTV excessively smoothened the spectral signatures around the band. In the magnified areas of Figs.~\ref{fig:SSTV_results_on_MoffettField_stripe_only} (c), (d), and (e), we see that the land shapes of the red and green bands are removed as Gaussian and stripe noise. TV-SSTV also resulted in the low PSNRs and SSIMs of the band 95 and eliminated some edges in addition to the stripe noise (see Fig.~\ref{fig:SSTV_results_on_MoffettField_stripe_only} (f)). S-ATV, GS-ATV, LR-ATV, and TV-ATV removed bats as stripe noise, resulting in poor performance (see Figs.~\ref{fig:TV1_results_on_test1_stripe_only} (c), (d), (e), and (f)). Figures~\ref{fig:bw_PSNR_SSIM_TNN} (c) and (d) show that the PSNRs and SSIMs of S-ATV, GS-ATV, LR-ATV, and TV-ATV vary according to frame numbers. The reason is that the results are worse as the number of unrestored bats increases. In contrast, FC-SSTV recovers the land shapes and edges (see Fig.~\ref{fig:SSTV_results_on_MoffettField_stripe_only} (f)) and FC-ATV accurately removed stripe noise, leading to high PSNRs and SSIMs. The SSIM results for Figs.~\ref{fig:bw_PSNR_SSIM_TNN} (e) and (f) were better for LR than FC and TV, but the PSNRs were better for FC and TV than LR. In particular, from 30 to 150 bands, FC and TV achieved $10$ [dB] better PSNRs and $0.01$ worse SSIMs than LR. In the magnified area of the stripe noise by LR-TNN (Fig.~\ref{fig:TNN_results_on_Salinas} (e)), the yellow line appears along with a field shape. This indicates that LR-TNN restores the image structure but does not recover the contrast. The three results verify that FC consistently achieves high performance due to its accurate capturing ability for stripe noise.

Figure~\ref{fig:all_PSNR_SSIM_mean_case} shows the means of MPSNRs and MSSIMs in each noise case. In Case (i), LR and FC accurately captured stripe noise, leading to better performances than TV. In Case (ii), FC achieved the best performance. This is because FC captures the temporal flatness while the other characterizations do not. In Case (iii), LR captured horizontal lines as a stripe noise component to remove Gaussian noise by the intersections between vertical stripe noise and the horizontal lines, leading to worse results. On the other hand, TV and FC obtained better results than LR without capturing the horizontal lines.

Figure~\ref{fig:all_PSNR_SSIM_mean} plots the means of MPSNRs and MSSIMs in each stripe noise intensity. LR dropped its MPSNRs as the stripe noise intensities increased. This is due to the fact that LR removes the meaningful image components as stripe noise components if stripe noise intensity is high. The MPSNRs and MSSIMs of TV did not decrease depending on the stripe noise intensities but were lower than FC overall. Compared with these existing stripe noise characterizations, FC accurately eliminated stripe noise, resulting in high destriping performances regardless of the stripe noise intensity.

Figure~\ref{fig:all_PSNR_SSIM_mean_reg} shows the means of MPSNRs and MSSIMs in each image regularization. FC resulted in $0.5$ [dB] worse MPSNRs than LR for the ASSTV and SSTV+TNN cases. This is because FC-ASSTV and FC-SSTV+TNN stop the iterations before the stripe noise components satisfy the flatness constraint, leading to slightly dropping their MPSNRs and MSSIMs. On the other hand, FC did obtain a $2$ [dB] better MPSNR and $0.05$ better MSSIM than LR for the TNN case. Compared with TV, the performances of FC were similar for HTV, TNN, SSTV+TNN, and $l_{0}$-$l_{1}$HTV and better for SSTV and ASSTV. Moreover, FC stably performed better than the other characterizations for ATV, ITV, and ATV+NN. These reveal that our framework achieves good performance on average, whatever image regularizations are used.

\subsection{Experiments in Real Noise Cases}

In the real noise-case experiments, the parameter $\lambda$ (Tab.~\ref{tab:various_stripe_noise_characterization}) for each method was determined manually to balance the tradeoff between the visual quality (e.g., over-smoothed or not) and destriping performance (e.g., stripe noise is sufficiently removed or not). For the data fidelity parameter $\varepsilon$, we adjusted it to an appropriate value after empirically estimating the intensity of the noise in the real data. Specifically, it was set to 200 for \textit{Suwannee} and 0 for \textit{Bats2}. The stopping criterion of Alg.~\ref{algo:DP_PDS_for_zero_gradient_constraint_vt} was set as $\frac{\|\mathcal{U}^{(n+1)}-\mathcal{U}^{(n)}\|_{F}}{\|\mathcal{U}^{(n)}\|_{F}}<1.0\times 10^{-4}$.

We show the \textit{Suwannee} destriping results for a real noise case in Fig.~\ref{fig:HTV_TNN_results_on_Suwannee}. The destriping result by S-HTV (Fig.~\ref{fig:HTV_TNN_results_on_Suwannee} (a1)) includes residual stripe noise. The results by S-SSTV (Fig.~\ref{fig:HTV_TNN_results_on_Suwannee} (a2)), GS-SSTV (Fig.~\ref{fig:HTV_TNN_results_on_Suwannee} (b2)), S-ASSTV (Fig.~\ref{fig:HTV_TNN_results_on_Suwannee} (a3)), GS-ASSTV (Fig.~\ref{fig:HTV_TNN_results_on_Suwannee} (b3)), S-TNN (Fig.~\ref{fig:HTV_TNN_results_on_Suwannee} (a4)), GS-TNN (Fig.~\ref{fig:HTV_TNN_results_on_Suwannee} (b4)), and S-$l_{0}$-$l_{1}$HTV (Fig.~\ref{fig:HTV_TNN_results_on_Suwannee} (a6)) have brighter areas than the original image (Fig.~\ref{fig:real_data} (a)), and some of the land shapes in the magnified areas were removed as the stripe noise components. These suggest that S and GS are less capable of capturing the vertical continuity of stripe noise. LR-ASSTV (Fig.~\ref{fig:HTV_TNN_results_on_Suwannee} (c3)) recovered the narrow river that lies along with the vertical direction in the magnified areas. On the other hand, LR-SSTV (Fig.~\ref{fig:HTV_TNN_results_on_Suwannee} (c2)) and LR-TNN (Fig.~\ref{fig:HTV_TNN_results_on_Suwannee} (c4)) removed part of the global structure in the image as stripe noise. This may be due to the fact that LR allows for changes in the overall luminance level so that it does not prevent spectral oversmoothing caused by the image regularizations. In the results by TV-SSTV (Fig.~\ref{fig:HTV_TNN_results_on_Suwannee} (d2)), TV-ASSTV (Fig.~\ref{fig:HTV_TNN_results_on_Suwannee} (d3)), TV-SSTV+TNN (Fig.~\ref{fig:HTV_TNN_results_on_Suwannee} (d5)), and TV-$l_{0}$-$l_{1}$HTV (Fig.~\ref{fig:HTV_TNN_results_on_Suwannee} (d6)), land shape was also partially removed as stripe noise. For example, TV-ASSTV (Fig.~\ref{fig:HTV_TNN_results_on_Suwannee} (d3)) completely removed the narrow river in the magnified area. This is because there is a conflict between SSTV, ASSTV, SSTV+TNN, and $l_{0}$-$l_{1}$HTV, used as image regularizations, and TV, used as a stripe noise characterization. Compared with these existing stripe noise characterizations, for FC-HTV, FC-SSTV, FC-TNN, FC-$l_{0}$-$l_{1}$HTV, its strong ability of stripe noise characterization allows us to achieve desirable destriping. However, our results do not satisfy the flatness constraint and slightly include land shapes in the stripe noise components only for FC-ASSTV and FC-SSTV+TNN (Figs.~\ref{fig:HTV_TNN_results_on_Suwannee} (e3) and (e5)). This indicates that FC-ASSTV and FC-SSTV+TNN need more iterations to preclude the land shapes from their stripe noise components.

%%%%%%%%%%%%%%%%%%%%%%%%%%%%%%%%%%%%%%%%%%%%%%%%%%%%%%%
%% real TIV data destriping results
%%%%%%%%%%%%%%%%%%%%%%%%%%%%%%%%%%%%%%%%%%%%%%%%%%%%%%%
\begin{figure}[!t]
	
	\centerline{ATV}
	
	\vspace{1mm}
	%1st line
	\begin{minipage}{0.02\hsize}
		\rotatebox[origin=c]{90}{\scriptsize{Estimated IR video}}
	\end{minipage}
	\begin{minipage}{0.18\hsize}
		\centerline{\includegraphics[width=\hsize]{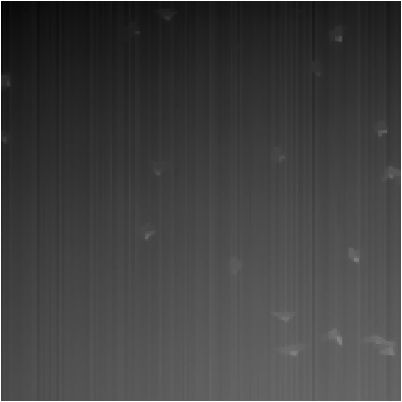}}
	\end{minipage}
	\begin{minipage}{0.18\hsize}
		\centerline{\includegraphics[width=\hsize]{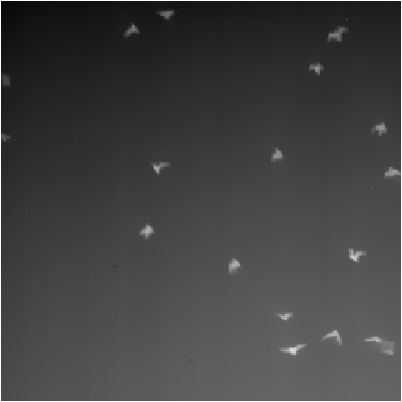}}
	\end{minipage}
	\begin{minipage}{0.18\hsize}
		\centerline{\includegraphics[width=\hsize]{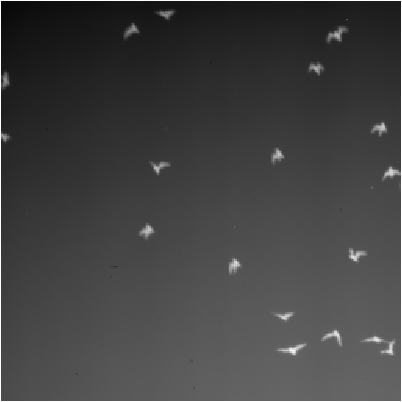}}
	\end{minipage}
	\begin{minipage}{0.18\hsize}
		\centerline{\includegraphics[width=\hsize]{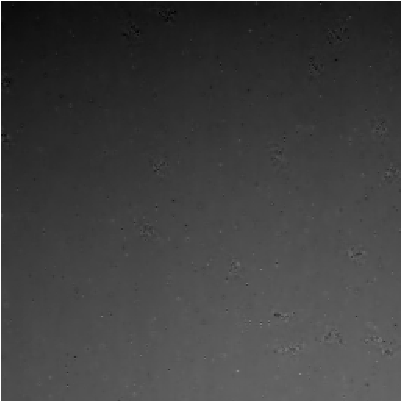}}
	\end{minipage}
	\begin{minipage}{0.18\hsize}
		\centerline{\includegraphics[width=\hsize]{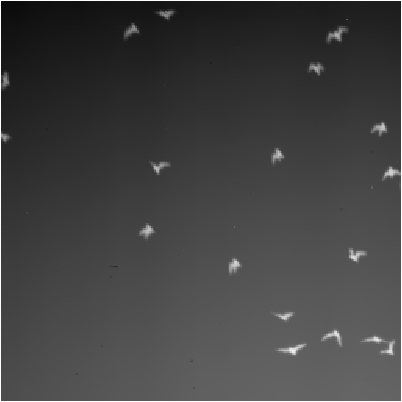}}
	\end{minipage}

	\begin{minipage}{0.02\hsize}
		\rotatebox[origin=c]{90}{\scriptsize{Stripe noise}}
	\end{minipage}
	\begin{minipage}{0.18\hsize}
		\centerline{\includegraphics[width=\hsize]{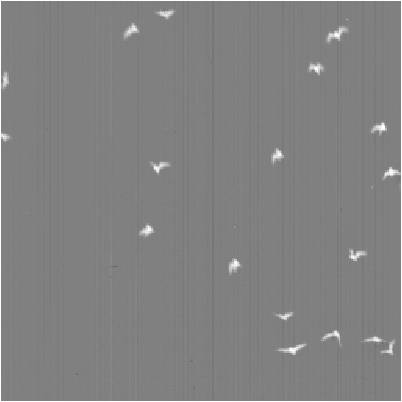}}
	\end{minipage}
	\begin{minipage}{0.18\hsize}
		\centerline{\includegraphics[width=\hsize]{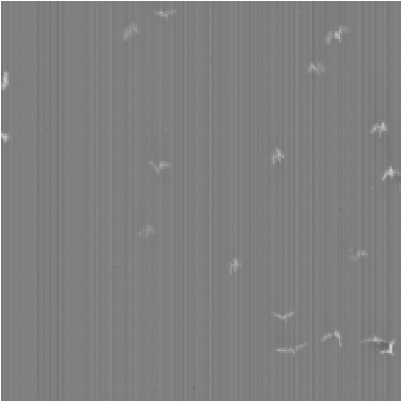}}
	\end{minipage}
	\begin{minipage}{0.18\hsize}
		\centerline{\includegraphics[width=\hsize]{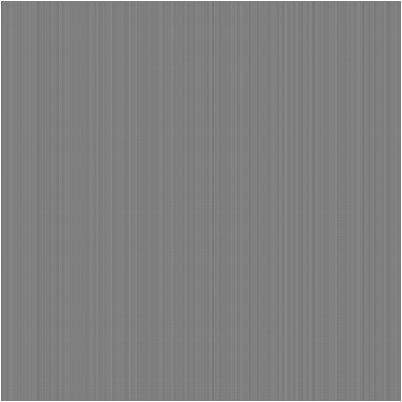}}
	\end{minipage}
	\begin{minipage}{0.18\hsize}
		\centerline{\includegraphics[width=\hsize]{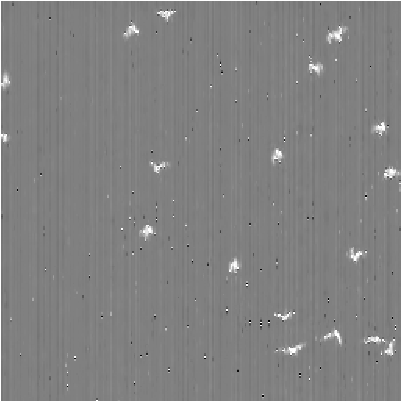}}
	\end{minipage}
	\begin{minipage}{0.18\hsize}
		\centerline{\includegraphics[width=\hsize]{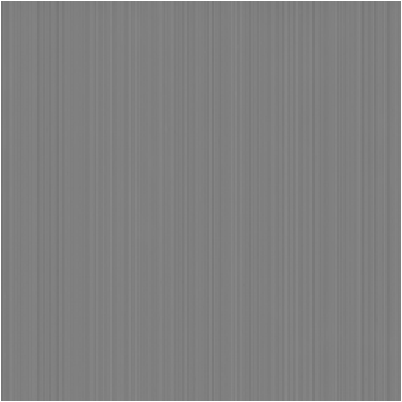}}
	\end{minipage}
	
	\vspace{1mm}
	
	\begin{minipage}{0.02\hsize} %この一つのminipageが一つの部屋って感じで，これをまず配置する
		~
	\end{minipage}
	\begin{minipage}{0.18\hsize}
		\centerline{\footnotesize{(a1) S~\cite{LRMR}}}
	\end{minipage}
	\begin{minipage}{0.18\hsize}
		\centerline{\footnotesize{(b1) GS~\cite{GLSSTV}}}
	\end{minipage}
	\begin{minipage}{0.18\hsize}
		\centerline{\footnotesize{(c1) LR~\cite{NN_char}}}
	\end{minipage}
	\begin{minipage}{0.18\hsize}
		\centerline{\footnotesize{(d1) TV~\cite{gradient_constraint}}}
	\end{minipage}
	\begin{minipage}{0.18\hsize}
		\centerline{\footnotesize{(e1) FC}}
	\end{minipage}
	
	\vspace{1mm}
	
	\centerline{ITV}
	
	\vspace{1mm}
	%1st line
	\begin{minipage}{0.02\hsize}
		\rotatebox[origin=c]{90}{\scriptsize{Estimated IR video}}
	\end{minipage}
	\begin{minipage}{0.18\hsize}
		\centerline{\includegraphics[width=\hsize]{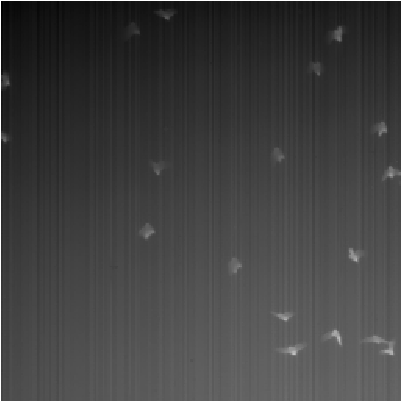}}
	\end{minipage}
	\begin{minipage}{0.18\hsize}
		\centerline{\includegraphics[width=\hsize]{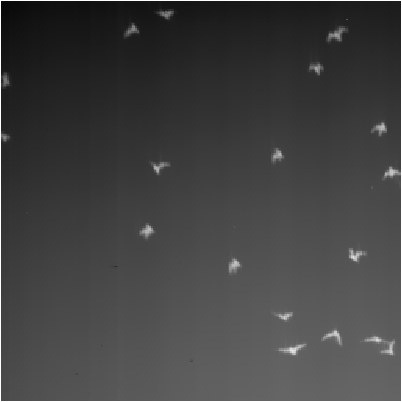}}
	\end{minipage}
	\begin{minipage}{0.18\hsize}
		\centerline{\includegraphics[width=\hsize]{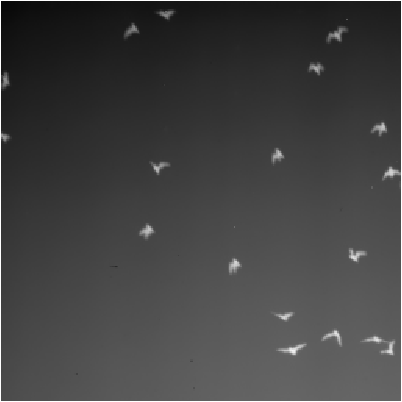}}
	\end{minipage}
	\begin{minipage}{0.18\hsize}
		\centerline{\includegraphics[width=\hsize]{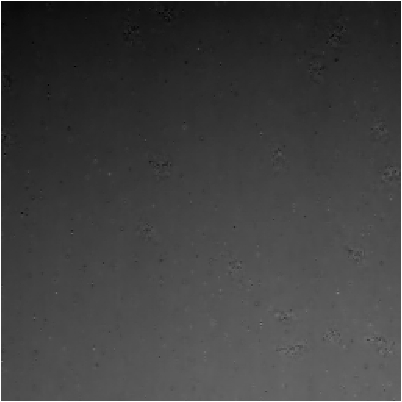}}
	\end{minipage}
	\begin{minipage}{0.18\hsize}
		\centerline{\includegraphics[width=\hsize]{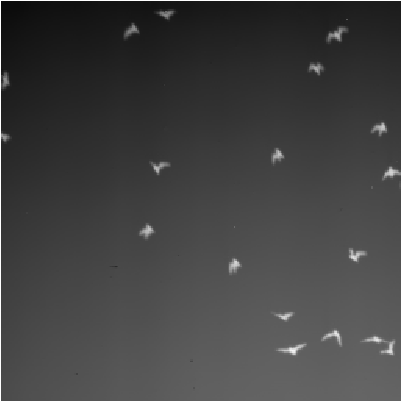}}
	\end{minipage}

	\begin{minipage}{0.02\hsize}
		\rotatebox[origin=c]{90}{\scriptsize{Stripe noise}}
	\end{minipage}
	\begin{minipage}{0.18\hsize}
		\centerline{\includegraphics[width=\hsize]{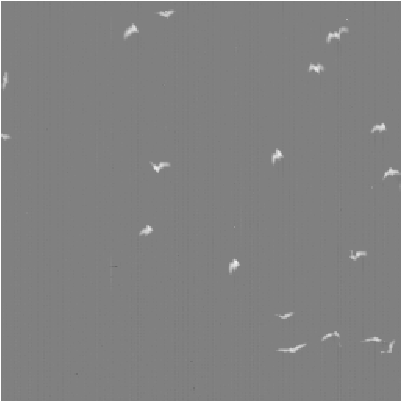}}
	\end{minipage}
	\begin{minipage}{0.18\hsize}
		\centerline{\includegraphics[width=\hsize]{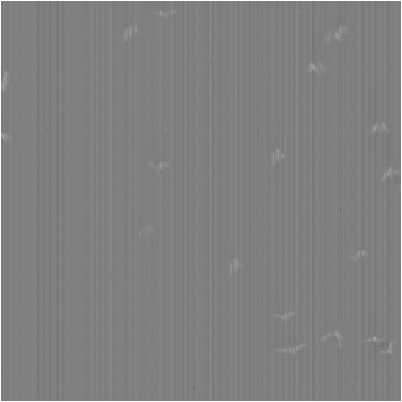}}
	\end{minipage}
	\begin{minipage}{0.18\hsize}
		\centerline{\includegraphics[width=\hsize]{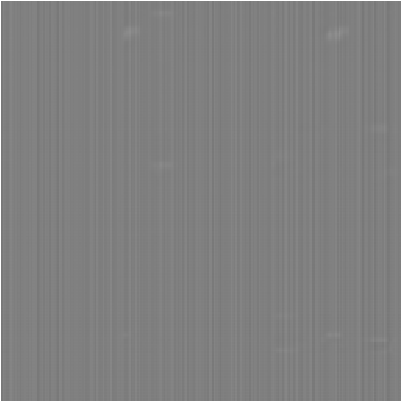}}
	\end{minipage}
	\begin{minipage}{0.18\hsize}
		\centerline{\includegraphics[width=\hsize]{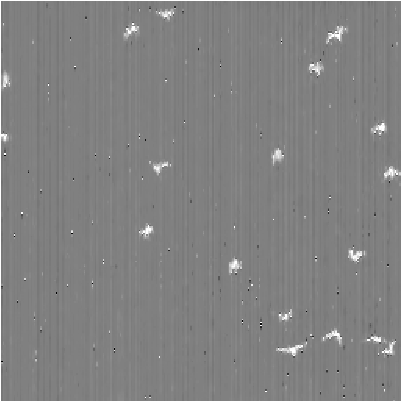}}
	\end{minipage}
	\begin{minipage}{0.18\hsize}
		\centerline{\includegraphics[width=\hsize]{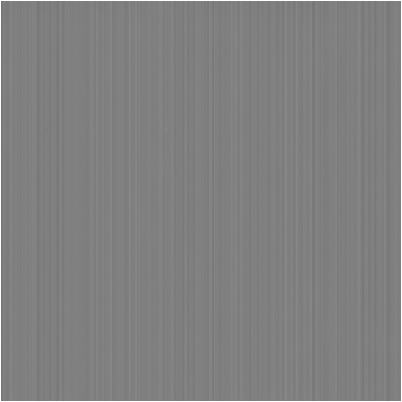}}
	\end{minipage}

	\vspace{1mm}
	
	\begin{minipage}{0.02\hsize} %この一つのminipageが一つの部屋って感じで，これをまず配置する
		~
	\end{minipage}
	\begin{minipage}{0.18\hsize}
		\centerline{\footnotesize{(a2) S~\cite{LRMR}}}
	\end{minipage}
	\begin{minipage}{0.18\hsize}
		\centerline{\footnotesize{(b2) GS~\cite{GLSSTV}}}
	\end{minipage}
	\begin{minipage}{0.18\hsize}
		\centerline{\footnotesize{(c2) LR~\cite{NN_char}}}
	\end{minipage}
	\begin{minipage}{0.18\hsize}
		\centerline{\footnotesize{(d2) TV~\cite{gradient_constraint}}}
	\end{minipage}
	\begin{minipage}{0.18\hsize}
		\centerline{\footnotesize{(e2) FC}}
	\end{minipage}

	\vspace{1mm}
	
	\centerline{ATV+NN}
	
	\vspace{1mm}
	%1st line
	\begin{minipage}{0.02\hsize}
		\rotatebox[origin=c]{90}{\scriptsize{Estimated IR video}}
	\end{minipage}
	\begin{minipage}{0.18\hsize}
		\centerline{\includegraphics[width=\hsize]{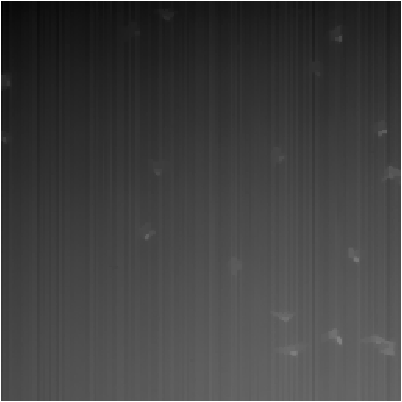}}
	\end{minipage}
	\begin{minipage}{0.18\hsize}
		\centerline{\includegraphics[width=\hsize]{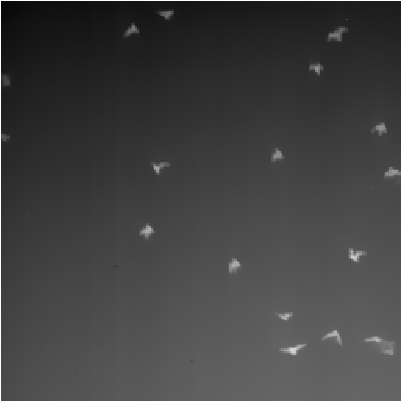}}
	\end{minipage}
	\begin{minipage}{0.18\hsize}
		\centerline{\includegraphics[width=\hsize]{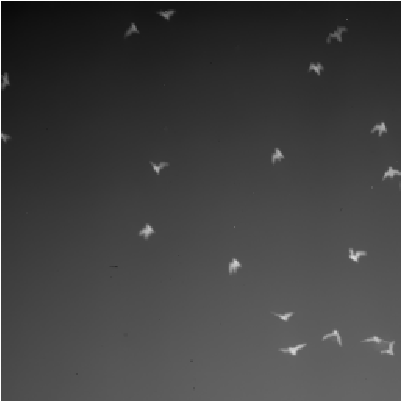}}
	\end{minipage}
	\begin{minipage}{0.18\hsize}
		\centerline{\includegraphics[width=\hsize]{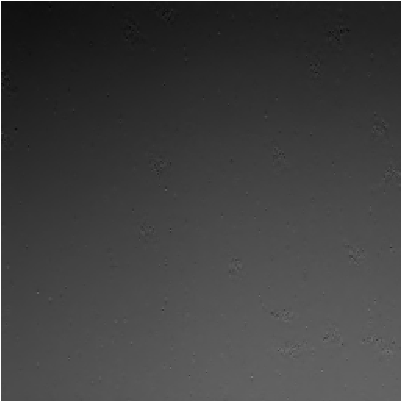}}
	\end{minipage}
	\begin{minipage}{0.18\hsize}
		\centerline{\includegraphics[width=\hsize]{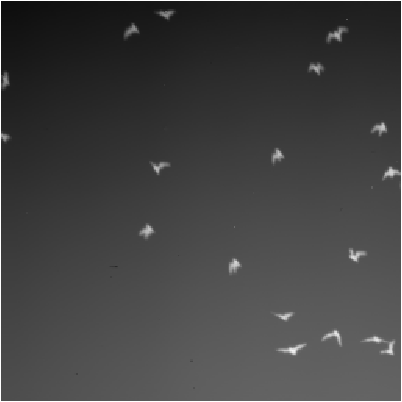}}
	\end{minipage}

	\begin{minipage}{0.02\hsize}
		\rotatebox[origin=c]{90}{\scriptsize{Stripe noise}}
	\end{minipage}
	\begin{minipage}{0.18\hsize}
		\centerline{\includegraphics[width=\hsize]{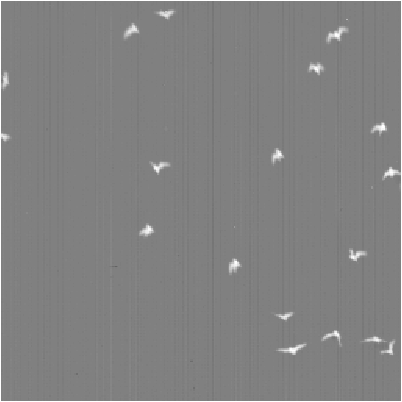}}
	\end{minipage}
	\begin{minipage}{0.18\hsize}
		\centerline{\includegraphics[width=\hsize]{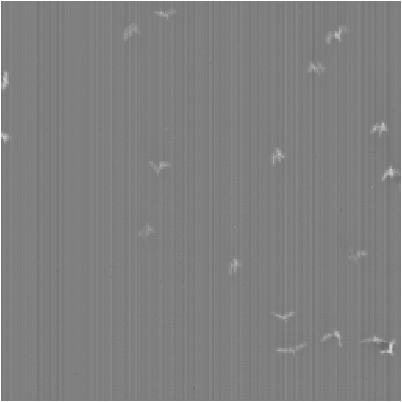}}
	\end{minipage}
	\begin{minipage}{0.18\hsize}
		\centerline{\includegraphics[width=\hsize]{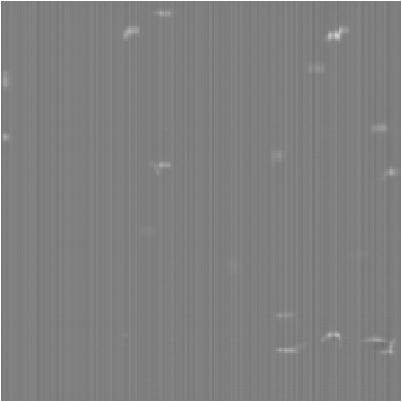}}
	\end{minipage}
	\begin{minipage}{0.18\hsize}
		\centerline{\includegraphics[width=\hsize]{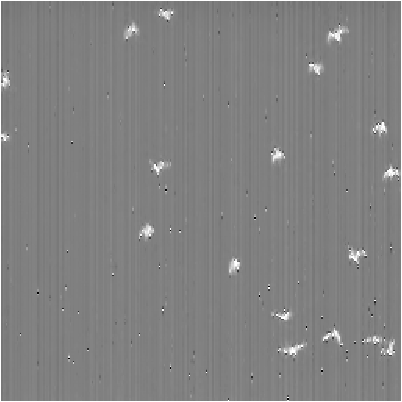}}
	\end{minipage}
	\begin{minipage}{0.18\hsize}
		\centerline{\includegraphics[width=\hsize]{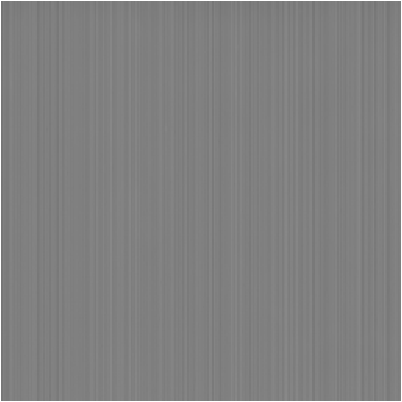}}
	\end{minipage}

	\vspace{1mm}
	
	\begin{minipage}{0.02\hsize} %この一つのminipageが一つの部屋って感じで，これをまず配置する
		~
	\end{minipage}
	\begin{minipage}{0.18\hsize}
		\centerline{\footnotesize{(a3) S~\cite{LRMR}}}
	\end{minipage}
	\begin{minipage}{0.18\hsize}
		\centerline{\footnotesize{(b3) GS~\cite{GLSSTV}}}
	\end{minipage}
	\begin{minipage}{0.18\hsize}
		\centerline{\footnotesize{(c3) LR~\cite{NN_char}}}
	\end{minipage}
	\begin{minipage}{0.18\hsize}
		\centerline{\footnotesize{(d3) TV~\cite{gradient_constraint}}}
	\end{minipage}
	\begin{minipage}{0.18\hsize}
		\centerline{\footnotesize{(e3) FC}}
	\end{minipage}
	
	\caption{IR video destriping results in real noise cases.}
	\label{fig:IR_video_destriping_results_on_real_data}
\end{figure}

Figure~\ref{fig:IR_video_destriping_results_on_real_data} shows the destriping results of the IR video \textit{Bats2}. S and TV removed bats (moving objects) as stripe noise. This is because the stripe noise components (Figs.~\ref{fig:IR_video_destriping_results_on_real_data} (a1), (a2), (a3), (d1), (d2), and (d3)) have sparse or vertical smoothness properties. GS and LR performed better than S and TV, but some of bats were regarded as stripe noise components (see Figs.~\ref{fig:IR_video_destriping_results_on_real_data} (b1), (b2), (b3),  (c1), (c2), and (c3)). In contrast to these stripe noise characterizations, our FC, when combined with any of the image regularizations, removed only the stripe noise while maintaining bats (see Figs.~\ref{fig:IR_video_destriping_results_on_real_data} (e1), (e2), and (e3)).

\begin{figure}[t]
	\centering
	\begin{minipage}{0.25\hsize}
		\centerline{\includegraphics[width=\hsize]{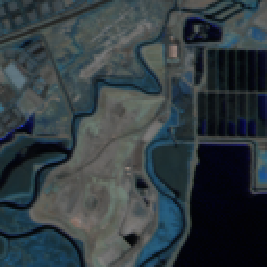}}
	\end{minipage}
	\begin{minipage}{0.25\hsize}
		\centerline{\includegraphics[width=\hsize]{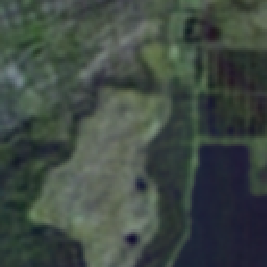}}
	\end{minipage}
	\begin{minipage}{0.25\hsize}
		\centerline{\includegraphics[width=\hsize]{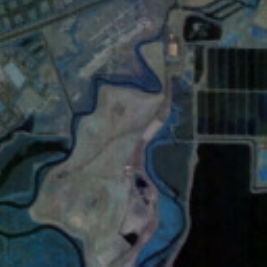}}
	\end{minipage}

	\vspace{1mm}
	
	\begin{minipage}{0.25\hsize}
		\centerline{(a)}
	\end{minipage}
	\begin{minipage}{0.25\hsize}
		\centerline{(b)}
	\end{minipage}
	\begin{minipage}{0.25\hsize}
		\centerline{(c)}
	\end{minipage}

	\begin{minipage}{0.25\hsize}
		\centerline{\small{(MPSNR, MSSIM)}}
	\end{minipage}
	\begin{minipage}{0.25\hsize}
		\centerline{\small{(20.03, 0.5271)}}
	\end{minipage}
	\begin{minipage}{0.25\hsize}
		\centerline{\small{(35.80, 0.8972)}}
	\end{minipage}

	\vspace{1mm}

	\begin{minipage}{0.25\hsize}
		\centerline{\includegraphics[width=\hsize]{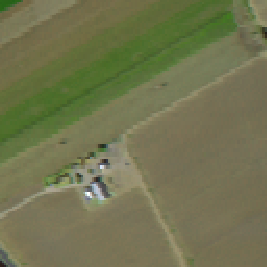}}
	\end{minipage}
	\begin{minipage}{0.25\hsize}
		\centerline{\includegraphics[width=\hsize]{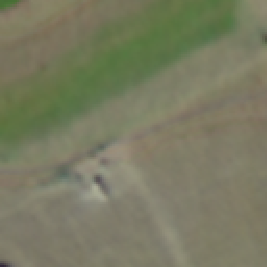}}
	\end{minipage}
	\begin{minipage}{0.25\hsize}
		\centerline{\includegraphics[width=\hsize]{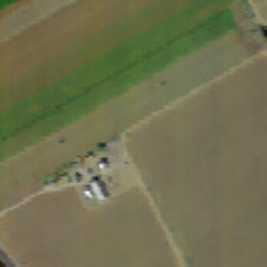}}
	\end{minipage}
	
	\vspace{1mm}

	\begin{minipage}{0.25\hsize}
		\centerline{(d)}
	\end{minipage}
	\begin{minipage}{0.25\hsize}
		\centerline{(e)}
	\end{minipage}
	\begin{minipage}{0.25\hsize}
		\centerline{(f)}
	\end{minipage}

	\begin{minipage}{0.25\hsize}
		\centerline{\small{(MPSNR, MSSIM)}}
	\end{minipage}
	\begin{minipage}{0.25\hsize}
		\centerline{\small{(23.71, 0.8059)}}
	\end{minipage}
	\begin{minipage}{0.25\hsize}
		\centerline{\small{(36.92, 0.9474)}}
	\end{minipage}
	
	\vspace{1mm}
	
	\caption{Comparison with a deep learning-based method~\cite{Sidorov_2019_ICCV}. (a) and (d) Ground-truth images of the \textit{Moffett Field} and \textit{Salinas}, respectively. (b) and (e) Denoising results of~\cite{Sidorov_2019_ICCV}. (c) and (f) are denoising results of our framework (FC-$l_{0}$-$l_{1}$HTV).}
	\label{fig:comparison_with_deep}
\end{figure}

\subsection{Comparison With A Deep Learning-Based Method}
We compare our framework with a deep learning-based method~\cite{Sidorov_2019_ICCV} \footnote{The code is available at \url{https://github.com/acecreamu/deep-hs-prior}.}, where we adjust the parameter so as to achieve the best MPSNR. As observed images, the \textit{Moffett Field} and \textit{Salinas} degraded by stripe noise with $[-0.3, 0.3]$ and Gaussian noise with $\sigma = 0.05$ are used. Figure~\ref{fig:comparison_with_deep} shows the destriping results, which validate the effectiveness of our framework compared to a deep learning-based method. The method in~\cite{Sidorov_2019_ICCV} did not recover edges and objects (Figs.~\ref{fig:comparison_with_deep} (b) and (e)), leading to worse MPSNRs and MSSIMs. This is due to the limitation of deep learning-based methods in capturing textures and singular features, as also mentioned in~\cite{P_Liu_2019,T_X_Jiang_2022}.

\subsection{Discussion}
From the above experiments, we summarize the advantages and limitations of our framework as follows:
\begin{itemize}
	\item FC accurately captures various intensities of stripe noise for any target images without image components.
	\item In particular, FC eliminates high intensities of stripe noise.
	\item Our framework consistently removes stripe noise, whatever image regularizations are combined.
	\item When using some image regularization such as ASSTV and SSTV+TNN, our framework requires many iterations to converge.
\end{itemize}

\section{Conclusion}
\label{sec:Conclusion}
In this paper, we have proposed a general destriping framework for remote sensing images. Specifically, we formulated the destriping as a convex optimization problem equipped with the flatness constraint. Thanks to the strong characterization of stripe noise, our framework is compatible with various regularization functions and achieves effective destriping. Then, we develop a solver for the problem based on DP-PDS, which allows us to avoid stepsize adjustment. Through destriping experiments using HSI and IR video data, we found that our framework is advantageous on average compared to existing methods, whatever image regularizations are used. For future work, our framework needs the extension to consider the various degradation such as the spectral variability and the effectiveness demonstration in remote sensing image applications such as classification, unmixing, compressed sensing reconstruction, and target recognition.

\appendices
\section{Convergence of DP-PDS}
\label{app:convergence_of_DP_PDS}

Consider a convex optimization problem of the following form:
\begin{equation}
	\label{eq:convex_optimization}
	\min_{\mathcal{Z},\mathcal{Y}}f_{1}(\mathcal{Z})+f_{2}(\mathcal{Y})\quad\mathrm{s.t.}\quad\mathcal{Y}=\mathfrak{K}(\mathcal{Z}),
\end{equation}
where $\mathcal{Z}=(\mathcal{Z}_{1},\cdots,\mathcal{Z}_{N_{0}})\in\prod_{i=1}^{N_{0}}\mathbb{R}^{n_{i,1}\times \cdots \times n_{i,N_{i}}}$ and $\mathcal{Y}=(\mathcal{Y}_{1},\cdots,\mathcal{Y}_{M_{0}})\in\prod_{i=1}^{M_{0}}\mathbb{R}^{m_{i,1}\times \cdots \times m_{i,M_{i}}}$ are variables that include $N_{0}$ tensors and $M_{0}$ tensors, respectively, $f_{1}$ and $f_{2}$ are proper lower semi-continuous convex functions, and $\mathfrak{K}$ is a linear operator.

We consider the following iterative procedures:
\begin{equation}
	\label{eq:update_step}
	\begin{array}{l}
		\mathcal{Z}^{(n+1)} \leftarrow \mathrm{prox}_{\mathcal{G}_{1}^{-1},f_{1}}\left(\mathcal{Z}^{(n)}-\mathcal{G}_{1}\odot\mathfrak{K}^{*}(\mathcal{Y}^{(n)})\right), \\
		\mathcal{Y}^{(n+1)} \leftarrow \mathrm{prox}_{\mathcal{G}_{2}^{-1},f_{2}^{*}}\left(\mathcal{Y}^{(n)}+\mathcal{G}_{2}\odot\mathfrak{K}(2\mathcal{Z}^{(n+1)}-\mathcal{Z}^{(n)})\right),
	\end{array}
\end{equation}
where $f_{2}^{*}$ is the Fenchel--Rockafellar conjugate function of $f_{2}$, and $\mathcal{G}_{1}=(\mathcal{G}_{1,1}, \cdots, \mathcal{G}_{1,N_{0}})\in\prod_{i=1}^{N_{0}}\mathbb{R}_{++}^{n_{i,1}\times \cdots \times n_{i,N_{i}}}$ and $\mathcal{G}_{2}=(\mathcal{G}_{2,1}, \cdots, \mathcal{G}_{2,M_{0}})\in\prod_{i=1}^{M_{0}}\mathbb{R}_{++}^{m_{i,1}\times \cdots \times m_{i,M_{i}}}$ are preconditioners. For any $\mathcal{Z}^{(0)}\in\prod_{i=1}^{N_{0}}\mathbb{R}^{n_{i,1}\times \cdots \times n_{i,N_{i}}}$ and $\mathcal{Y}^{(0)}\in\prod_{i=1}^{M_{0}}\mathbb{R}^{m_{i,1}\times \cdots \times m_{i,M_{i}}}$, the sequence generated by~\eqref{eq:update_step} converges to the optimal solution of Prob.~\eqref{eq:convex_optimization} if the linear operator $\mathfrak{K}$ and preconditioners $\mathcal{G}_{1}, \mathcal{G}_{2}$ satisfy the following condition~\cite[Lemma 1]{DP-PDS}: for any $\mathcal{X}(\neq\mathcal{O})\in\Pi_{i=0}^{N_{0}}\mathbb{R}^{n_{i,1}\times\cdots\times n_{i,N_{i}}}$
\begin{equation}
	\label{eq:stepsize_condition}
	\|\mathcal{G}_{2}\odot\mathfrak{K}(\mathcal{G}_{1}\odot\mathcal{X})\|_{F}<\|\mathcal{X}\|_{F}.
\end{equation}
Note that matrix-vector multiplication between a diagonal matrix and a vector is equivalent to tensor-tensor Hadamard product. Therefore, Eq.~\eqref{eq:update_step} is identical to the algorithm described in~\cite{DP-PDS}.

DP-PDS sets $\mathcal{G}_{1}$ and $\mathcal{G}_{2}$ as follows.
Since $\mathfrak{K}^{*}$ is a linear operator, the $(i_{1},\cdots,i_{N_{i}})$th entry of $\mathcal{Z}_{i}$ is yielded by linear combinations of $\mathcal{Y}$ as follows:
\begin{align}
	& \mathcal{Z}_{i}(i_{1},\cdots,i_{N_{i}}) \nonumber \\
	& = \sum_{j}\sum_{j_{j,1},\cdots,j_{j,M_{j}}}k_{j,j_{j,1},\cdots,j_{j,M_{j}}}^{\prime}*\mathcal{Y}_{j}(j_{j,1},\cdots,j_{j,M_{j}}).
\end{align}
Then, the $(i_{1},\cdots,i_{N_{i}})$th entry of $\mathcal{G}_{1,i}$ is given as
\begin{equation}
	\label{eq:def_preconditioner1}
	\mathcal{G}_{1,i}(i_{1},\cdots,i_{N_{i}})= \frac{1}{\sum_{j}\sum_{j_{j,1},\cdots,j_{j,M_{j}}}|k_{j,j_{j,1},\cdots,j_{j,M_{j}}}^{\prime}|}.
\end{equation}
Similarly, the $(i_{1},\cdots,i_{M_{i}})$th entry of $\mathcal{Y}_{i}$ is given as
\begin{align}
	& \mathcal{Y}_{i}(i_{1},\cdots,i_{M_{i}}) \nonumber \\
	& =  \sum_{j}\sum_{j_{j,1},\cdots,j_{j,N_{j}}}k_{j,j_{j,1},\cdots,j_{j,N_{j}}}*\mathcal{Z}_{j}(j_{j,1},\cdots,j_{j,N_{j}}).
\end{align}
Then, the $(i_{1},\cdots,i_{M_{i}})$th entry of $\mathcal{G}_{2,i}$ is given as
\begin{equation}
	\label{eq:def_preconditioner2}
	\mathcal{G}_{2,i}(i_{1},\cdots,i_{M_{i}})= \frac{1}{\sum_{j}\sum_{j_{j,1},\cdots,j_{j,N_{j}}}|k_{j,j_{j,1},\cdots,j_{j,N_{j}}}|}.
\end{equation}
These preconditioners $\mathcal{G}_{1}$ and $\mathcal{G}_{2}$ satisfy the condition in~\cite[Lemma 2]{DP-PDS}, i.e., Eq.~\eqref{eq:update_step} computes the solution of Prob.~\eqref{eq:convex_optimization}.

\section{Convergence of our Algorithm}
\label{app:convergence_of_our_algorithm}
Let $\mathcal{Z}=(\mathcal{U},\mathcal{S})$ and $\mathcal{Y}  =  (\mathcal{Y}_{1,1},\ldots,\mathcal{Y}_{1,K},\mathcal{Y}_{2},\mathcal{Y}_{3},\mathcal{Y}_{4})$.
Then, by defining 
%\begin{align}
%	& f_{1}(\mathcal{Z}):=\|\mathcal{S}\|_{1}, \nonumber \\
%	& f_{2}(\mathcal{Y}) := R(\mathcal{Y}_{1}) + \iota_{\{\mathcal{O}\}}(\mathcal{Y}_{2}) + \iota_{\{\mathcal{O}\}}(\mathcal{Y}_{3}) + \iota_{B_{(\mathcal{V},\varepsilon)}}(\mathcal{Y}_{4}), \nonumber \\
%	& \mathfrak{K}(\mathcal{Z}):=(\mathfrak{L}(\mathcal{U}), \mathfrak{D}_{v}(\mathcal{S}), \mathfrak{D}_{t}(\mathcal{S}), \mathcal{U+S}),
%\end{align}
\begin{align}
	& f_{1}(\mathcal{Z}):=\|\mathcal{S}\|_{1}, \nonumber \\
	& f_{2}(\mathcal{Y}) := \sum_{k=1}^{K}R_{k}(\mathcal{Y}_{1,k}) + \iota_{\{\mathcal{O}\}}(\mathcal{Y}_{2}) + \iota_{\{\mathcal{O}\}}(\mathcal{Y}_{3}) + \iota_{B_{(\mathcal{V},\varepsilon)}}(\mathcal{Y}_{4}), \nonumber \\
	& \mathfrak{K}(\mathcal{Z}):=(\mathfrak{L}_{1}(\mathcal{U}), \ldots, \mathfrak{L}_{K}(\mathcal{U}), \mathfrak{D}_{v}(\mathcal{S}), \mathfrak{D}_{t}(\mathcal{S}), \mathcal{U+S}).
\end{align}
Prob.~\eqref{eq:DP_PDS_form_of_zero_gradient_optimization_vt} is reduced to Prob.~\eqref{eq:convex_optimization}, i.e., Prob.~\eqref{eq:DP_PDS_form_of_zero_gradient_optimization_vt} is a special case of Prob.~\eqref{eq:convex_optimization}. Therefore, our algorithm satisfies the convergence property of the original DP-PDS.

% or


%\appendices

%\section{}

%\section*{Acknowledgment}


\ifCLASSOPTIONcaptionsoff
  \newpage
\fi



%\bibliographystyle{IEEEtran}
\bibliographystyle{IEEEtran}

%\bibliography{./ZeroGradientConstraintForDestripingOfRemoteSensingData}


%\begin{thebibliography}{1}

%\end{thebibliography}


\begin{IEEEbiography}[{\includegraphics[width=1in,height=1.25in,clip,keepaspectratio]{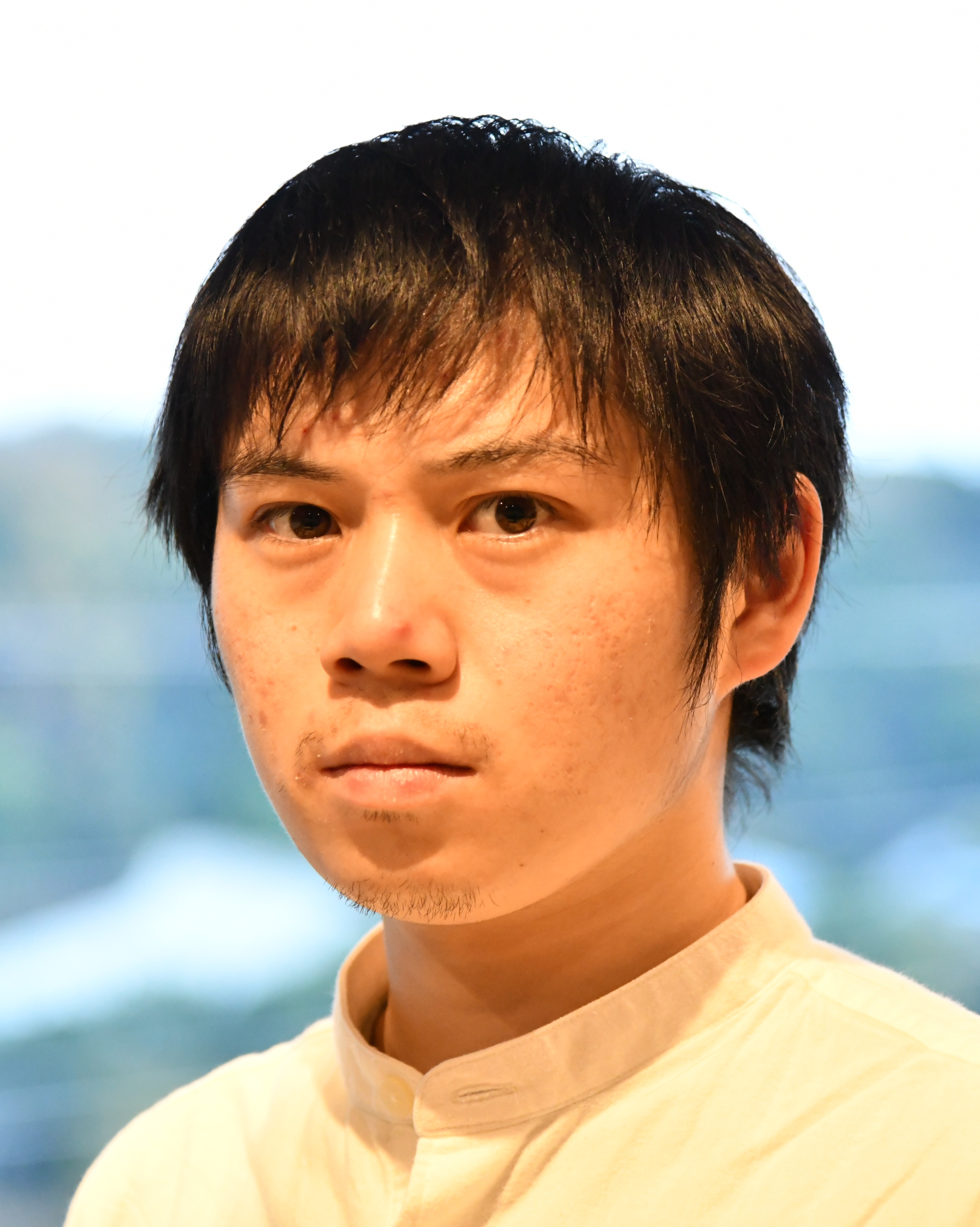}}]{Kazuki Naganuma}
(S’21) received a B.E. degrees in Information and Computer Sciences in 2020 from the Kanagawa Institute of Technology.

He is currently pursuing an M.E. degree at the Department of Computer Science in the Tokyo Institute of Technology. His current research interests are in signal and image processing and optimization theory.
\end{IEEEbiography}

\begin{IEEEbiography}[{\includegraphics[width=1in,height=1.25in,clip,keepaspectratio]{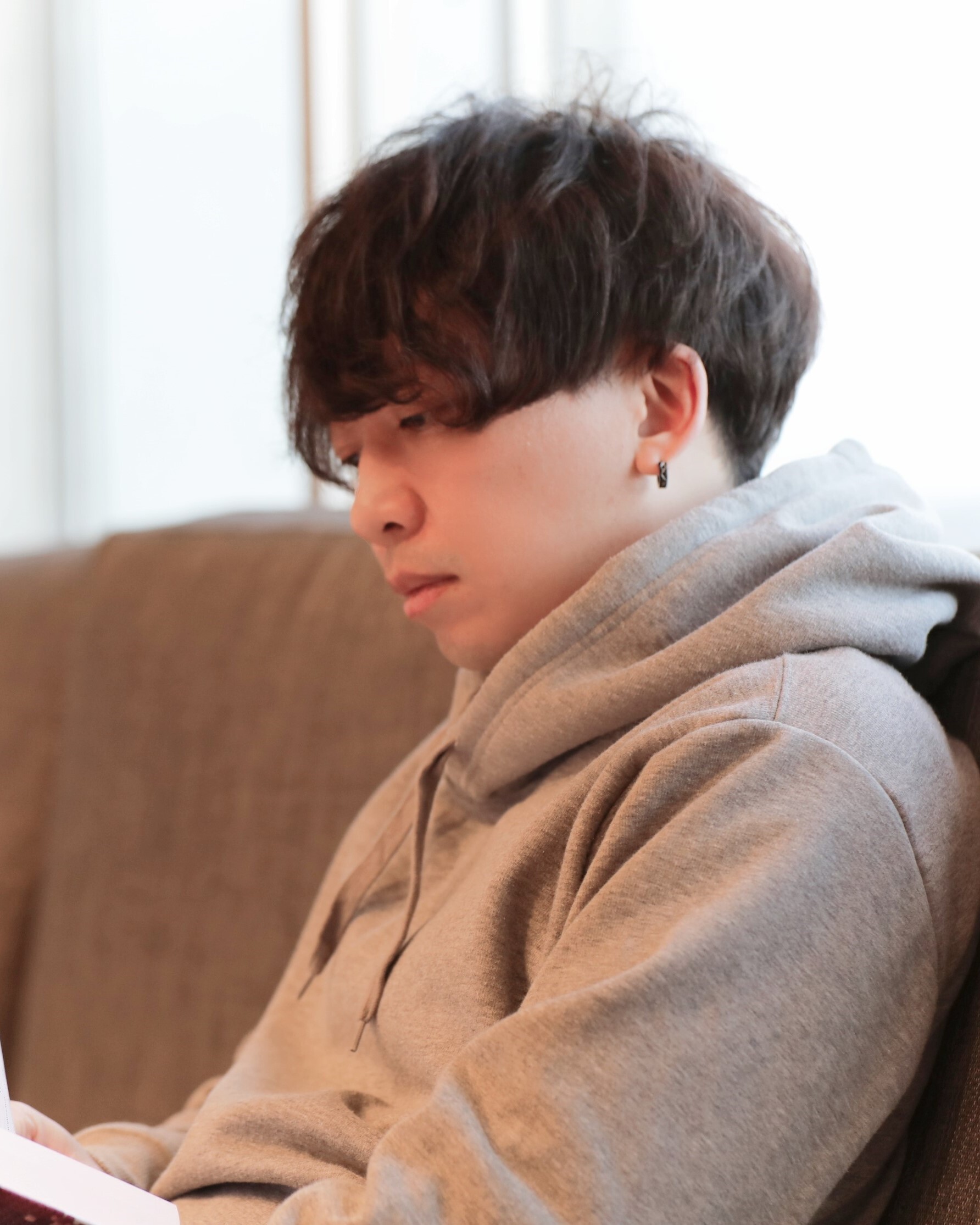}}]{Shunsuke Ono}
(S'11--M'15) received a B.E. degree in Computer Science in 2010 and M.E. and Ph.D. degrees in communications and computer engineering in 2012 and 2014 from the Tokyo Institute of Technology, respectively.

From April 2012 to September 2014, he was a Research Fellow (DC1) of the Japan Society for the Promotion of Science (JSPS). He is currently an Associate Professor in the Department of Computer Science, School of Computing, Tokyo Institute of Technology. From October 2016 to March 2020, he was a Researcher of Precursory Research for Embryonic Science and Technology (PRESTO), Japan Science and Technology Corporation (JST), Tokyo, Japan. His research interests include signal processing, computational imaging, hyperspectral imaging and fusion, mathematical optimization, and data science.

Dr. Ono received the Young Researchers' Award and the Excellent Paper Award from the IEICE in 2013 and 2014, respectively, the Outstanding Student Journal Paper Award and the Young Author Best Paper Award from the IEEE SPS Japan Chapter in 2014 and 2020, respectively, and the Funai Research Award from the Funai Foundation in 2017. He has been an Associate Editor of IEEE TRANSACTIONS ON SIGNAL AND INFORMATION PROCESSING OVER NETWORKS since 2019.
\end{IEEEbiography}

%\newpage


%\vfill


\end{document}